\documentclass[aps,prx,twocolumn,groupedaddress,showpacs,preprintnumbers,amsmath]{revtex4-1}
\usepackage{epsfig,amsmath,amssymb,graphics,color,calc,hyperref}
\usepackage{mathtools}

\newcommand{\ErRe}{Erd\"os-R\'enyi~}
\newcommand{\ER}{ER~}

\begin{document}

\title{Fully localized and partially delocalized states in the tails of \ErRe graphs in the critical regime}

\author{M. Tarzia\textsuperscript{1,2}}

\affiliation{
	\textsuperscript{1} \mbox{LPTMC, CNRS-UMR 7600, Sorbonne Universit\'e, 4 Pl. Jussieu, F-75005 Paris, France}
	\textsuperscript{2} \mbox{Institut  Universitaire  de  France,  1  rue  Descartes,  75231  Paris  Cedex  05,  France}}

\begin{abstract}
In this work we study the spectral properties of the adjacency matrix of critical \ErRe (ER) graphs, i.e. when the average degree is of order $\log N$. In a series of recent inspiring papers Alt, Ducatez, and Knowles have rigorously shown that these systems exhibit a ``semilocalized'' phase in the tails of the spectrum where the eigenvectors are exponentially localized on a sub-extensive set of nodes with anomalously large degree. We  propose two approximate analytical strategies to analyze this regime based respectively on the simple ``rules of thumb'' for localization and ergodicity and on an approximate treatment of the self-consistent cavity equation for the resolvent. Both approaches suggest the existence of two different regimes: a fully Anderson localized phase at the spectral edges, in which the eigenvectors are localized around a unique center, and an intermediate partially delocalized but non-ergodic phase, where the eigenvectors spread over many resonant localization centers. In this phase the exponential decay of the effective tunneling amplitudes between the localization centers is counterbalanced by the large number of nodes towards which tunneling can occur, and the system exhibits mini-bands in the local spectrum over which the Wigner-Dyson statistics establishes. We complement these results by 
a detailed numerical study of the finite-size scaling behavior of several observables that supports the theoretical predictions and allows us to determine the critical properties of the two transitions. Critical \ER graphs provide a pictorial representation of the Hilbert space of a generic many-body Hamiltonian with short range interactions. In particular we argue that their phase diagram can be mapped onto the out-of-equilibrium phase diagram of the quantum random energy model.
\end{abstract}


\maketitle

\section{Introduction}
Since Anderson's celebrated discovery of localization more than sixty years ago~\cite{Anderson}, a huge amount of work has been devoted to the study of  transport and  spectral properties of quantum particles in random environments~\cite{50years,lee,evers08}. These investigations have deeply influenced the development of many areas of condensed matter physics, such as transport in disordered quantum systems, random matrices, and quantum chaos, just to name a few, and are still in the focus of current research, continuing to reveal new facets and subtleties.

In this context, the study of Anderson localization (AL) on sparse random graphs has attracted a strong and renewed interest in the last few years: On the one hand these tree-like structures, which correspond to the infinite dimensional limit of the tight-binding model, allow in principle for an exact solution, making it possible to establish the transition point and the corresponding critical behavior~\cite{abou,efetov,efetov1,mirlin_fyodorov,fyodorov_mirlin,fyodorov_mirlin_sommers,fyod,mirlin1994,Zirn,tikhonov2019,ourselves,aizenmann,Verb}. On the other hand, the spectral properties of (weighted) adjacency matrices of sparse graphs encode the structural and topological features of many physical systems.  


AL on sparse random graphs has been first studied by Abou Chacra, Anderson and Thouless~\cite{abou}, and then by many others~\cite{mirlin_fyodorov,fyodorov_mirlin,fyodorov_mirlin_sommers,fyod,mirlin1994,Zirn,tikhonov2019,ourselves,aizenmann,Verb}. Most of these works focused on the localization transition induced by the random potential. In a series of recent inspiring works~\cite{Knowles,KnowlesL,Knowles1,KnowlesD,KnowlesE}, Alt, Ducatez, and Knowles studied instead the case in which localization is induced by the topology of the graph, and in particular by the strong fluctuations of the local connectivity. In particular, Alt {\it et al} 
studied the spectral properties of the adjacency matrix of \ErRe (ER) graphs in the critical regime (i.e., when the average degree is of order of the logarithm of the number of vertices) in absence of disorder in the local potential. In~\cite{Knowles} the authors first showed that the spectrum of these systems splits into (at least) two phases separated by a sharp transition transition: a fully GOE-like delocalized phase in the bulk of the spectrum, where the eigenvectors are completely delocalized~\cite{KnowlesD}, and a ``semilocalized'' phase near the edges of the spectrum, where the wave-functions are exponentially localized on a sub-extensive number of vertices of anomalously large degree. In a subsequent paper~\cite{KnowlesL} the same authors went a step further and proved the existence of a fully localized phase near the spectral edges.

These findings are particularly interesting at least for two reasons: First,  \ER graphs in the critical regime provide a natural representation of the topological features of the Hilbert space of generic interacting Hamiltonians with finite-range interactions~\cite{A97}. Specifically, basis states of a many-body system chosen as eigenstates of the non-interacting part of the Hamiltonian (which can be straightforwardly diagonalized) correspond to vertices (or site orbitals) of the sparse graph, while interaction-induced couplings between them gives rise to the links between the nodes. Take for concreteness a quantum spin-$1/2$ chain of $n$ spins with nearest neighbor interactions. By choosing as a basis of the Hilbert space the simultaneous eigenstates of the operators $\sigma_i^z$, the Hilbert space results in a $n$-dimensional hypercube of $N = 2^n$ sites (in absence of any symmetry on the global magnetization). Each configuration of $n$ spins corresponds to a corner of the hypercube by considering $\{ \sigma_i^z = \pm 1 \}$ as the top/bottom face of the cube's $n$-th dimension. The interacting part of the Hamiltonian, e.g. of the form of a transverse field $\Gamma \sum_i \sigma_i^x$, acts as single spin flips on the configurations $\{ \sigma_i^z \}$, and plays the role the hopping rates connecting ``neighboring'' sites in the configuration space. The quantum many-body dynamics can thereby be seen as single-particle diffusion on a very high-dimensional graph with a average degree equal to $n = \log_2 N$. Based on this analogy, for instance, it has been argued that AL on sparse random graphs offers a paradigmatic and intuitive representation of the so-called Many Body Localization (MBL) transition~\cite{BAA}. In fact, during the last 15 years it was indeed established that quantum systems of interacting particles subject to sufficiently strong disorder will fail to come to thermal equilibrium when they are not coupled to an external bath even though prepared with extensive amounts of energy above their ground states~\cite{Gornyi2005,Altman2015Review,Nandkishore2015,AbaninPapic2017,AletLaflorencie2018,Abanin2019RMP}. To the extent that one of the most successful theories of physics, namely thermodynamics, is founded on the assumption of ergodicity, it is evident that whether or not many-body quantum systems constitute a heat bath for themselves, and hence are able to thermalize, is a very fundamental question. The analogy of this problem with single-particle AL was put forward in the seminal work of~\cite{A97}, where the decay of a hot quasiparticle in a quantum dot (at zero temperature) was mapped onto an appropriate non-interacting tight binding model on a disordered tree-like graph, and then further analyzed by later works in a more general context~\cite{A97,BAA,Gornyi2005,jacquod,scardicchioMB,roylogan,mirlinreview}. In this respect, a deep understanding of the spectral properties of critical \ER graphs could give useful insight to make sense of more complex problems. In particular, below we will put forward a direct analogy between the phase diagram of  critical \ER graphs and the out-of-equilibrium phase diagram of the Quantum Random Energy Model (QREM), which is the simplest toy model featuring a many-body localized phase~\cite{qrem1,qrem2,qrem3,qrem4,qrem5}.

The second reason is that the appearance of states which are neither fully localized nor fully ergodic and occupy a sub-extensive part of the whole accessible volume has emerged as a fundamental property of many physical problems, including Anderson~\cite{wegner,noiCT,mirlinCT} and many-body localization~\cite{mace,alet,laflorencie,war,resonances1,resonances2,tarzia,ros,deluca,serbyn,luitz,qrem1,qrem2,qrem3,qrem5}, random matrix theory~\cite{kravtsov,kravtsov1,khay,monthus-LRP,LRP,barlev,dynLNRP,floquet1,floquet2,floquet3,floquet4,pwave,nosov,duthie,kutlin,motamarri,tang}, Josephson junction chains~\cite{jj}, quantum information~\cite{boixo,qrem4}, and even quantum gravity~\cite{syk}. Simple solvable dense random matrix models with independent and identically distributed (iid) entries, such as the the paradigmatic Gaussian Rosenzweig-Porter (RP) model~\cite{kravtsov} and its generalizations~\cite{kravtsov1,khay,monthus-LRP,LRP,barlev,dynLNRP} feature the appearance of fractal wave-functions in an intermediate region of the phase diagram sandwiched between the fully ergodic and the fully localized phases. In these models, which have been intensively investigated over the past few years~\cite{warzel,facoetti,bogomolny,bera,pino,truong,amini,berkovits}, every site of the reference space, represented by a matrix index, is connected to every other site with the transition amplitude distributed according to some probability law. In the latest years other class of random matrix models featuring multifractal phases have emerged: These are one-dimensional systems with quasiperiodic potential in presence of a periodic drive~\cite{floquet1,floquet2,floquet3,floquet4} as well as in the static setting with $p$-wave superconducting order~\cite{pwave}, and one-dimensional power-law random
banded matrix models with strongly correlated translation-invariant long-range hopping~\cite{nosov,floquet4,duthie,kutlin,motamarri,tang}. In this context, \ER graphs in the critical regime could provide yet another mechanism responsible for the appearance of partially delocalized but non-ergodic states which complement the physical pictures provided by the families of models described above.

In this paper we investigate 
the spectral properties of the adjacency matrix of critical \ER graphs  using both numerical methods and analytical arguments. The two main questions that we address are: (i) What are the critical properties of the transition between the fully delocalized GOE-like phase in the bulk of the spectrum and the semilocalized phase near the spectral edges highlighted in Refs.~\cite{Knowles,KnowlesL}? How does the critical behavior compare to the one corresponding to standard AL on sparse matrices induced by the randomness of the local potential? (ii) What are the spectral properties of such semilocalized phase? Is there a region of the phase diagram where eigenvectors localized around far away rare localization centers hybridize due to the exponentially small effective matrix elements between them? In order to address these questions we apply simple rules of thumb for localization and ergodicity and put forward an approximate treatment of the self-consistent cavity equations for the resolvent. These approaches provide a rough estimation of the phase diagram of the model. Our analysis suggests that the tails of the spectrum split in two phases separated by a mobility edge which separates fully localized eigenstates at the spectral edges (whose existence has been already rigorously proven in~\cite{KnowlesL}), from an intermediate partially delocalized but non-ergodic phase in which the wave-functions hybridize (at least partially) around many resonating localization centers. In this region the exponentially decaying tunnelling amplitudes between localization centres are counterbalanced by an the large number of possible localization centers towards which  tunnelling can occur. We complement this analysis by extensive numerical calculations showing that the finite-size scaling behavior of several observables related to the statistics of the gaps and of the wave-functions' amplitudes fully support the validity 
of the theoretical results and allow one to determine the critical properties of the transitions.

The paper is organized as follows: In the next section we define the adjacency matrix of \ER graphs and provide a brief historical perspective on their study; In Sec.~\ref{sec:semi} we review the recent exact results of Alt, Ducatez, and Knowles~\cite{Knowles,KnowlesL,Knowles1} on the semilocalized phase that emerges in the critical regime. 
In Sec.~\ref{sec:PD} we discuss the phase diagram of the model using two complementary analytical approaches; In Sec.~\ref{sec:numerics} we present several numerical results on the finite-size behavior of several observables related to the statistics of the energy gaps and of the wave-functions' amplitudes and discuss the critical properties of the transitions between the different phases; In Sec.~\ref{sec:lmax} we characterize the statistics of the fluctuations of the largest eigenvalue of the spectrum; In Sec.~\ref{sec:qrem} we propose a mapping between critical \ER graphs and the out-of-equilibrium phase diagram of the QREM; Finally, in Sec.~\ref{sec:conclusions} we present some concluding remarks and perspectives for future investigations. In the Appendix section we  present  some  supplementary  information  that  complement the results discussed in the main text.

\section{The model}
The adjacency matrix of \ER graphs is a real, symmetric $N \times N$ matrix ${\cal H}$ whose elements ${\cal H}_{ij}$ are (up to the symmetry ${\cal H}_{ij} = {\cal H}_{ji}$) iid 
random variables, with a Bernoulli probability distribution
\begin{equation} \label{eq:H}
p({\cal H}_{ij}) = \left(1 - \frac{c}{N} \right ) \delta ({\cal H}_{ij}) + \frac{c}{N} \delta \! \left ( {\cal H}_{ij} - \frac{1}{\sqrt{c}} \right) 
\end{equation}
for $i \neq j$ and ${\cal H}_{ii} = 0$. (The off-diagonal elements are rescaled by $\sqrt{c}$ in order to have eigenvalues of order 1 for $N\gg 1$.) In the thermodynamic limit and for $c \ll N$, the degree of a given node, $k_i = \sqrt{c} \sum_j {\cal H}_{ij}$, is a random variable which follow a Poisson distribution $P(k) = e^{-c} c^k/k!$ of average $\langle k \rangle = c$ and variance $\langle k^2 \rangle - \langle k \rangle = c$. 

The adjacency matrices of sparse random graphs encode the structural and topological features of many complex systems~\cite{albert}. For instance, for random walks on graphs, the eigenvalue spectrum is directly connected to the relaxation time spectrum~\cite{lovasz}. From the condensed matter perspective, 
the spectra of such matrices have been used for the characterisation of many physical systems such as the study of gelation transition in polymers~\cite{broderix} and of the instantaneous normal modes in supercooled liquids~\cite{cavagna}. 

\ER graphs undergo a dramatic change in behaviour at the critical scale $c \sim \log N$, which is the scale at and below which the vertex degrees do not concentrate: For $c \gg \log N$, all degrees are approximately equal and the graph is homogeneous. In this regime \ER graphs share the spectral properties of the GOE ensemble and the density of states (DoS) is given by the semicircle law. On the other hand, for $c \lesssim \log N$ the degrees do not concentrate and the graph becomes highly inhomogeneous: it contains nodes of exceptionally large degree, leaves (i.e. nodes of degree 1), and isolated vertices (i.e. nodes of degree 0). As long as $c > 1$, the graph has a unique giant component.

Historically, the study of the spectrum of sparse symmetric \ER adjacency matrices was pioneered by Bray and Rodgers in~\cite{rodgers88} (and in a similar context in~\cite{bray88} and later on in~\cite{rodgers05}) using the Edwards-Jones recipe. In their formulation the evaluation of the average DoS $\rho(\lambda)$ relies on the replica method, which yield a very complicated integral equation. The same integral equation has been derived independently with a supersymmetric approach in~\cite{fyodorov91} and later obtained in a rigorous manner in~\cite{khorunzhy04}. A variety of approximation schemes~\cite{kuhn08}, such as the single defect approximation (SDA)~\cite{biroli99} and the effective medium approximation (EMA)~\cite{semerjian02}, were proposed to deal with the difficulty of solving the exact integral equation for the DoS. Almost in parallel, the cavity method~\cite{mezard} (see below) started to be employed for the determination of the spectral density of \ER graphs~\cite{rodgers2008}. A nice recent review of these studies can be found in Ref.~\cite{vivo}

In this paper we will focus on \ER graphs in the critical regime, which, as discussed above, are particularly relevant as they represent a toy model for the Hilbert space of interacting Hamiltonian with finite-range interactions. Hence throughout the rest of the paper we will set $c = b \log N$. Most of the numerical results presented below are obtained for $b=0.5$.

\section{The semilocalized phase}  \label{sec:semi}

In their recent insightful work~\cite{Knowles}, Alt, Ducatez, and Knowles showed that the spectrum of \ER graphs in the critical regime splits into (at least) two phases separated by a sharp transition at $\lambda = \lambda_{\rm GOE} = 2$: a GOE-like phase in the middle of the spectrum, $\lambda \in [-2,2]$, where the eigenvectors are completely delocalized~\cite{KnowlesD}, and a ``semilocalized'' phase near the edges of the spectrum, $\lambda \in (-\lambda_{\rm max},-2) \cup (2, \lambda_{\rm max})$, where the eigenvectors are essentially localized on a small number of vertices of anomalously large degree. In the semilocalized phase the mass of an eigenvector is concentrated in a small number of disjoint balls centred around resonant vertices, in each of which it is a radial exponentially decaying function. (Throughout the following, we always exclude the largest 
eigenvalue of ${\cal H}$ associated to the flat eigenvector $1/\sqrt{N}(1, \ldots, 1)$, which is an outlier separated from the rest of the spectrum, see e.g. Ref.~\cite{ourselves} for more details).

Both phases are amenable to rigorous analysis. The semilocalized phase only exists when $b < b_\star = 1/(2 \log 2 - 1)$, while above $b_\star$ one retrieves the spectral properties of the homogeneous regime. For $b < b_\star$ the average DoS in the interval $\lambda \in (- \lambda_{\rm max},-2) \cup (2, \lambda_{\rm max})$ is given asymptotically by $\rho^{\infty} (\lambda) \propto N^{\tau (\lambda) - 1}$, where $\tau (\lambda)$ is an exponent whose the explicit expression has been obtained rigorously in Ref.~\cite{Knowles}. In particular $\tau (\lambda)$ jumps discontinuously at the transition between the delocalized and the semilocalized phase from $\tau (\lambda) = 1$ for $\lambda \in [-2,2]$ to $\tau  = 1 - b/b_\star$ for $\lambda \to 2^+$ and $\lambda \to -2^-$.

The eigenvalues in the semilocalized phase were already analysed in~\cite{Knowles1} (see also Refs.~\cite{Knowles2,Knowles3,Tikhomirov}), where it was proved that they arise precisely from vertices of abnormally large degree~\cite{sda}. More precisely, it was proved that each vertex with degree $k> 2c$ gives rise to two eigenvalues of ${\cal H}$ near $\pm \Lambda(k/c)$, where
\begin{equation} \label{eq:Lambda}
\Lambda(x) = \frac{x}{\sqrt{x-1}} \, .
\end{equation}
One can rigorously show that the number of those resonant vertices at energy $|\lambda| > 2$ is sub-extensive and asymptotically equal to the number of eigenvalues, i.e. $N \rho^{(\infty)} (\lambda) = N^{\tau (\lambda)}$. In other words there is an approximate bijection between vertices of degree greater than $2 c$ and eigenvalues larger than 2. Hence, in the limit of very large graphs one than has that:
\[
\rho^{\infty} ( \Lambda ( k/c )  ) \left[ \Lambda \! \left( \frac{k}{c} \right) - \Lambda \! \left( \frac{k-1}{c} \right) \right] 
\approx \frac{e^{-c} \, c^k}{2 k!} \, .
\]
After expanding for large $k$ and changing variables $k \to \lambda$ one finds that:
\begin{equation} \label{eq:dos}
\rho^{\infty} (\lambda) \approx \frac{e^{-c} \, c^{1 + c \tilde{\kappa}(\lambda)}}{2 (c \tilde{\kappa} (\lambda))!} \tilde{\kappa}^\prime (\lambda) \, ,
\end{equation}
where 
\[
\begin{aligned}
\tilde{\kappa} (\lambda) & = \frac{\lambda}{2} \left( \lambda + \sqrt{\lambda^2 - 4} \right) \, , \\
\tilde{\kappa}^\prime (\lambda) & = \frac{{\rm d} \tilde{\kappa}}{{\rm d} \lambda} = \frac{\left( \lambda + \sqrt{\lambda^2 - 4} \right)^2}{2 \sqrt{\lambda^2 - 4}} \, .
\end{aligned}
\]
The function $\tilde{\kappa} (\lambda)$ above is the inverse of the function $\Lambda(x)$ given in Eq.~\eqref{eq:Lambda}, and gives the degree $c \tilde{\kappa} (\lambda)$ corresponding to an eigenvalue $\lambda$ in the tails of the spectrum. The exponent $\tau (\lambda)$ is then simply defined as
\begin{equation} \label{eq:tau}
\begin{aligned}
\tau (\lambda) -1 & = \! \lim_{N \to \infty} \! \frac{\log \rho^{\infty} (\lambda)}{\log N} \\
& = -b \left[\tilde{\kappa}(\lambda) \log \tilde{\kappa}(\lambda) - \tilde{\kappa}(\lambda) + 1 \right] \, .
\end{aligned}
\end{equation}
The maximum eigenvalue $\lambda_{\rm max} (b)$ in the thermodynamic limit (which correspond to the largest degree of ${\cal H}$, see Ref.~\cite{KnowlesL} and Sec.~\ref{sec:lmax}) is thus given by the value of $\lambda$ at which $\tau (\lambda)$ vanishes, 
\begin{equation} \label{eq:lmax}
\tilde{\kappa} (\lambda_{\rm max}) \left[ \log \tilde{\kappa} (\lambda_{\rm max}) - 1 \right] = \frac{1}{b} - 1\, ,
\end{equation}
and $b_\star$ is given by the condition $\lambda_{\rm max} (b_\star) = 2$.

In Ref.~\cite{Knowles} Alt, Ducatez, and Knowles also investigated the structure of the eigenvectors and proved that in the semilocalized phase the wave-functions associated to the eigenvalue at energy $\lambda$ is highly concentrated around resonant vertices $i$ such that $\Lambda(k_i /c)$ is close to $\lambda$, 
while the mass far away from the resonance vertices is an asymptotically vanishing proportion of the total mass. More precisely Alt \& {\it al} also obtained an exact bound on the anomalous dimension $D_\infty$ of the eigenvectors in the semilocalized regime. We recall that the anomalous dimensions are defined from the asymptotic behavior of the $\ell^{2q}$-norm of the eigenvectors as
\[
|| \psi ||_{2 q}^2 = \left ( \sum_i | \psi(i) |^{2 q} \right)^{\! \! \frac{1}{q}} \! \propto N^{D_q \left (\frac{1}{q} - 1 \right ) } \, ,
\]
and fully characterize the geometric structure of the wave-functions, allowing one to discriminate between ergodic, localized, and multifractal states: In the fully delocalized regime $\psi(i) \sim N^{-1/2}$ uniformly on all the sites, and $D_q=1$ for all $q$'s; In the localized phase instead, the eigenstates are essentially concentrated in a small number $O(1)$ of vertices, and $D_q = 0$; In an intermediate multifractal phase, e.g. if the mass of $\psi$ is uniformly distributed over some subset $N^D$ of the sites, the $D_q$'s take values between $0$ and $1$. Focusing on the $q \to \infty$ limit 
in the semilocalized regime in Ref.~\cite{Knowles} it has been proven that the fractal dimension $D_\infty$ is bounded by $\tau (\lambda)$ in the interval $D_\infty (\lambda) \in [0, \tau (\lambda)]$. This also implies that $D_\infty$ exhibit a discontinuity in the thermodynamic limit as a function of the energies at $|\lambda| = \lambda_{\rm GOE} = 2$.

In a more recent paper~\cite{KnowlesL} Alt, Ducatez, and Knowles went a step further and 
proved that the statistics of the eigenvalues near the spectral edges is described by the Poisson statistics and the associated eigenvectors are exponentially localized around a {\it unique} center (i.e. $D_\infty=0$). In other words they proved the existence of a fully localized phase in the edge of the spectrum of ${\cal H}$.
However, this still leaves the possibility of the existence of an intermediate partially delocalized but non-ergodic  region sandwiched between the fully delocalized one 
and the fully localized one.

As a consequence of the analysis of Ref.~\cite{KnowlesL}, Alt, Ducatez, and Knowles also identify the asymptotic distribution of the largest (non trivial) eigenvalue of ${\cal H}$~\cite{KnowlesE}, which is given by a law that does not match with any previously known universal distribution 
(see Sec.~\ref{sec:lmax}).

\begin{figure}
\includegraphics[width=0.48\textwidth]{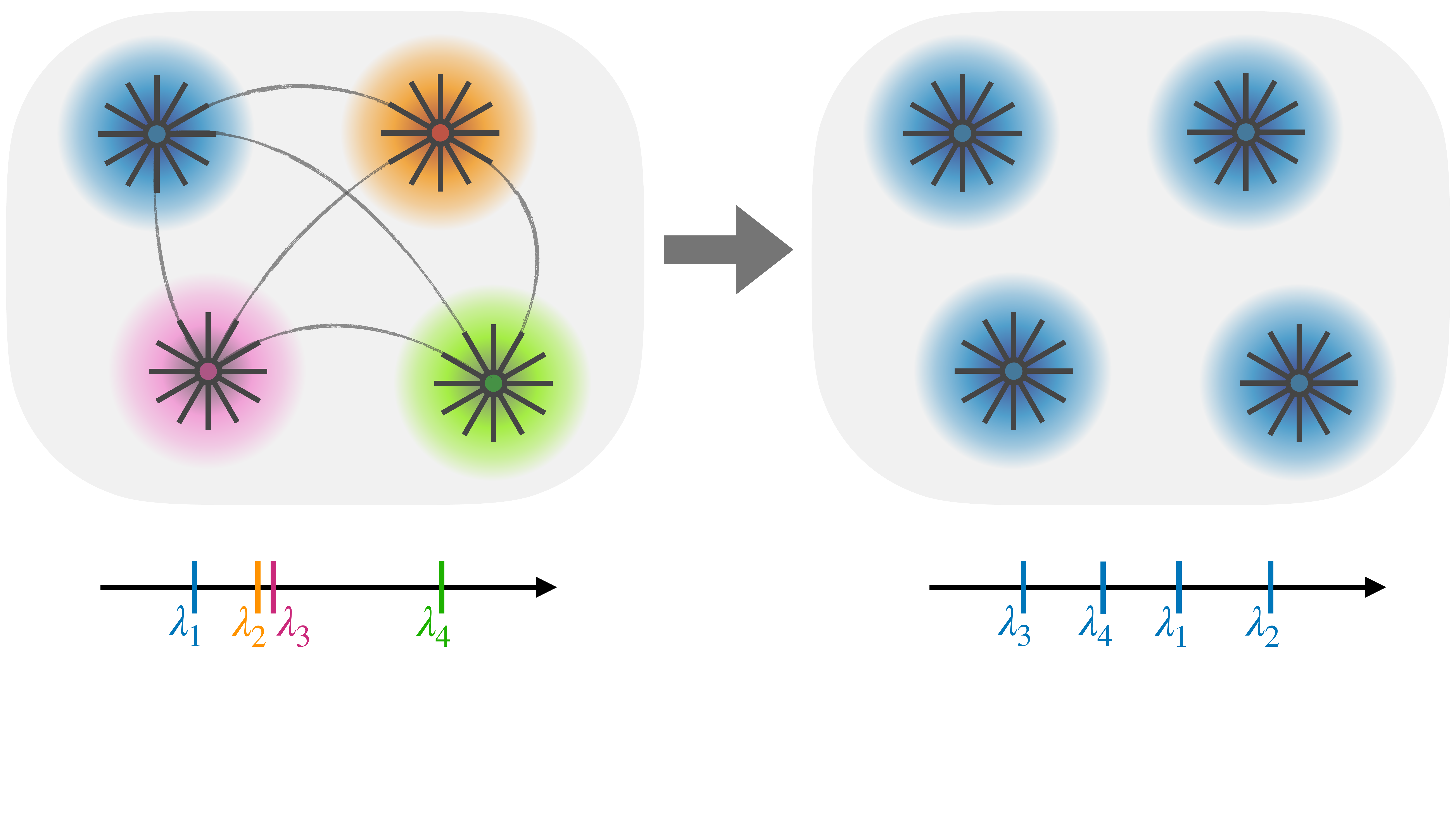}
	 	\vspace{-1.5cm}
	\caption{(color online) Schematic illustration of the possible structure of the eigenvectors in the tails of the spectrum of critical \ER graphs. Vertices of abnormally large degree $k = c \tilde{\kappa} (\lambda) > 2 c$ play the role of localization centers. The wave-functions decay exponentially around each vertex (shaded regions) and are connected by exponentially small effective tunneling amplitudes. Two situations are possible: these wave-functions might hybridize  (at least partially) around many resonant localization centers (right) or might stay fully localized around a single localization center (left). In the former case eigenstates close in energy occupy the same nodes and one has mini-bands in the local spectrum composed of $N^D \le N \rho$ consecutive energy levels within which the Wigner-Dyson statistics is locally established due to level repulsion. In the latter situation instead  eigenfunctions nearby in energy do not overlap and the level statistics is of Poisson type.
\label{fig:hyb}}
\end{figure}

At this point several key questions remain still open. Probably the two most important ones are: 
\begin{itemize}
    \item [(i)] What is the nature of the semilocalized phase? Two scenarios are in principle possible. All eigenstates in the tails of the spectrum could be fully localized around a {\it unique} vertex (i.e. $D_\infty = 0$), or there could be a region of the phase diagram where eigenvectors are partially delocalized around {\it many} resonant vertices with the same degree (i.e. $0 < D_\infty < \tau$) due to the hybridization of the exponentially decaying part of the wave-functions around each vertex, as schematically depicted in Fig.~\ref{fig:hyb}; In the first case the level statistics 
should be of Poisson type, while in the second case it is reasonable to expect that level repulsion should arise among nearby energy levels which should form mini-bands in the local spectrum, giving rise to  Wigner-Dyson statistics at least on the scale of the mean level spacing; 
\item[ii)] What are the critical properties of the transition(s) for the spectral statistics?  
What are the similarities and the differences compared to the standard localization transition observed in the Anderson tight-binding model on sparse graphs~\cite{abou,mirlin_fyodorov,fyodorov_mirlin,fyodorov_mirlin_sommers,fyod,mirlin1994,Zirn,tikhonov2019,ourselves,aizenmann,Verb}?
\end{itemize}
In the following sections we attempt to provide a tentative answer to these questions.

\section{The phase diagram} \label{sec:PD}

As explained above, in the tails of the spectrum we have $N \rho (\lambda)$ vertices of abnormally large degree $k = c \tilde{\kappa} (\lambda) > 2 c$ that play the role of localization centers. The wave-functions decay exponentially around each vertex. In this section we attempt to determine whether it exists a region of the phase diagram where these wave-functions hybridize  (at least partially)  due to the exponentially small tunneling amplitudes~\cite{Combes} between them (see Fig.~\ref{fig:hyb} for a schematic illustration). In this case wave-functions close in energy occupy the same sets of nodes. Since the effective matrix elements between different localization centers and their energies are essentially uncorrelated, it is natural to expect that, in analogy with RP-type models with iid entries~\cite{kravtsov,kravtsov1,khay,LRP,facoetti,bogomolny}, the system forms mini-bands in the local spectrum composed of $N^D \le N \rho$ consecutive energy levels within which the Wigner-Dyson statistics is locally established. Alternatively, all eigenstates in the tails of the spectrum can remain exponentially localized around a unique vertex. 
In this case nearby eigenfunctions do not overlap and the level statistics is of Poisson type. Below we present two analytical arguments to address this question.




\begin{figure*}
\includegraphics[width=0.53\textwidth]{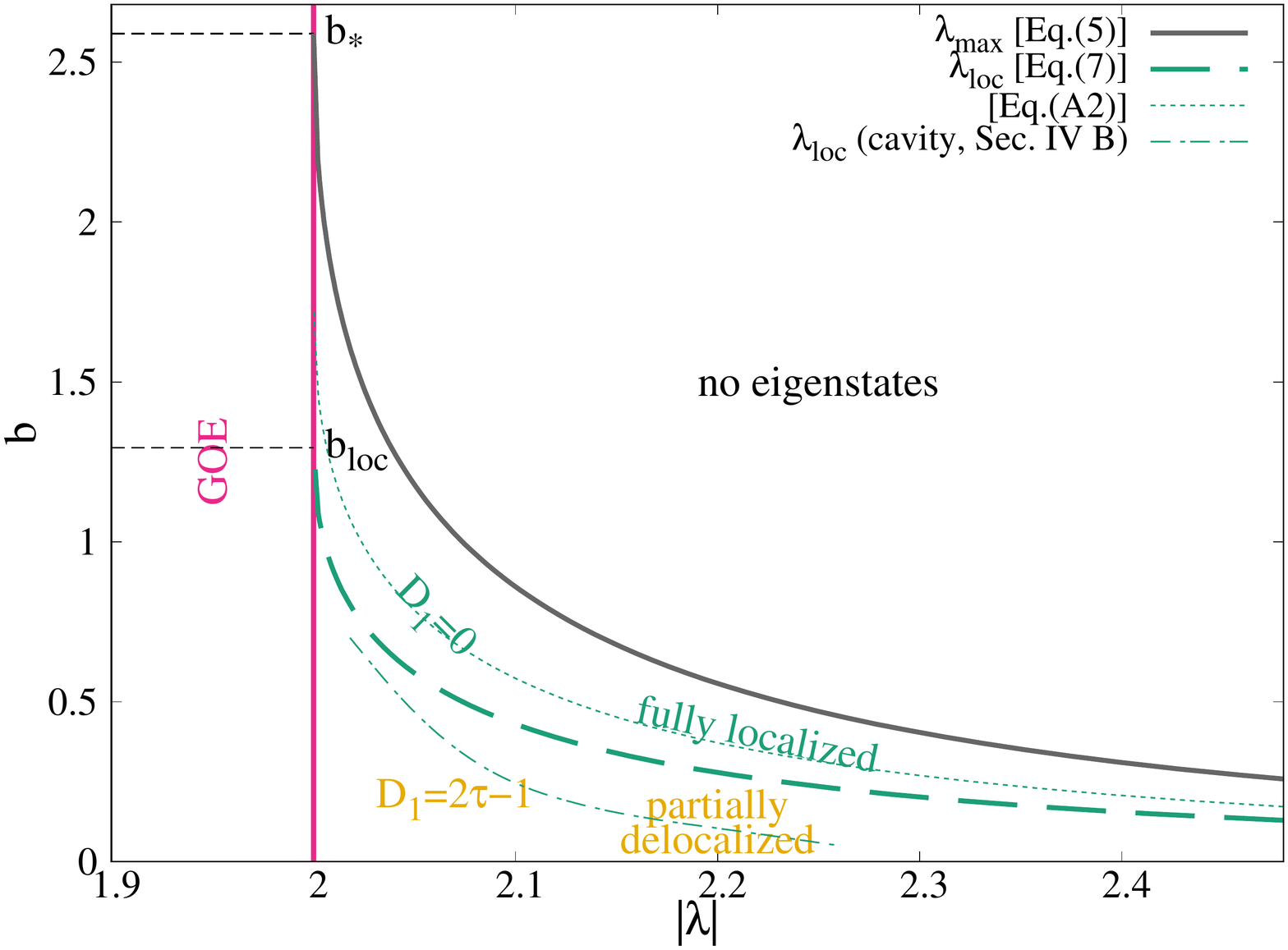} \hspace{0.05cm}
\includegraphics[width=0.44\textwidth]{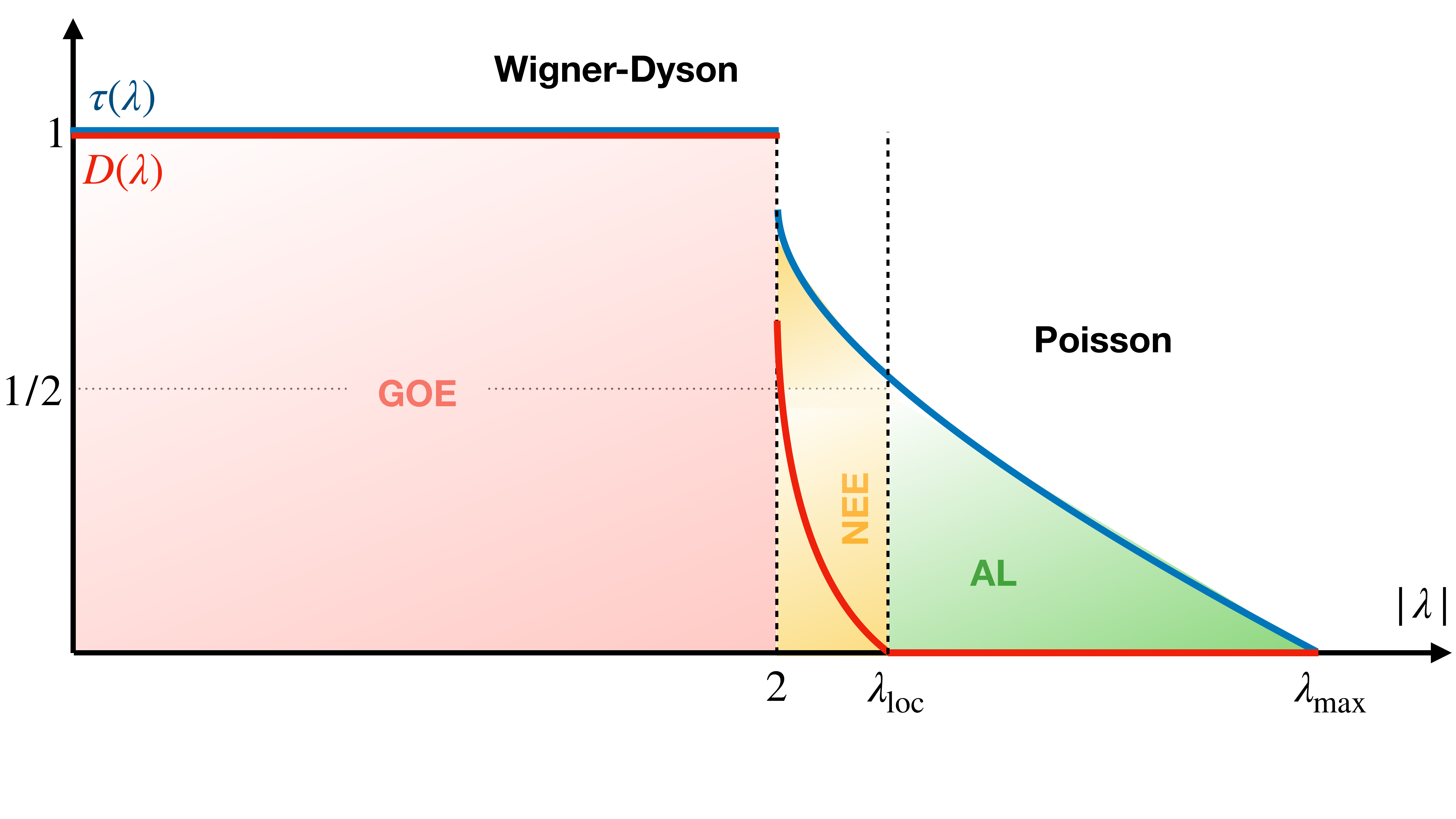}
	 	\vspace{-0.1cm}
	\caption{(color online) Left: Rough estimation of the phase diagram of the adjacency matrix of \ER graphs in the critical regime, $c = b \log N$, in the $(|\lambda|,b)$ plane obtained applying the Mott's criteria for localization and ergodicity. For $|\lambda|<2$ in the bulk of the spectrum the eigenvectors are fully delocalized and the DoS is given by the semicircle law. For $b<b_\star$ the spectrum is confined below the line $\lambda_{\rm max}$, given in Eq.~\eqref{eq:lmax}. For $|\lambda| \in (\lambda_{\rm loc}, \lambda_{\rm max})$ 
	the eigenvectors are fully localized around a unique localization center and the level statistics is of Poisson type~\cite{KnowlesL}. For $|\lambda| \in (2, \lambda_{\rm loc})$ the wave-functions partially delocalize around many resonant localization centers, due to the hybridization of energy levels. In this regime the system exhibits mini-bands in the local spectrum and the Wigner-Dyson statistics is established locally up to the Thouless energy scale $E_{\rm Th} \equiv \Gamma \propto N^{D (\lambda)-\tau (\lambda)}$ much larger than the mean level spacing $\Delta \propto N^{- \tau (\lambda)}$. 
	The thick dashed line represents the estimation of the mobility edge obtained from the Mott's criterion, Eq.~\eqref{eq:loc}. The thinner dashed dotted line corresponds to the estimation of $\lambda_{\rm loc}$ obtained from the approximate treatment of the self-consistent cavity equations discussed in Sec.~\ref{sec:cavity}. The dotted line shows the position of an upper bound for the mobility edge obtained in App.~\ref{app}. Right: Illustration of the behavior of the exponents $\tau$ and $D$ as a function of the energy $\lambda$ for fixed $b$ in the interval $b \in (0,b_{\rm loc})$. In the fully delocalized GOE-like phase in the bulk of the spectrum $\tau$ and $D$ are identically equal to $1$. In the tails of the spectrum the exponent $\tau$, which controls the asymptotic scaling behavior of the average DoS, is a decreasing function of $\lambda$ exhibiting a finite jump from $1$ to $1 - b/ b_\star$ at $|\lambda|=2$ and vanishing at $\pm \lambda_{\rm max}$ (see  Eq.~\eqref{eq:tau}). According to the Mott criterion, AL around a unique localization center occurs when $\tau < 1/2$ (localization nodes are too rarefied to be hybridized), implying that $D=0$ for $|\lambda| > \lambda_{\rm loc}$. Conversely, if $\tau>1/2$ the eigenstates are partially delocalized around many resonant localization centers and the exponent $D$ can be estimated from the Fermi Golden Rule, Eq.~\eqref{eq:D1}. At the transition in $|\lambda|=2$, $D$ is predicted to jump from $1$ to $1-2b/b_\star$.
\label{fig:PD}}
\end{figure*}

\subsection{Rules of thumb criteria for localization and ergodicity} \label{sec:mott}

The first approach is based on the so-called ``rules of thumb'' criteria for localization and ergodicity which have been formulated in the context of dense random matrix with uncorrelated entries~\cite{bogomolny,kravtsov1,khay,LRP}, and have been successfully adapted and applied in the latest years in the context of the MBL transition, where the subjacent adjacency matrix in the corresponding Hilbert space is sparse~\cite{tarzia,qrem1,qrem5}. Here the basic idea is that at a given energy $|\lambda|>2$ 
we can restrict ourselves to the $N \rho(\lambda) \propto N^{\tau (\lambda)}$ localization centers of degree $c \tilde{\kappa}(\lambda)$ and build an effective RP random matrix model where $N \rho(\lambda)$ independent levels with average energy separation of order $\Delta = 1/(N \rho)$ are coupled by exponentially small off-diagonal matrix elements which we estimate perturbatively (see in particular Ref.~\cite{qrem5} for a very similar mapping in the context of the QREM). Notice that the mapping onto an effective RP model here seems justified by the fact that the fluctuations of the energies of the localization centers of a given degree depend mostly on the fluctuations of the degrees of their neighbors~\cite{Knowles1,Knowles} and are essentially uncorrelated from the effective tunneling amplitudes between them, which depend mostly on their distances (see below).

The first criterion, known as the Mott's criterion for localization, states that AL around a single localization centre occurs when the level spacing $\Delta = 1/(N \rho)$ is much larger than the tunnelling amplitude between localization centres. The second criterion~\cite{bogomolny,nosov,kravtsov1,khay,LRP,qrem5,tarzia}, known as the Mott's criterion for ergodicity, is a sufficient condition for ergodicity. The idea is to estimate the average escape rate $\Gamma$ of a particle sitting on a localization center using the Fermi Golden rule and compare it to the spread of energy levels: When the average spreading width $\Gamma$ is much larger than the spread of energy levels, then the different localization centers are fully hybridized since starting from a given site the wave-packet spreads to any other localization center at the same energy in times of order one. 

In order to apply these two criteria we thus need to estimate the effective transition rates between two localization centers, which depend on their energy $\lambda$ and on their distance $r$. This can be done at the lowest order of the perturbative expansion starting from the insulating phase. The nodes of abnormally large connectivity $k = c \tilde{\kappa} (\lambda)$ produce localization centers at energy $\lambda = \Lambda(k/c)$. 
Since on a \ER graph in the large $N$ limit the shortest path connecting two nodes is with high probability unique, we can estimate the matrix elements between two  localization centers at distance $r$ as:
\begin{equation} \label{eq:Gfsa}
{\cal G}_r (\lambda) \approx \left( \frac{1}{\sqrt{c} \lambda} \right)^{\!r} \, .
\end{equation}
Since the $N \rho(\lambda) \propto N^{\tau (\lambda)}$ localization centers occupy random positions on the graph, the average distance between them is (asymptotically) given by the typical distance between two randomly chosen nodes, $r_{\rm typ} = \log N/\log c$. (This can be also checked numerically, as shown in Fig.~\ref{fig:RTYP} of App.~\ref{app}.) 

We then obtain that according to the Mott criterion, full localization around a unique vertex occurs when $|G_{r_{\rm typ} (\lambda)}(\lambda)| < (N \rho(\lambda))^{-1}$, i.e.
\[
\begin{aligned}
N^{\left[ \tau(\lambda) - \frac{1}{2} - \frac{\log \lambda}{\log c} \right] } < 1 \, .
\end{aligned}
\]
In the thermodynamic limit (and in the critical regime, $c = b \log N$) this condition is only fulfilled provided that $\tau (\lambda) < 1/2$. (Note that the finite size corrections to the Mott's criterion decay very slowly, as $1/\log \log N$.) Using the asymptotic expression for the exponent $\tau$ given in Eq.~\eqref{eq:dos}, one then finally obtains an implicit equation for the mobility edge $\lambda_{\rm loc}$ which separates fully localized eigenstates, from an intermediate partially delocalized phase in which the wave-functions hybridize (at least partially) around many resonating  localization centers:
\begin{equation} \label{eq:loc}
\tilde{\kappa} (\lambda_{\rm loc}) \left[ \log \tilde{\kappa} (\lambda_{\rm loc}) - 1 \right] = \frac{1}{2 b} - 1\, .
\end{equation}
Hence, for $|\lambda| \in (2,\lambda_{\rm loc})$ the exponentially decaying tunnelling amplitudes between localization centres are counterbalanced by an the large number of possible localization centers towards which  tunnelling can occur and the eigenvectors are delocalized across many resonant localization centres. One should keep in mind however that Eq.~\eqref{eq:loc} only provides a rough estimation of the mobility edge, since the analysis neglects the effect of the loops on the \ER graphs as well as higher order terms in the perturbative expansion. 

Since $\tau (\lambda)$ is a decreasing function of $\lambda$ which tends to $1 - b / b_\star$ for $|\lambda| \to 2^+$, the existence of the intermediate non-ergodic phase is only possible if $b < b_{\rm loc} =  b_\star/2 = 1/(\log 16 - 2)$. In App.~\ref{app} we will come back to this analysis suggesting a way to estimate an upper bound for the position of the mobility edge.

At this point, one can also wander whether in this intermediate partially delocalized phase the wave-functions occupy all the $N^{\tau(\lambda)}$ localization centers at the corresponding energy or spread only over a subset $N^D$ of them. In order to address this question one can estimate the average escape rate of a particle sitting on a localization center and compare it to the spectral bandwidth at the same energy. Using the Fermi Golden Rule the escape rate is approximately given by:
\[
\Gamma (\lambda) \approx 2 \pi N \rho (\lambda) |{\cal G}_{r_{\rm typ} (\lambda)}(\lambda)|^2 \propto N^{\left[ \tau (\lambda) - 1 - \frac{2 \log \lambda}{\log c} \right]} \, .
\]
This quantity corresponds to the average spreading of the energy levels due to the exponentially small hopping amplitudes between different localization centers. Assuming for simplicity that mini-bands are locally compact (as in the Gaussian RP model~\cite{kravtsov,facoetti,bogomolny}), this energy scale, usually called the Thouless energy $E_{\rm Th}$, coincides with the number of hybridized states within a mini-band times the mean level spacing, yielding a direct estimation of the fractal exponent $D$:
\[
\frac{1}{N \rho(\lambda)} N^{D(\lambda)} \propto \Gamma (\lambda) \, .
\]
In the thermodynamic limit one finds $D = 2 \tau - 1$ (note that $D=0$ at the localization threshold where $\tau(\lambda_{\rm loc}) = 1/2$):
\begin{equation} \label{eq:D1}
D (\lambda) = \left \{
\begin{array}{ll}
2 \tau(\lambda) - 1 & \textrm{for~} 2 < |\lambda| < \lambda_{\rm loc} \, , \\
0 & \textrm{for~} |\lambda| > \lambda_{\rm loc} \, .
\end{array}
\right.
\end{equation}
This function is plotted in Fig.~\ref{fig:D1} below for $b=0.5$, and also pictorially illustrated in the left panel of Fig.~\ref{fig:PD}. In Sec.~\ref{sec:numerics} we will provide a quantitative numerical test of the validity of this result. In the following for simplicity we will make the (questionable) assumption that the mini-bands in the local spectrum are fractal and not {\it multi}fractal, meaning that all the fractal dimensions are equal in the thermodynamic limit, $D_q (\lambda) = D(\lambda)$.

The resulting phase diagram of the adjacency matrix of \ER graphs in the critical regime obtained applying these simple arguments for localization and ergodicity is summarized in Fig.~\ref{fig:PD}, showing the transition lines between the different phases. As discussed in Sec.~\ref{sec:qrem} this phase diagram is, {\it mutatis mutandis}, qualitatively very similar to the one of the QREM recently obtained in Refs.~\cite{qrem1,qrem2,qrem3,qrem4,qrem5}. 

\begin{figure*}
 \includegraphics[width=0.97\textwidth]{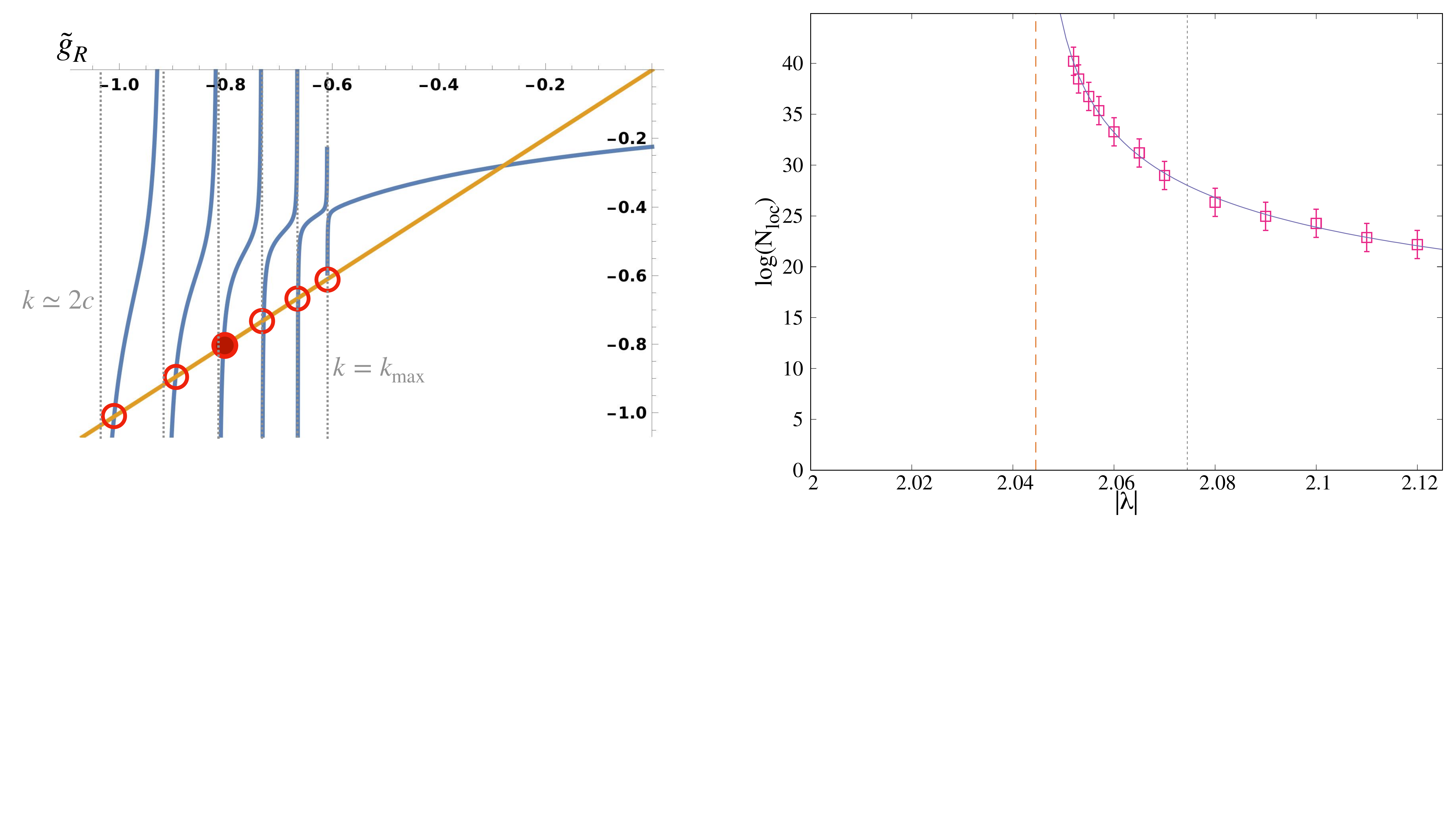}
	 	\vspace{-3.6cm}
	\caption{(color online) Left: Illustration of the procedure used to find the solution of Eq.~\eqref{eq:gR}. For each value of $k$ between $2c$ and $k_{\rm max}$ the right-hand side of the equation has a pole at $\tilde{g}_R = - c \lambda/k$. Each one of these singularities produces a solution of the self-consistent equation (circles). We choose the solution associated to the value of the singularity in $- c \lambda/k_\star$, where $k_\star$ is the closest integer to $c \tilde{\kappa} (\lambda)$, which corresponds to the connectivity of the localization centers at energy $\lambda$ in the thermodynamic limit. Right: Logarithm of the smallest value of the system size for which Eq.~\eqref{eq:stability} ceases to be satisfied, $N_{\rm loc} (\lambda)$, as a function of the energy $\lambda$ in the tails of the spectrum of critical \ER graphs with $b=0.5$. The continuous curve is a fit of the data of the form $\log N_{\rm loc} (\lambda) \propto (\lambda - \lambda_{\rm loc})^{-\zeta}$, with $\lambda_{\rm loc} \approx 2.045$ and $\zeta \approx 0.258$. The estimation of the mobility edge obtained by this fit is represented by the dashed vertical line. The dotted vertical line shows the estimation of $\lambda_{\rm loc}$ obtained from the Mott criterion, $\lambda_{\rm loc} \approx 2.074$  (see also Fig.~\ref{fig:PD}).
\label{fig:nloc}}
\end{figure*}

\subsection{Estimation of the localization transition from the cavity approach} \label{sec:cavity}

A complementary approximate analytical strategy to tackle the localization transition and identify the position of the mobility edge as a function of the parameters of the model is based on the approximate treatment of the self-consistent cavity equations for the Green's functions.

In fact the Green's function of the adjacency matrix of \ER graphs satisfy an exact self-consistent equation in the thermodynamic limit~\cite{abou,rodgers2008}. The recursive equations are obtained by introducing the (cavity) Green's functions of auxiliary graphs,
$G_{i \to j} (z) = [{\cal H}_{i \to j} - z {\cal I}]^{-1}_{ii}$, i.e., the $i$-th diagonal element of the resolvent matrix of the modified Hamiltonian ${\cal H}_{i \to j}$  where one of the neighbors of $i$, 
say node $j$, has been removed. On an infinite tree, all neighbors $\{ j_1, \ldots, j_{k_i} \}$ of a given vertex $i$ with degree $k_i$ are in different connected components of ${\cal H}$. By removing 
one of its neighbors $j_a$, one then obtains (by direct Gaussian integration or using the block matrix inversion formula) the following iteration relations for the diagonal elements of the cavity Green’s functions on a given node $i$ in absence of one if its neighbors as a function of the diagonal elements of the cavity Green's functions on the neighboring nodes in absence of $i$:
\begin{equation} \label{eq:green}
G_{i \to {j_a} }^{-1} (z) = -z - \frac{1}{c} \!\! \sum_{j_b \in \partial i / j_a} \!\! G_{{j_b} \to i} (z)\, ,
\end{equation}
where $j_a$ with $a = 1, \ldots, k_i$ denote the excluded neighbor of $i$, $z = \lambda + i \eta$, $\eta$ is an infinitesimal imaginary regulator which smoothens out the pole-like singularities in the right hand sides, and $\partial i / l$ denotes the set of all $k_i$ neighbors of $i$ except $l$. Note that for each site with $k_i$ neighbors one can define $k_i$ cavity Green’s functions and $k_i$ recursion relations of this kind, and hence on a finite \ER graph of $N$ nodes and average connectivity $\langle k \rangle = c$, Eq.~\eqref{eq:green} represents in fact a system of $\sim c N$ coupled equations. 

After that the solution of Eqs.~\eqref{eq:green} has been found, one can finally obtain the diagonal elements of the resolvent matrix of the original problem on a given vertex $i$ as a function of the cavity Green’s functions for all the neighboring nodes in absence of $i$:
\begin{equation} \label{eq:greenF}
G_{ii}^{-1} (z) = -z - \frac{1}{c} \! \sum_{j_b \in \partial i} \! G_{{j_b} \to i} (z) \, .
\end{equation}
Although \ER graphs are not loop-less infinite trees, in the large $N$ limit the neighborhood of $i$ is, with high probability, a tree since the typical length of the loops grows as $\log N/ \log c \propto \log N / \log \log N$. One can then expect that if $N$ is large enough Eqs.~\eqref{eq:green} and~\eqref{eq:greenF} provide a very good approximation of the true Green's functions~\cite{Bordenave}. 

The statistics of the diagonal elements of the resolvent encodes the spectral properties of ${\cal H}$. In particular, the local density of states (LDoS) is given by
\[
\rho_i (\lambda) = \sum_{\alpha = 1}^N \left \vert \psi_\alpha (i) \right \vert^2 \delta (\lambda - \lambda_\alpha) = \lim_{\eta \to 0^+} \frac{1}{\pi} {\rm Im} G_{ii} (\lambda) \, .
\]
From the LDoS one can compute the average DoS, which is simply given by $\rho(\lambda)  = (1/N) \sum_i \rho_i (\lambda) = 1/(N\pi) {\rm Tr} \, {\rm Im} G$. We will be also interested in the typical LDoS, defined as $\rho_{\rm typ} = e^{\langle \log {\rm Im} G \rangle} / \langle {\rm Im} G \rangle$. 

Note that in principle the statistics of the LDoS allows one to distinguish between a localized and a delocalized phase. In fact in a localized regime the probability distribution of the LDoS is singular in the $\eta \to 0^+$ limit and characterized by power-law tails, while in a delocalized regime the LDoS is unstable with respect to the imaginary regulator $\eta$ and its probability distribution converges to stable non-singular $\eta$-independent distribution functions (provided that $\eta$ is sufficiently small). 

In the tails of the spectrum of the adjacency matrix of critical \ER graphs, where $c \gg 1$ and the main contribution to the local DoS comes from the vertices of abnormally large degree $k>2c$, it is very tempting to write an approximate equation for the Green's function in the spirit of the SDA~\cite{biroli99,semerjian02}, in which one uses the central limit theorem to evaluate the sums over the 
neighbors 
appearing in the right hand side of Eqs.~\eqref{eq:green} and~\eqref{eq:greenF}. In fact, at least in the delocalized regime where the elements of the resolvent are described by a stable non-singular distribution function in the $\eta \to 0^+$ limit, $(1/c) \sum_{j \in \partial i} G_{j \to i} (z)$ tends to a Gaussian random variable of mean proportional to $k_i/c$ (which is of order 1) and variance proportional to $\sqrt{k_i}/c$ (which is of order $1/\sqrt{\log N}$). Hence, neglecting completely the fluctuation of the local degrees, in the large $N$ limit one can write an approximate equation for the average value of the Green's function restricted on the nodes of  degree $k$,  $\langle G \rangle_k= 1/(N P(k)) \sum_{i : k_i=k} G_{ii}$:
\begin{equation} \label{eq:gapprox}
\langle G (z) \rangle_k \approx \frac{1}{-z - \frac{k}{c} \tilde{g} (z)} \, ,
\end{equation}
where $\tilde{g} (z)$ is defined as the average Green's function, $\tilde{g} (z) = \sum_k P(k) \langle G \rangle_k$. Summing over all degrees $k$ with the corresponding probability $P(k)$, Eq.~\eqref{eq:gapprox} finally leads to the following self-consistent equation for $\tilde{g} (z)$:
\begin{equation} \label{eq:dosapprox}
\tilde{g} (z) \approx - \sum_k \frac{P(k)}{z + \frac{k}{c} \tilde{g} (z)} \, .
\end{equation}
Once the solution of the equation above is found, using Eq.~\eqref{eq:gapprox} one can obtain an approximate expression for the whole probability distribution of the elements of the Green's function as 
\begin{equation} \label{eq:qgapprox}
Q(G) \approx \sum_k P(k) \, \delta \left( G + \frac{1}{z + \frac{k}{c} \tilde{g} (z)}\right) \, .
\end{equation}
In the thermodynamic limit and for $|\lambda| \in (2 , \lambda_{\rm max})$ one expects that the sum over $k$ is dominated by the nodes of connectivity $k = c \tilde{\kappa} (\lambda)$. In Figs.~\ref{fig:dos}, \ref{fig:igk}, \ref{fig:Pldos}, and~\ref{fig:imgtyp} we discuss the quality of this approximation with respect to the exact solution of the model for several observables and for several values of $\lambda$ and $N$ (and for $b=0.5$).

At this point, in order to determine the position of the mobility edge, one can seek for solutions of Eq.~\eqref{eq:gapprox} in absence of the imaginary part of $\tilde{g}$, and then study the stability of these solutions with respect to the addition of a small imaginary part. The approximate self-consistent equation for the real part of $\tilde{g}$ is
\begin{equation} \label{eq:gR}
\tilde{g}_R = - \sum_k \frac{P(k)}{\lambda + \frac{k}{c} \tilde{g}_R} \, ,
\end{equation}
where the sum over $k$ is in fact cut-off at $k_{\rm max}$ which is the largest degree on a graph of $N$ nodes. The right-hand side of the equation above has poles at all values of $\tilde{g}_R$ such that $\tilde{g}_R = - c \lambda/k$. As illustrated in the left panel of Fig.~\ref{fig:nloc}, each one of these singularities produces a crossing between the right-hand side and the left-hand side of the equation and gives rise to a solution of~\eqref{eq:gR}. We assume that in the thermodynamic limit the relevant solution is the one associated to the value of the singularity in $- c \lambda/k_\star$, where $k_\star$ is the closest integer to $c \tilde{\kappa} (\lambda)$, which corresponds to the connectivity of the localization centers at energy $\lambda$ (see also Fig.~\ref{fig:igk}). Adding now a small imaginary part to the average Green's function and linearizing with respect to it, one obtains the self-consistent equation describing the exponential decay or the exponential growth of the imaginary part starting from the real solution. The stability condition of the localized phase is thus simply given by:
\begin{equation} \label{eq:stability}
\frac{1}{c} \sum_k \frac{k P(k)}{\left( \lambda + \frac{k}{c} \tilde{g}_R \right)^2} < 1 \, .
\end{equation}
We have solved numerically Eqs.~\eqref{eq:gR} and~\eqref{eq:stability} for several values of $b$, varying the energy $\lambda$ and the system size $N$ (and choosing the solution of Eq.~\eqref{eq:gR} which is the closest to the pole in $- c \lambda/k_\star$). This can be done for $N \lesssim 2^{50}$, since for $N$ too large the exponentially small probability in the numerator and the  poles in the denominator cannot be handled with a sufficient degree of numerical precision to yield reliable results. For every value of $\lambda$ at fixed $b$ we determine the value of $N_{\rm loc} (\lambda)$, which corresponds to the smallest value of $N$ such that Eq.~\eqref{eq:stability} is satisfied. This procedure is illustrated in Fig.~\ref{fig:nloc}  for $b=0.5$. 
One observes that $N_{\rm loc}(\lambda)$ increases very rapidly when $\lambda$ is decreased and seems to diverge for $\lambda \approx 2.045$. This analysis suggests that in the thermodynamic limit and for $b=0.5$ the mobility edge is located around $\lambda_{\rm loc} \approx 2.045$, which is in fact not too far from the estimation of $\lambda_{\rm loc}$ obtained from the Mott's criterion (Eq.~\eqref{eq:loc} of Sec.~\ref{sec:mott}) for the same value of $b$, $\lambda \approx 2.074$. 
A similar behavior is found for other values of $b$. The estimation of $\lambda_{\rm loc}$ obtained from this analysis is plotted as a dashed dotted line in the $(|\lambda|,b)$ plane on the phase diagram of Fig.~\ref{fig:PD}. Although the prediction for the mobility edge obtained from the approximate treatment of the self-consistent cavity equations does not coincide quantitatively with the one obtained from the Mott's criterion, the two lines have a similar qualitative shape.

Note that, 
similarly to the Mott criterion, within this approach delocalization occurs due to a trade-off between the exponential decrease of $P(k)$ 
and the accumulation of singularities in the denominator of 
Eq.~\eqref{eq:stability} 
which become closer and closer to each other 
and make the sums over $k$ blow. 


\section{Numerics}  \label{sec:numerics}

\begin{figure*}
	 \hspace{-0.22cm}\includegraphics[width=0.49\textwidth]{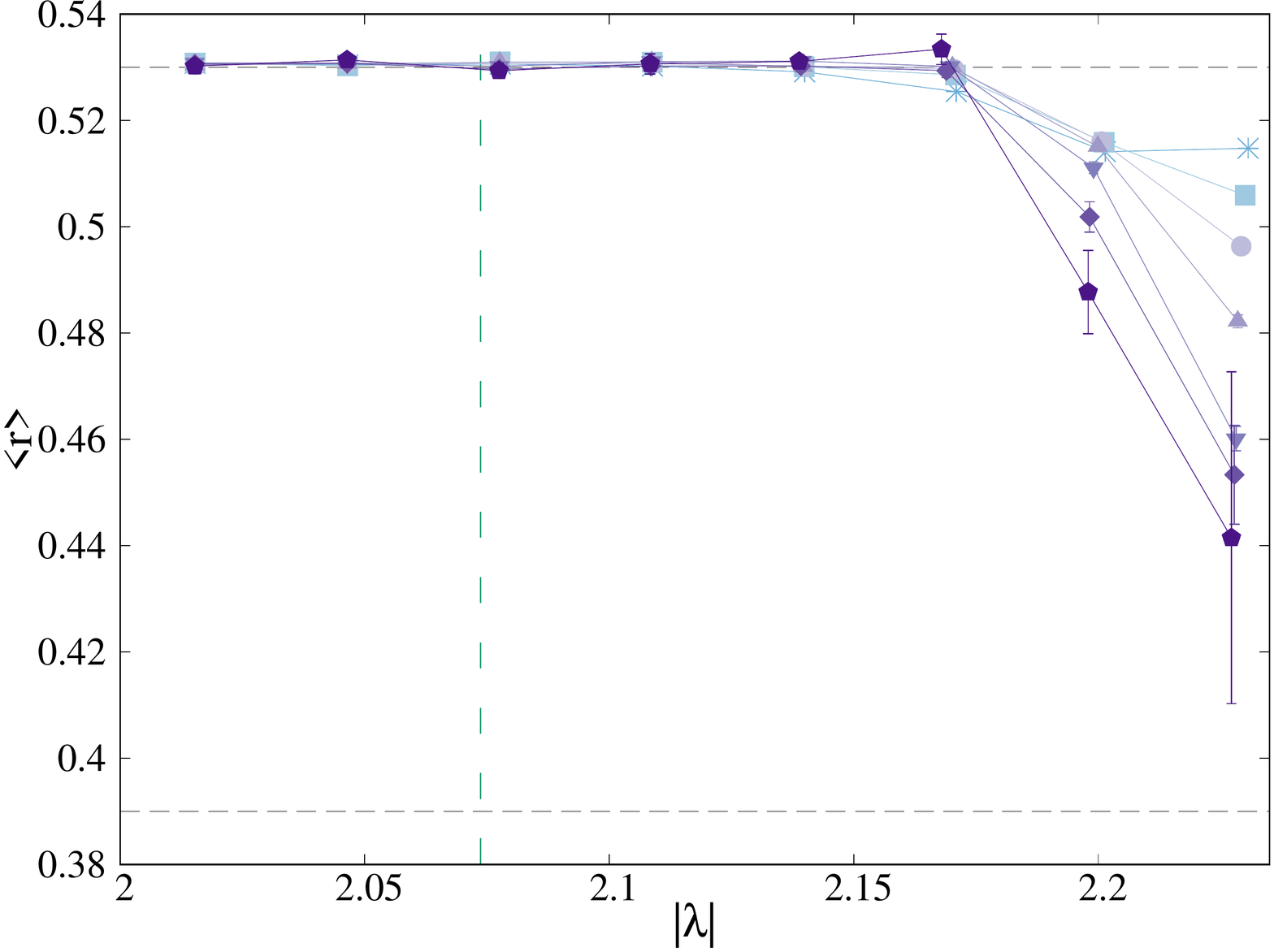}\hspace{-.35cm}
	 \includegraphics[width=0.51\textwidth]{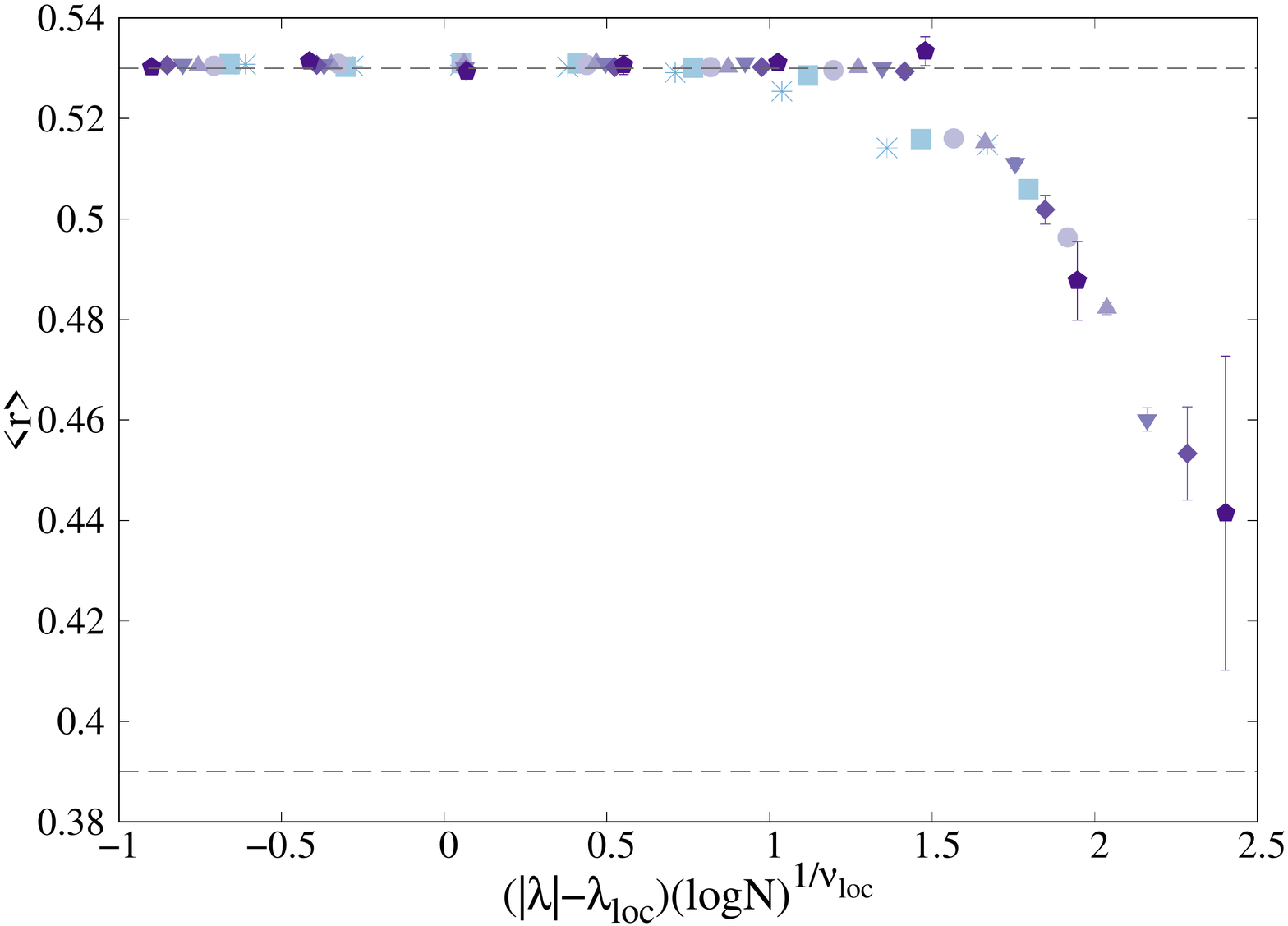} \vspace{-0.1cm}
	 
	 \hspace{-0.22cm}\includegraphics[width=0.487\textwidth]{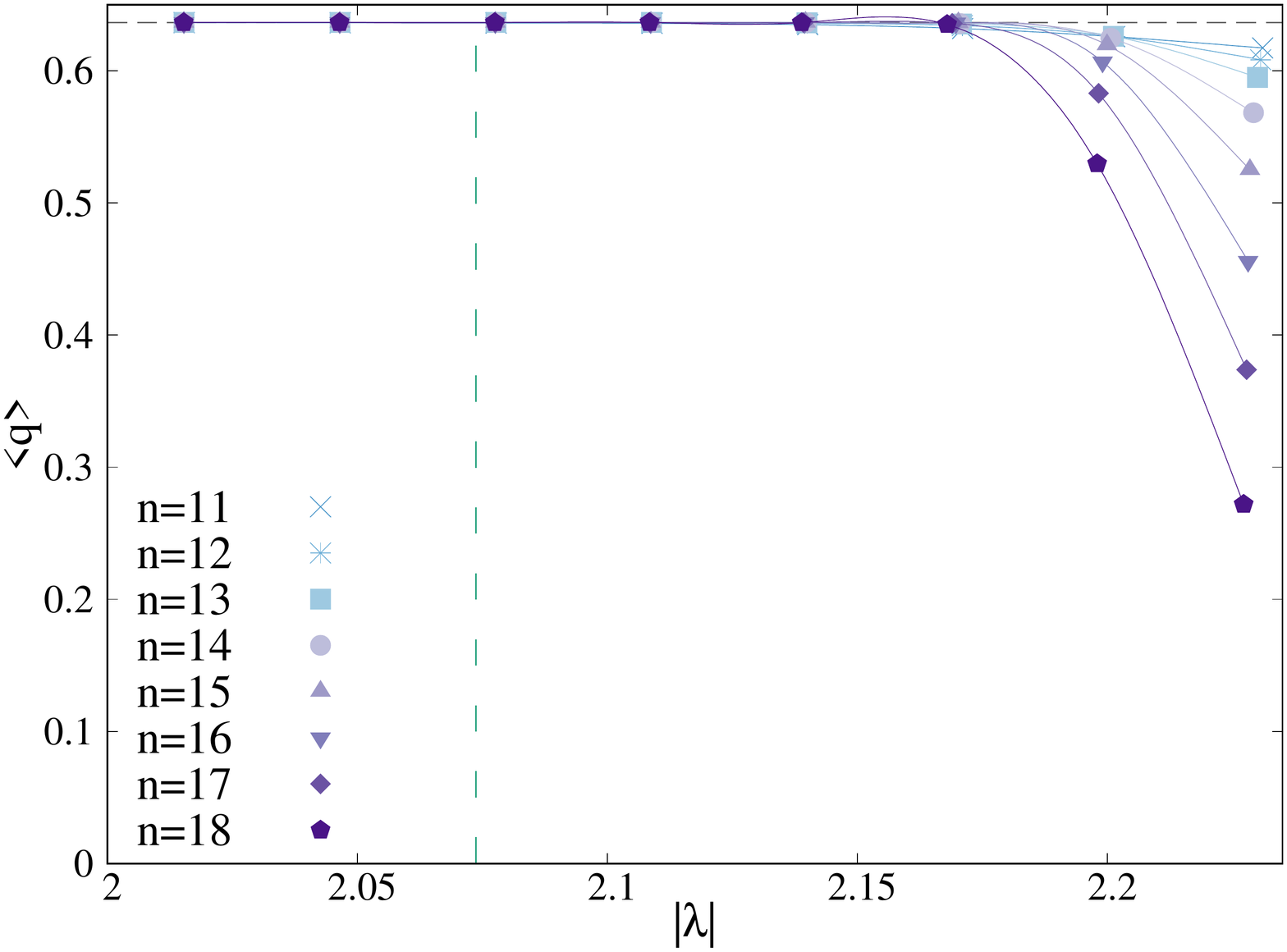}\hspace{-.3cm}
	 \includegraphics[width=0.505\textwidth]{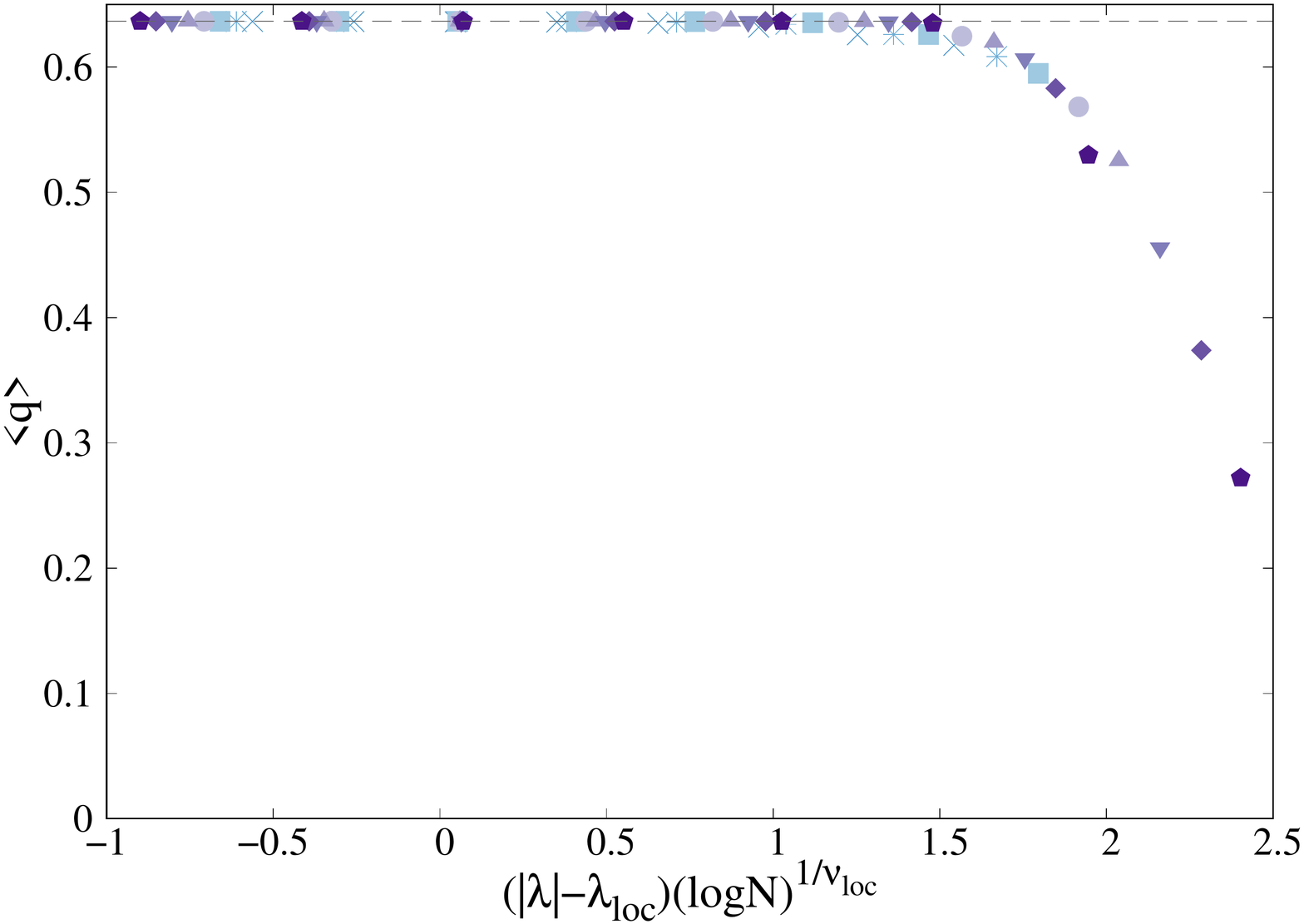} 
	 	\vspace{-0.2cm}
	\caption{(color online) $\langle r \rangle$ (top) and $\langle q \rangle$ (bottom) as a function of $|\lambda| \in (2, \lambda_{\rm max})$ in the tails of the spectrum of the adjacency matrix of critical \ER graphs for $b=0.5$ ($\lambda_{\rm max} \approx 2.231$). Different system sizes $N = 2^n$ (with $n$ from $11$ to $18$) correspond to different symbols and colors as indicated in the legend. The right panels show that a good data collapse is obtained for both $\langle r \rangle$ and $\langle q \rangle$ in terms of the scaling variable $(\lambda - \lambda_{\rm loc}) (\log N)^{1/\nu_{\rm loc}}$, with $\lambda_{\rm loc} \approx 2.074$ (Eq.~\eqref{eq:loc}) and $\nu_{\rm loc} \approx 1$. 
	The horizontal dashed grey lines correspond the GOE and Poisson universal values.}
\label{fig:level_statistics}
\end{figure*}

In this section we present a numerical verification of the theoretical predictions for the phase diagram discussed above. We set $b=0.5$ throughout. According to the phase diagram of Fig.~\ref{fig:PD}, at $b=0.5$ one should cross two phase transitions as the energy is increased: 1) A transition at $\lambda_{\rm GOE} = 2$ from the fully delocalized GOE-like phase to the partially delocalized but non-ergodic phase, where the statistics of the wave-functions' amplitudes should exhibit a dramatic change (in particular the fractal dimension should display a discontinuous jump at the transition);
2) A AL transition at $\lambda_{\rm loc}$ where the gap statistics likely undergoes a transition from Wigner-Dyson statistics to Poisson statistics. In fact, as shown in~\cite{KnowlesL}, in the fully localized part of the spectrum above $\lambda_{\rm loc}$, the exponentially decaying eigenvectors around unique localization centers do not interact and the statistics of level spacing is Poisson. Conversely, as explained above, 
below $\lambda_{\rm loc}$ eigenstates close in energy 
are partially deolcalized around many resonant localization centers and, in analogy with RP-type models with iid entries, the system is expected to exhibit mini-bands in the local spectrum, within which the Wigner-Dyson statistics is established up to an energy scale much larger than the mean level spacing~\cite{kravtsov,kravtsov1,khay,LRP,pino,facoetti,bogomolny}.

We employ two complementary numerical strategies to investigate these two transitions: The first approach consists in performing exact diagonalizations of the adjacency matrix of critical \ER graphs of size $N = 2^n$ with $n$ ranging from $9$ to $18$. Since we are interested in the  properties of the tails of the spectrum, we only focus on a sub-extensive set of eigenvalues and eigenvectors in the spectral edges, $2<|\lambda|<\lambda_{\rm max}$. The number of these eigenstates scales approximately as $\sim N^{1-b/b_\star}$ and the Lanczos algorithm works efficiently up to moderately large sizes. Averages are performed over (at least) $2^{43-2n}$ different independent realizations of the graph and over eigenstates in the same energy window.

 The second strategy consists instead in solving directly the self-consistent cavity equations~\eqref{eq:green} and~\eqref{eq:greenF} on random instances of critical \ER graphs of large but finite sizes $N=2^n$, from $n=12$  to $n=28$. In practice, we first generate the graph according to the Bernoulli distribution~\eqref{eq:H}. Then we find the fixed point of Eqs.~\eqref{eq:green}, which represent a system of $\sim cN$ coupled equation for the cavity Green's functions. Finally, using Eqs.~\eqref{eq:greenF} we obtain the diagonal elements of the resolvent matrix on each vertex. We repeat this procedure $2^{32-n}$ times to average over different realizations of the graph. The advantage of this method over EDs is that the solution of the cavity equations can be obtained with arbitrary precision by iteration in a time that scales as $c N \propto N \log N$, which is much faster than the computational time needed to diagonalize the Hamiltonian, which scales roughly as $N^{3-b/b_\star}$, thereby allowing one to access system sizes about $10^3$ times larger.

\subsection{Level statistics and statistics of the wave-functions' amplitudes} \label{sec:level}

 Here we start by focusing on the AL transition. To this aim we perform a finite-size scaling analysis of the behavior of two observables related to the level statistics of neighboring eigenvalues.  
The first is the average ratio of adjacent gaps:
\[
r_n = {\rm min} \left \{ \frac{\lambda_{n+2} - \lambda_{n+1}}{ \lambda_{n+1} - \lambda_n} , \frac{ \lambda_{n+1} - \lambda_{n}}{\lambda_{n+2} - \lambda_{n+1}}  \right \}  \, ,
\]
whose probability distribution displays a universal form depending on the level statistics, with $\langle r \rangle$ equal to $0.53$ in the GOE ensemble and to $0.39$ for Poisson statistics~\cite{huse}.

The second observable which captures the transition from Wigner-Dyson to Poisson statistics is given by the mutual overlap between two subsequent eigenvectors, defined as
\[
q_n = \sum_{i=1}^N \left \vert \psi_n(i) \right \vert \left \vert \psi_{n+1}(i) \right \vert \, ,
\]
In the Wigner-Dyson phase $\langle q \rangle$ converges to $2/\pi$ (as expected for random vector on a $N$-dimensional sphere), while in the localized phase two successive eigenvector are typically peaked around  different sites and do not overlap and $\langle q \rangle \to 0$ for $N \to \infty$. At first sight this quantity seems to be related to the statistics of wave-functions' coefficients rather than to energy gaps. Nonetheless, in all the random matrix models that have been considered in the literature so far, one empirically finds that $\langle q \rangle$ is directly associated to the statistics of gaps between neighboring energy levels~\cite{notaRP,LRP,Levy,large_deviations}. 

In the left panels of Fig.~\ref{fig:level_statistics} we plot $\langle r \rangle$ (top) and $\langle q \rangle$ (bottom) as a function of $\lambda$ for $b=0.5$, showing that both observables take their Wigner-Dyson universal values for $\lambda \le \lambda_{\rm loc}$, while they depart from the Wigner-Dyson values for $\lambda> \lambda_{\rm loc}$, in a way that is more pronounced when the system size is increased. The right panels demonstrate that a 
good collapse of the data (especially for $\langle q \rangle$ which turns out to be much less noisy than $\langle r \rangle$) corresponding to different sizes is obtained in terms of the scaling variable $(\lambda - \lambda_{\rm loc}) (\log N)^{1/\nu_{\rm loc}}$, with 
$\nu_{\rm loc} \approx 1$. (Such value of the exponent is the same found for the Gaussian RP model at the AL transition~\cite{pino}.) Here for concreteness we have used the estimation of $\lambda_{\rm loc}$ given by the Mott criterion, Eq.~\eqref{eq:loc}, i.e.  $\lambda_{\rm loc} \approx 2.074$ for $b=0.5$. A reasonably good collapse can be also obtained setting the mobility edge to the value given by the linear stability analysis of the approximate cavity equations, $\lambda_{\rm loc} \approx 2.045$ (see Sec.~\ref{sec:cavity}) and using $\nu_{\rm loc} \approx 0.75$. It is also possible to collapse the data for different values of $N$ on the same curve assuming that transition from Wigner-Dyson to Poisson statistics takes place at $\lambda_{\rm loc} = 2$ and setting $\nu_{\rm loc} \approx 0.5$. This situation would be realized either if the intermediate partially delocalized but non-ergodic phase is only a finite-size crossover and  eventually all eigenvalues in the tails of the spectrum become fully localized in the thermodynamic limit, or if the structure of the fractal states is different from the one of RP-type models, as it happens for instance in correlated random matrix models having a fractal phase that does not feature mini-bands in the local spectrum within which the Wigner-Dyson statistics establishes~\cite{kutlin,motamarri,tang}. 
However the quality of the collapse in this case is slightly less good than the one achieved in the right panels of Fig.~\ref{fig:level_statistics}. To sum up, the finite-size scaling analysis of the level statistics is fully compatible with a transition from Wigner-Dyson to Poisson statistics at $\lambda_{\rm loc}>2$, corroborating the results of the previous section. Yet, our numerical data are limited to too small sizes to rule out definitely other possible scenarios and to be fully conclusive on the nature of the transition.

Finally, the plots Fig.~\ref{fig:level_statistics} call attention on an important difference with the standard AL transition on sparse graphs induced by the random local potential. In fact in this case it is well established that the critical point is in the localized phase and it is thus described by the Poisson statistics~\cite{mirlinrrg,efetov,efetov1,tikhonov2019,fyod,Zirn,Verb}, while in the present case the critical point lies clearly in the Wigner-Dyson phase. This latter behavior is also observed in random matrix models of the RP type featuring an intermediate non-ergodic extended phase sandwiched between the fully ergodic one and the fully localized one~\cite{LRP,pino}. This observation thus provides another hint of the existence of a genuine partially delocalized but non-ergodic phase in the tails of critical \ER graphs.

\begin{figure*}
	 \hspace{-0.22cm}\includegraphics[width=0.345\textwidth]{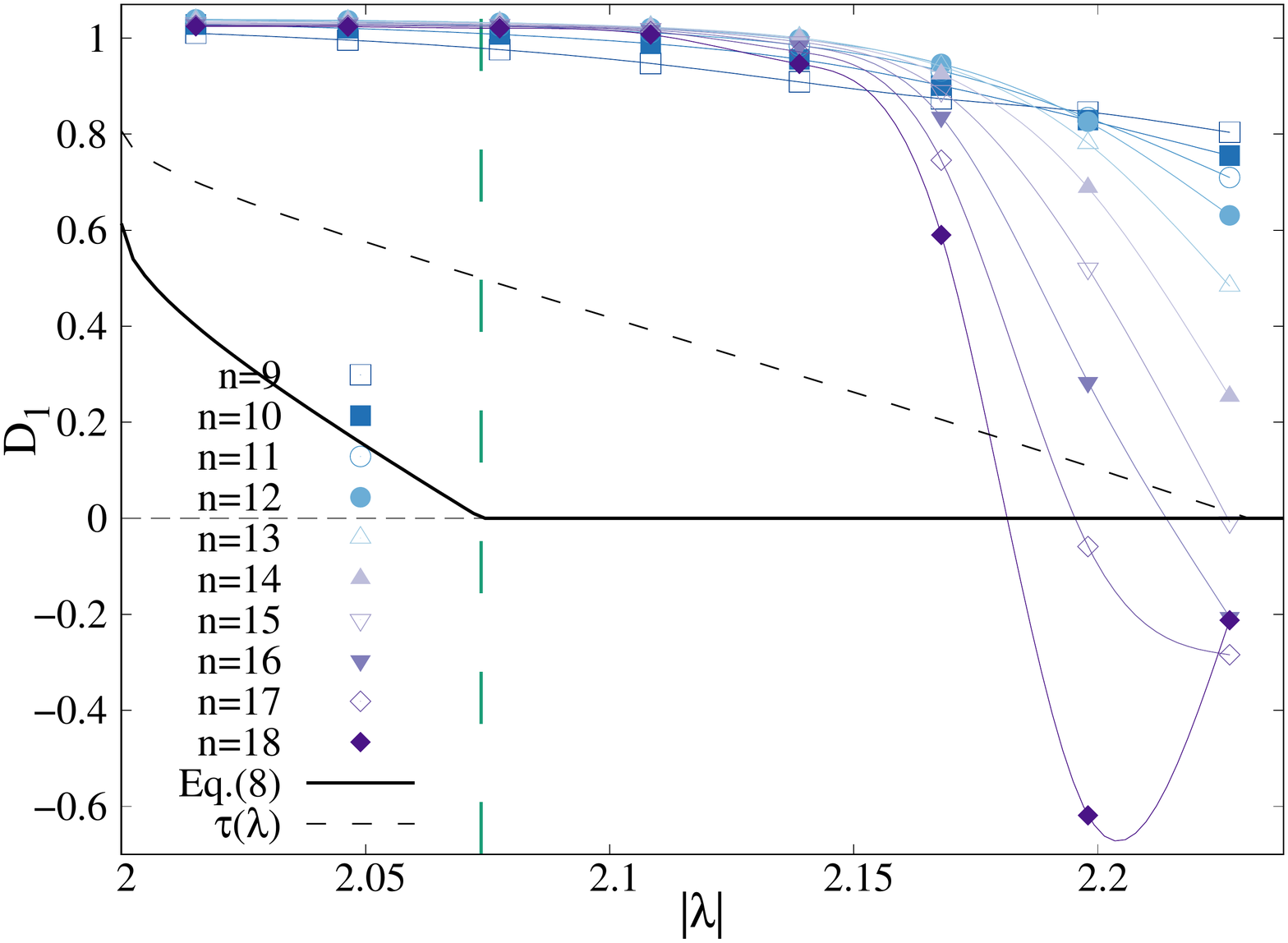}\hspace{-.4cm}
	 \includegraphics[width=0.345\textwidth]{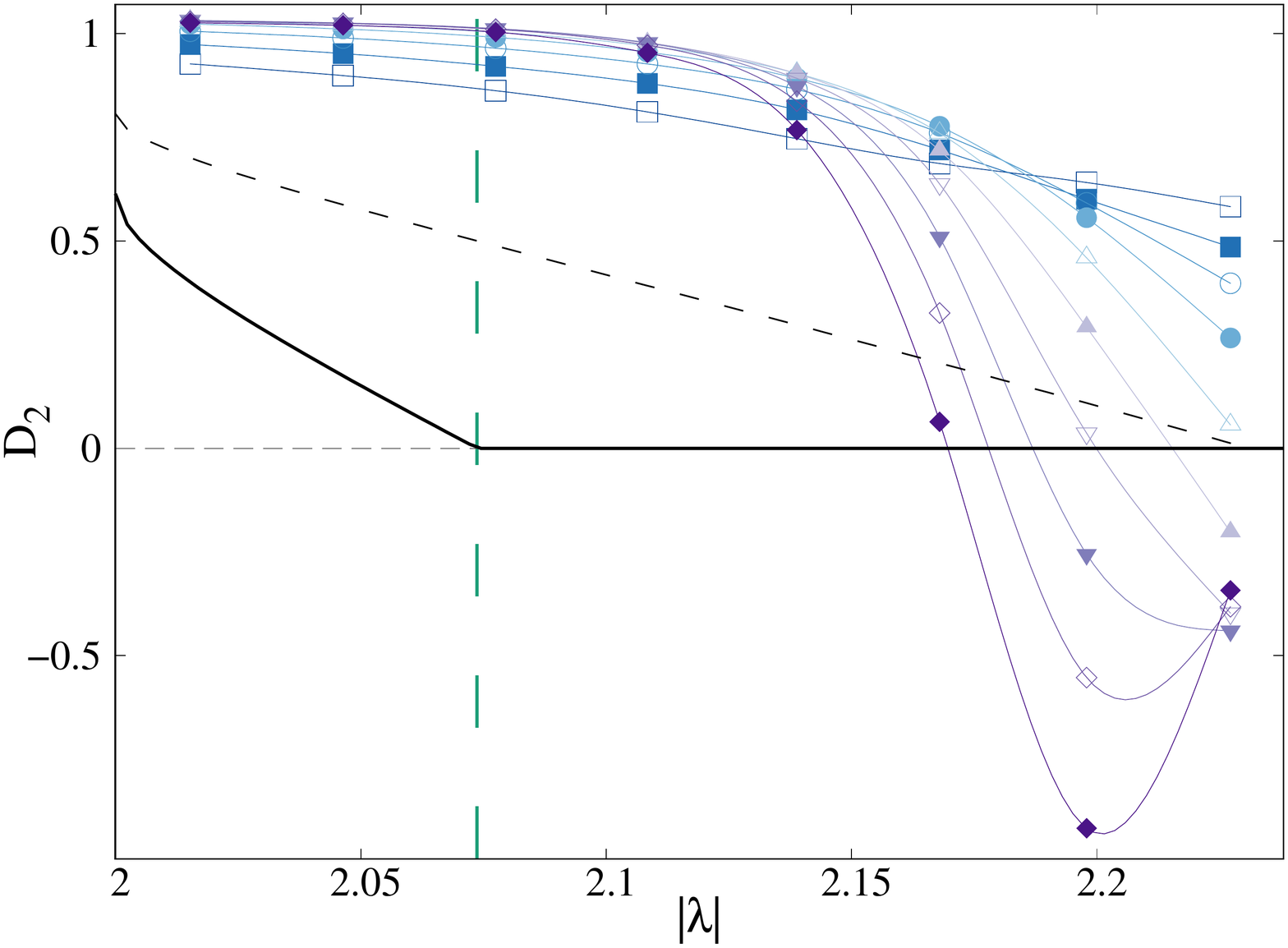} \hspace{-.4cm}
	 \includegraphics[width=0.345\textwidth]{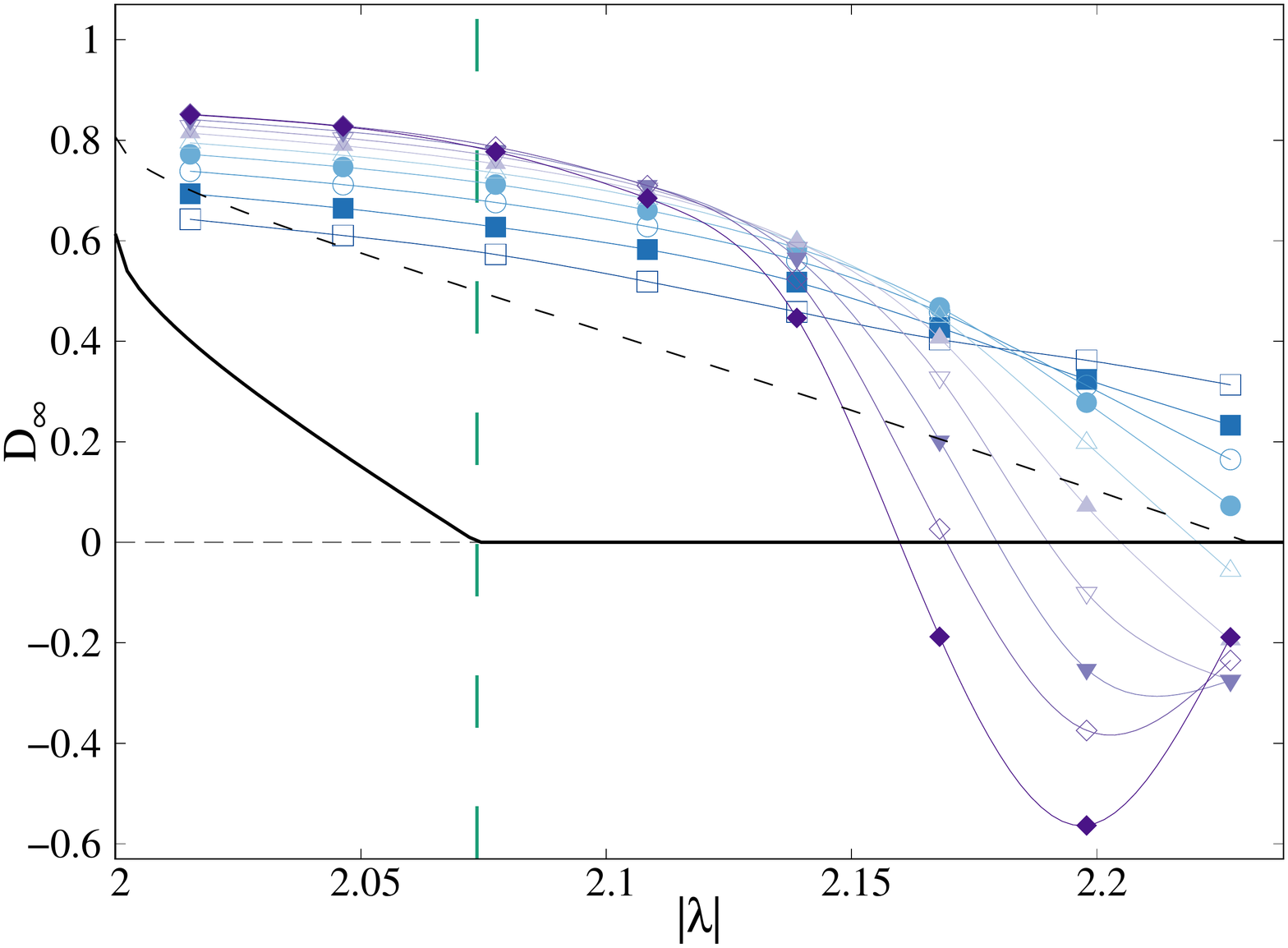}
	 	\vspace{-0.1cm}
	\caption{(color online) Flowing $N$-dependent fractal dimensions $D_1$ (left), $D_2$ (middle), and $D_\infty$ (right) as a function of $|\lambda| \in (2, \lambda_{\rm max})$ in the tails of the spectrum of the adjacency matrix of critical \ER graphs for $b=0.5$ ($\lambda_{\rm max} \approx 2.231$). $N =2^n$ with $n= 9 , \ldots, 18$ (different values of $n$ correspond to different symbols and colors as indicated in the legend). $D_q$  are computed via Eq.~\eqref{eq:D} from the scaling of the $q$-th moment of the wave-functions’ amplitudes measured from EDs. The solid line corresponds to the analytic estimation based on the Fermi Golden Rule, Eq.~\eqref{eq:D1}. The vertical dashed line 
	represents the position of $\lambda_{\rm loc} 
	\approx 2.074$. The dashed curve shows the value of the exponent $\tau(\lambda)$ associated to the scaling of the DoS, which gives an upper bound for the anomalous dimensions~\cite{Knowles}.
\label{fig:D1}}
\end{figure*}

\begin{figure*}
	 \hspace{-0.22cm}\includegraphics[width=0.341\textwidth]{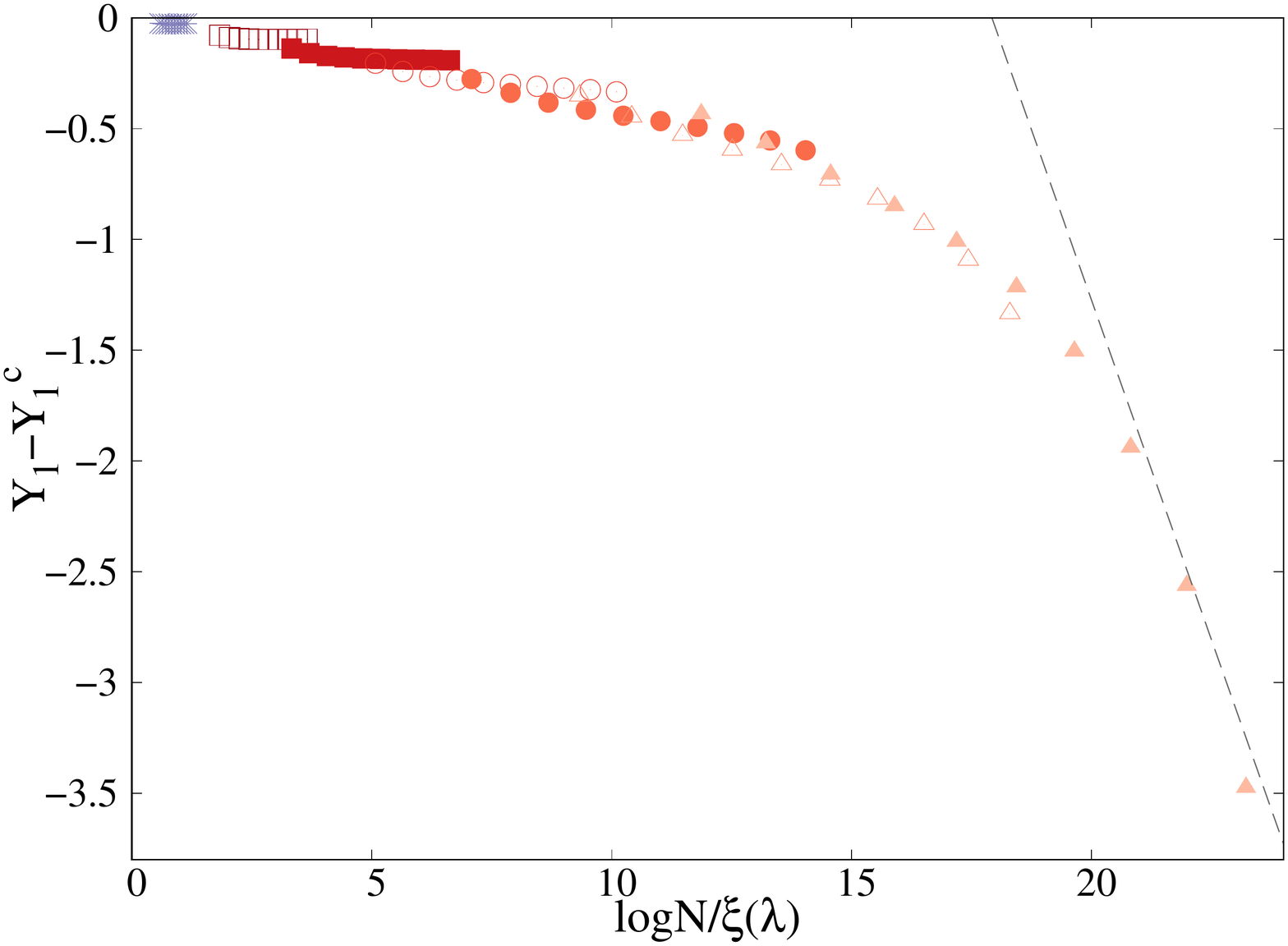}\hspace{-.3cm}
	 \includegraphics[width=0.341\textwidth]{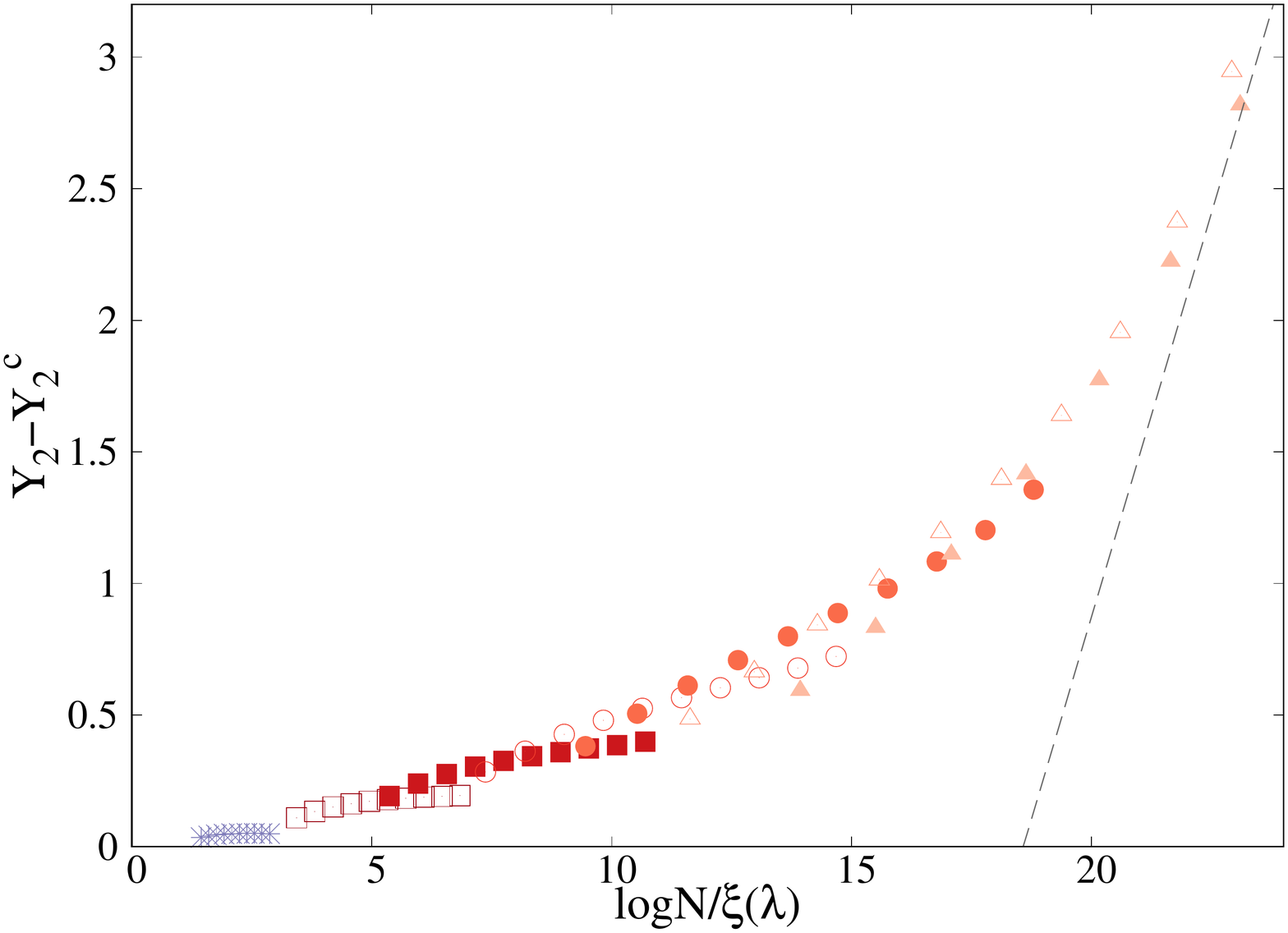} \hspace{-.3cm}
	 \includegraphics[width=0.341\textwidth]{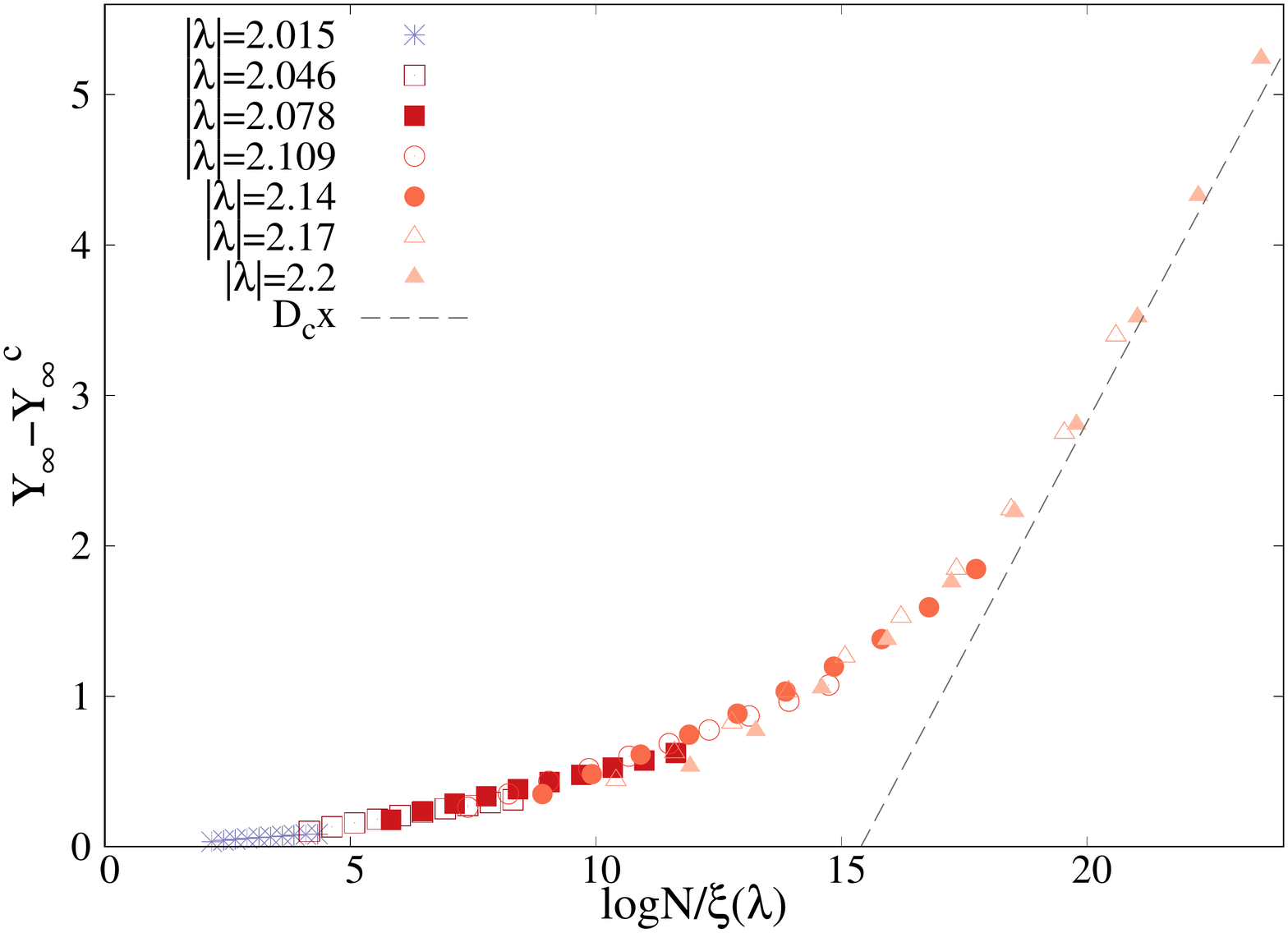}
	 	\vspace{-0.1cm}
	\caption{Scaling curves for the $q$-th moments of the wave-functions' amplitudes for $q=1$ (left), $q=2$ (middle), and $q=\infty$ (right) varying $\lambda$ in the tails of the spectrum of the adjacency matrix of critical \ER graphs for $b=0.5$. Different colors and symbols correspond to different values of the energy. $\Upsilon_q (N,\lambda) - \Upsilon_q (N,\lambda=2)$ are plotted as a function of the scaling variable $\log N / \xi(\lambda)$,  see Eq.~(\ref{eq:scaling}) with $\xi$ given by Eq.~(\ref{eq:xi}). The gray dashed lines correspond to the theoretical asymptotic behavior of the scaling functions that are predicted to exhibit a slope equal to $D_c = 1 - 2 b/b_\star$.
\label{fig:upsilon}}
\end{figure*}

We start now focusing on the transition for the statistics of the wave-functions' amplitudes taking place at $|\lambda|=\lambda_{\rm GOE} = 2$. To this aim we study the scaling behavior of the moments
\begin{equation} \label{eq:upsilon}
\begin{aligned}
\Upsilon_q (N,\lambda) &= \left \langle \log \left( \sum_{i=1}^N | \psi(i)|^{2q} \right) \right \rangle_{\!\lambda} \, , \\
\Upsilon_1 (N,\lambda) &= - \left \langle \sum_{i=1}^N | \psi(i)|^{2} \log | \psi(i)|^{2}  \right \rangle_{\!\lambda} \, ,
\end{aligned}
\end{equation}
where the averages $\langle \cdots \rangle_\lambda$ are done over the eigenfunctions of energy $\lambda$ and over different realizations of the graph. The flowing fractal dimensions are then obtained as logarithmic derivatives of the moments $\Upsilon_q$ with respect to $\log N$ (hereafter  the  logarithmic  derivatives  are  computed as discrete derivatives involving the five  available values of the system size closest to $N$~\cite{remark}): 
\begin{equation} \label{eq:D}
\begin{aligned}
D_q (N, \lambda) & = \frac{1}{1-q} \, \frac{\partial \Upsilon_q (N,\lambda) } {\partial \log N} \ , \\
D_1 (N, \lambda) & = \frac{\partial\Upsilon_1 (N, \lambda) } {\partial \log N} \, .
\end{aligned}
\end{equation}
In Fig.~\ref{fig:D1} we plot our numerical results for the flowing fractal exponent $D_q(N,\lambda)$ computed numerically according to Eq.~\eqref{eq:D}, and contrast it with the theoretical prediction of the Mott's argument based on the generalization of the Fermi Golden rule, Eq.~\eqref{eq:D1}. The figure shows that for $|\lambda| > \lambda_{\rm loc}$ 
the exponents $D_1$, $D_2$, and $D_\infty$ start to decrease rapidly as the system size is increased (and even take negative values). Conversely, for $|\lambda| < \lambda_{\rm loc}$, $D_q$ are still quite close to $1$ (and are still larger than $\tau$). Attempting a finite-size scaling analysis of these data is problematic due to the fact that $D_q$ should converge to a $\lambda$-dependent function. 

This kind of behavior 
is somewhat similar to the one observed in the insulating side of the MBL transition~\cite{mace,tarzia} (and also in the intermediate phase of the L\'evy RP ensemble~\cite{LRP}), in which the asymptotic values of $D_q$ depend continuously on the parameters of the model such as the disorder strength. We therefore perform a finite-size scaling analysis inspired by the one proposed in Refs.~\cite{mace,gabriel,LRP} to deal with this situation, which consists in positing that in the partially delocalized but non-ergodic region, $|\lambda| \in (\lambda_{\rm GOE}, \lambda_{\rm loc})$, the moments of the wave-functions' amplitudes $\langle \Upsilon_q \rangle$ [defined in Eq.~\eqref{eq:upsilon}] behave as:
\begin{equation} \label{eq:scaling}
	\begin{aligned}
		\Upsilon_1 (N,\lambda) - \Upsilon_1 (N,\lambda=2) &= -D_{1,c} \frac{\log N}{\xi (\lambda)} \, ,\\
		\Upsilon_q (N,\lambda) - \Upsilon_q (N,\lambda=2) &= (1 - q) D_{q,c} \frac{\log N}{\xi (\lambda)} \, ,
	\end{aligned}
\end{equation}
with $D_{q,c}$ being the fractal dimensions at the transition point. The length scale $\xi$ (i.e. the logarithm of a correlation volume $N_c (\lambda)$) depends on the distance to the critical point $\lambda_{\rm GOE} = 2$. 
The scaling ansatz above implies that in the limit $\log N \gg \xi$ the leading terms follows $\Upsilon_1 \simeq D_{1,c}( 1 - 1 / \xi (\lambda)) \log N$ and $\Upsilon_q \simeq - (q-1) D_{q,c}( 1 - 1 / \xi (\lambda)) \log N$, while in the opposite limit,  $\log N \ll \xi$, one retrieves the critical scaling. For simplicity here, in analogy with the RP model~\cite{kravtsov}, we assume that the mini-bands in the local spectrum in the partially delocalized but non-ergodic phase are fractal but non {\it multi}fractal, i.e. $D_q = D$ for all $q$. In order for Eq.~(\ref{eq:D1}) to be satisfied one then needs to have:
\begin{equation} \label{eq:xi}
	\xi(\lambda) = \frac{D_c}{D_c - D(\lambda)} \, ,
\end{equation}
where $D_c = 1 - 2 b/b_\star$ and $D(\lambda)$ is given in Eq.~\eqref{eq:D1}. As shown in Fig.~\ref{fig:upsilon} for $b=0.5$, a reasonably good data collapse is obtained in the partially delocalized phase  when the $q$-th moments of the wave-functions amplitudes for different values of the energy are plotted as a function of the scaling variable $\log N / \xi (\lambda)$, where  $\xi(\lambda)$ is chosen as in Eq.~(\ref{eq:xi}). Note that the quality of the data collapse is especially good since there is in fact {\it no adjustable parameter} in this procedure. Since $\tilde{\kappa} (2 + \epsilon) \approx 2(1 + \sqrt{\epsilon})$, in the vicinity of the transition to the fully delocalized GOE like phase one has that $\tau(2 + \epsilon) - \tau(2) \approx 2 b \log 2 \sqrt{\epsilon}$. Hence the scaling analysis of Fig.~\ref{fig:upsilon} indicates that:
\[
\xi (\lambda) = 
\frac{b_\star - 2 b}{2 b (b_\star + 1) \sqrt{|\lambda|-2}} \, , 
\]
i.e. $\nu_{\rm GOE} = 0.5$. 

\begin{figure*}
	 \hspace{-0.22cm}\includegraphics[width=0.345\textwidth]{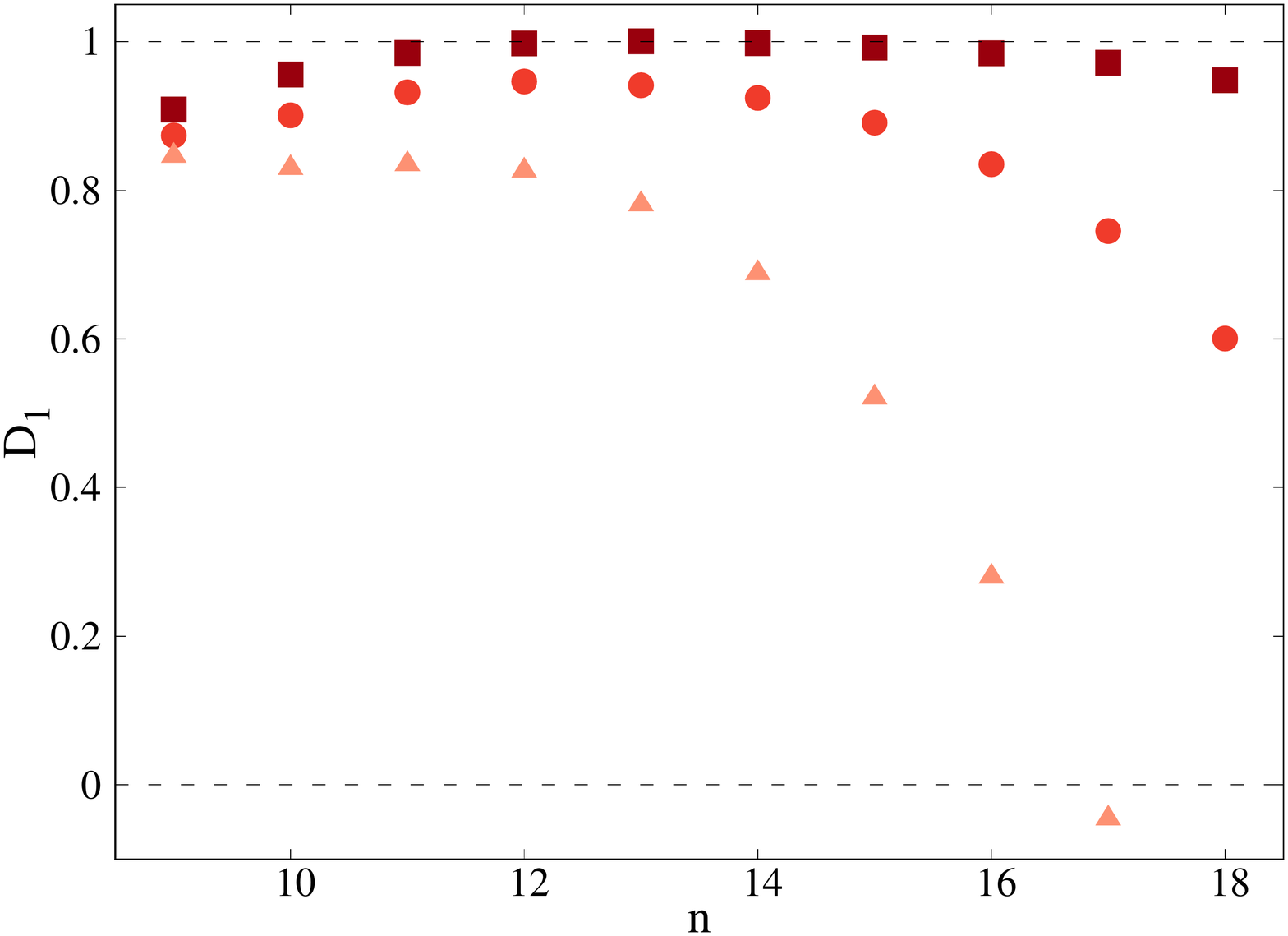}\hspace{-.4cm}
	 \includegraphics[width=0.345\textwidth]{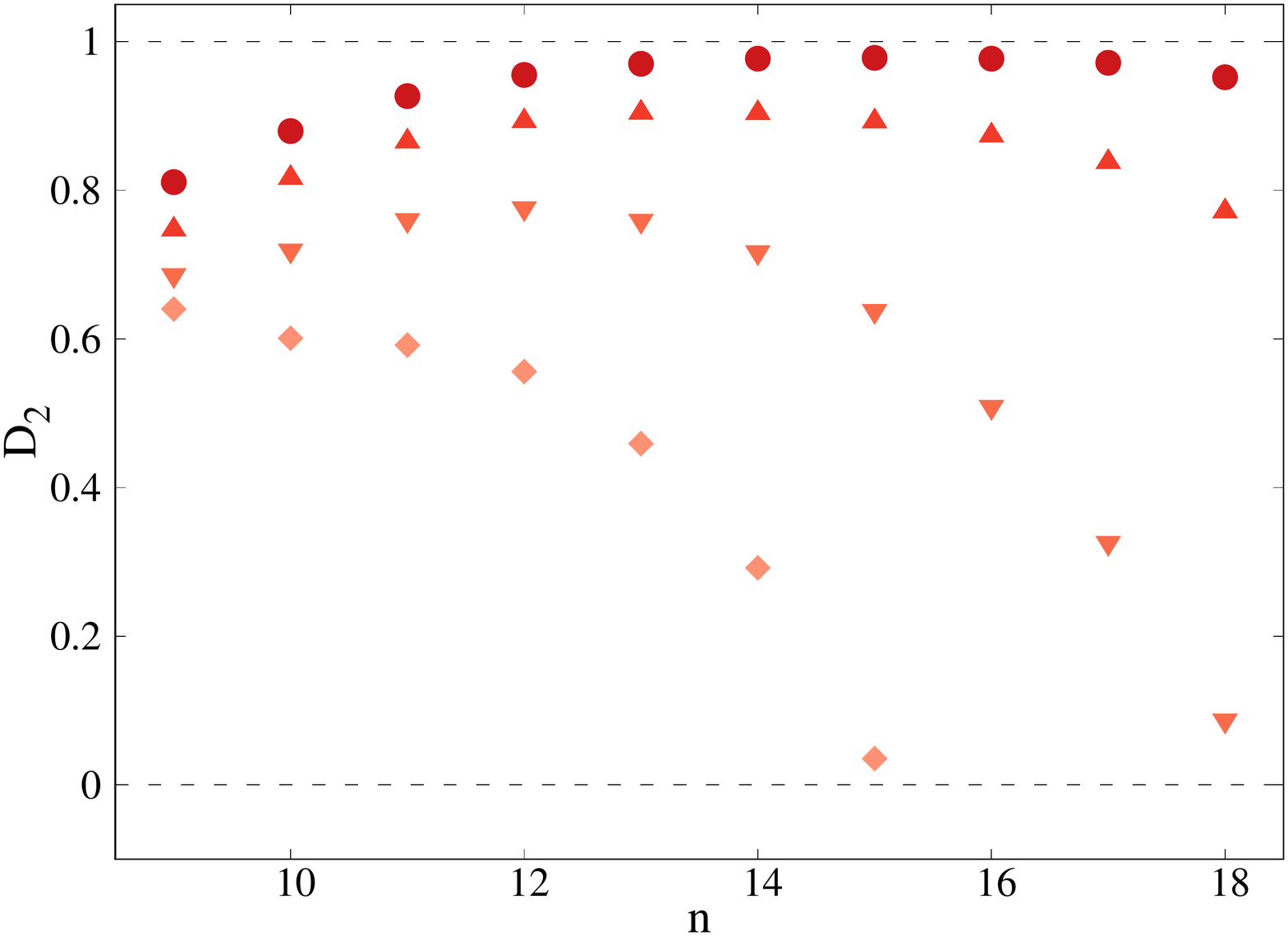} \hspace{-.4cm}
	 \includegraphics[width=0.345\textwidth]{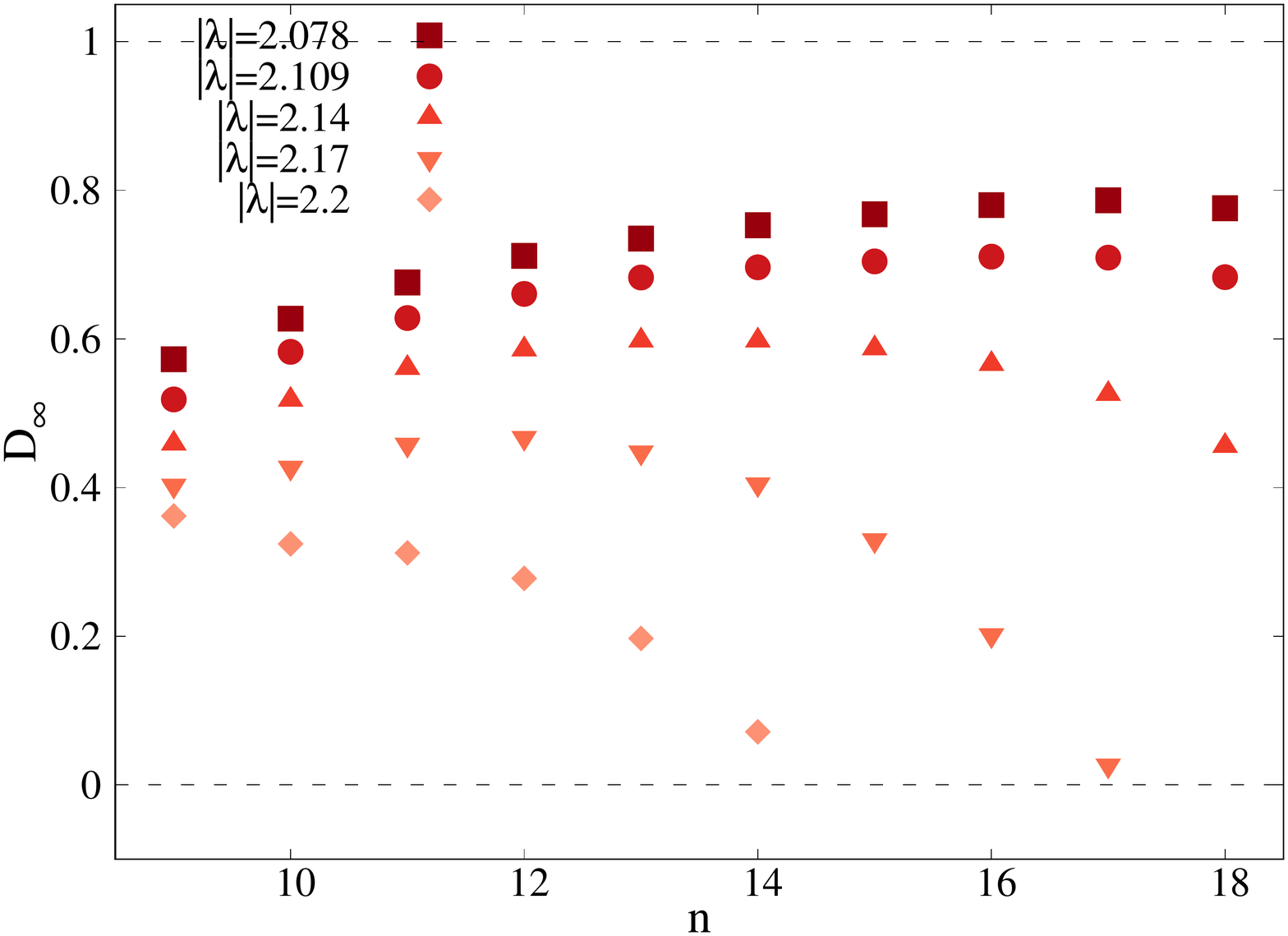}
	 	\vspace{-0.1cm}
	\caption{(color online) Flowing fractal dimensions $D_1$ (left), $D_2$ (middle), and $D_\infty$ (right) as a function of $n=\log_2 N$ for different values of $|\lambda| \in (2, \lambda_{\rm max})$ in the tails of the spectrum of critical \ER graphs of $N$ vertices and average degree $c=b \log N$ with $b=0.5$. Different symbols and colors correspond to different values of the energy  as indicated in the legend. The fractal exponents exhibit a clear non-monotonic dependence on $n$.
\label{fig:maxD}}
\end{figure*}

\begin{figure*}
\hspace{-0.22cm}\includegraphics[width=0.345\textwidth]{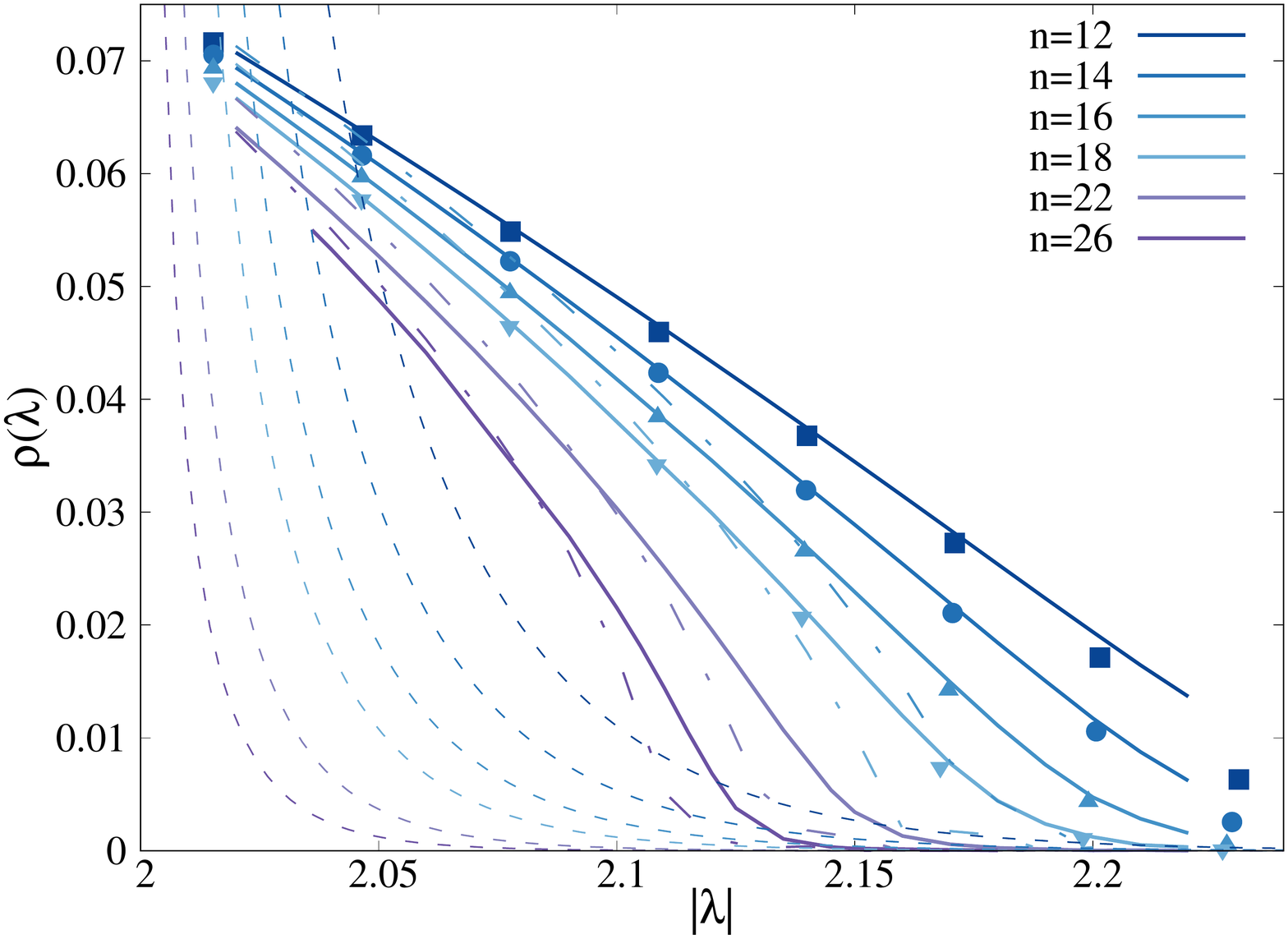}\hspace{-.4cm}
	 \includegraphics[width=0.345\textwidth]{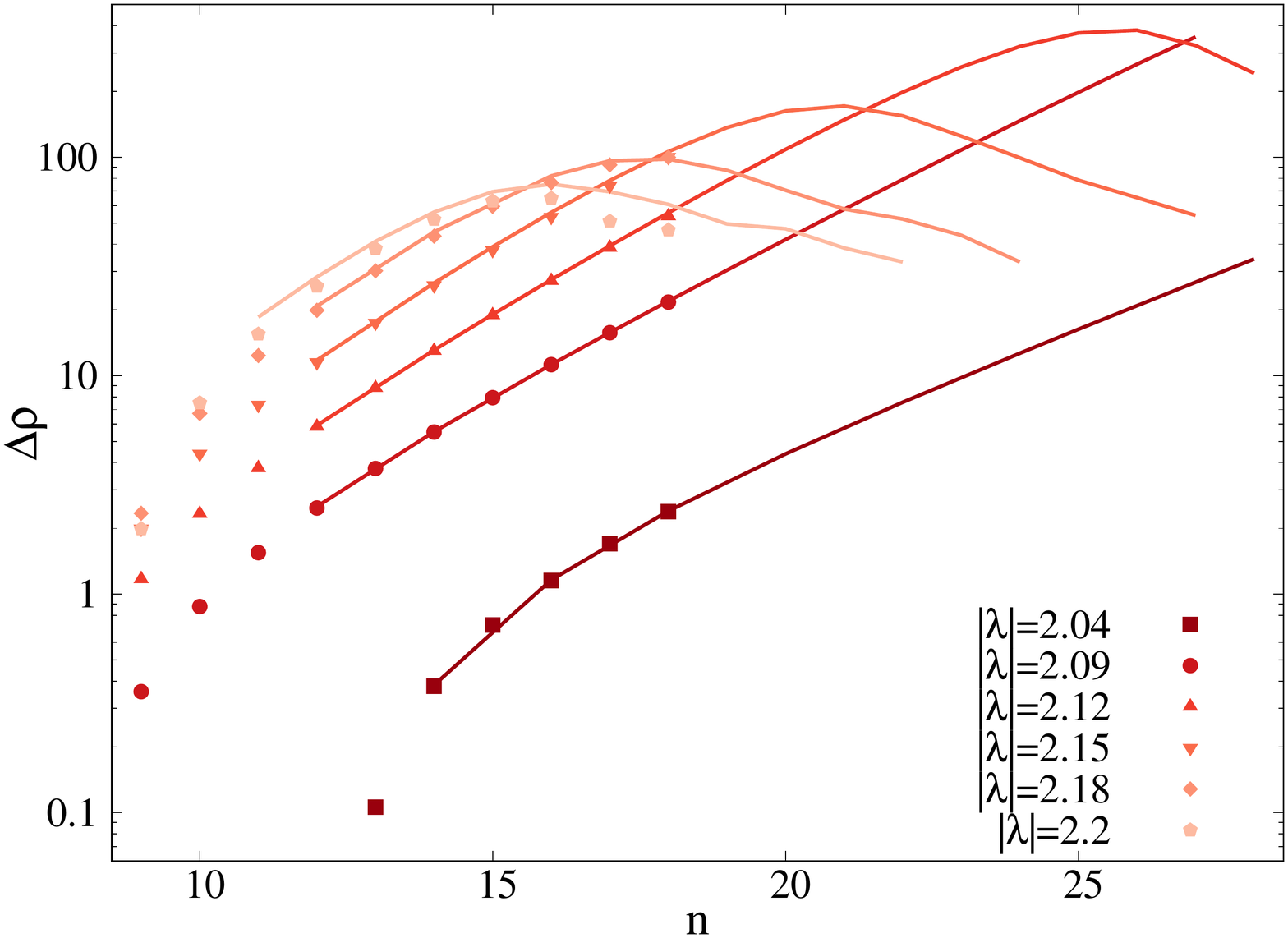} \hspace{-.4cm}
	 \includegraphics[width=0.345\textwidth]{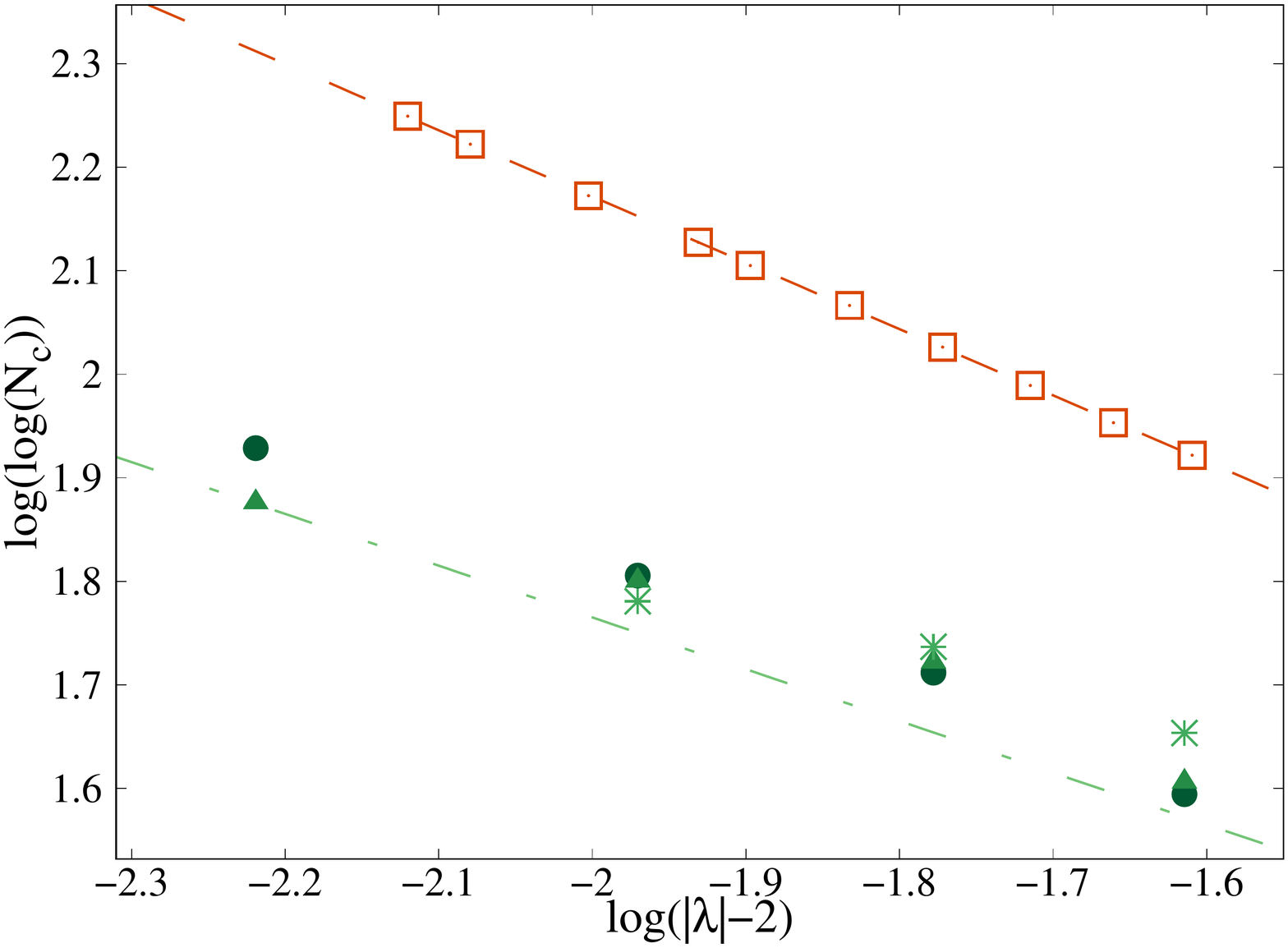}
	 	\vspace{-0.1cm}
	\caption{(color online) Left panel: Average DoS of critical \ER graphs (with $b=0.5$) in the semilocalized phase $|\lambda| \in (2, \lambda_{\rm max})$ computed using EDs (symbols) and the numerical solution of Eqs.~\eqref{eq:green} and~\eqref{eq:greenF} (thick continuous lines). $\lambda_{\rm max} \approx 2.231$ for $b=0.5$. Different system sizes $N = 2^n$ (with $n$ from $12$ to $26$) correspond to different symbols and colors as indicated in the legend. The dashed lines correspond to the asymptotic value of the DoS $\rho^{\infty}$ given in Ref.~\cite{Knowles} and in Eq.~\eqref{eq:dos}. The dashed-dotted lines corresponds to the approximate average DoS obtained from the solution of Eq.~\eqref{eq:dosapprox}, which is in reasonably good agreement with the average DoS obtained from EDs and from the solution of the exact cavity equations. Middle panel: Relative distance at finite $N$ of the DoS from its asymptotic scaling behavior $\Delta \rho = (\rho - \rho^{\infty})/\rho^{\infty}$ for several values of the energy in the interval $|\lambda| \in (2, \lambda_{\rm max})$ as a function of $n = \log_2 N$. Filled symbols correspond to ED results and solid lines to the results obtained from the solution of the cavity equations. Different symbols and colors correspond to different values of the energy  as indicated in the legend. The curves exhibit a non-monotonic behavior with a maximum at a characteristic volume $N_c(\lambda)$. Right panel: $\log \log (N_c)$ as a function of the log of the distance from the transition point, $\log(|\lambda|-2)$. The empty squares correspond to the values of $N_c$ extracted from the maximum of $\Delta \rho$, and are very well fitted by $\log \log N_c = a - \nu_{\rm GOE} \log (\lambda - 2)$ with $a\approx 0.891$ and $\nu_{\rm GOE} \approx 0.64$ (dashed straight line). The filled circles, up triangles, and down triangles correspond to the values of $N_c$ estimated from the non-monotonic behavior of the flowing fractal dimensions $D_1$, $D_2$, and $D_\infty$ respectively (see Fig.~\ref{fig:maxD}). The dashed-dotted line represents the estimation of $\xi \propto (|\lambda|-2)^{-1/2}$ given in Eq.~\eqref{eq:xi} (with $\nu_{\rm GOE} = 0.5$) obtained from the scaling analysis of the moments of the wave-functions amplitudes proposed in Sec.~\ref{sec:level}.
\label{fig:dos}}
\end{figure*}

An independent estimation of the exponent $\nu_{\rm GOE}$ which describes how the correlation length scale $\xi(\lambda)$ diverges when the critical point is approached can be obtained  from the non-monotonic behavior of the flowing fractal dimensions $D_q$ at fixed energy and as a function of the system size. In Fig.~\ref{fig:maxD} we plot the numerical estimations of $D_1$, $D_2$, and $D_\infty$ as a function of $n = \log_2 N$ for several values of $\lambda$ within the interval $2<|\lambda| < \lambda_{\rm max}$. The plots show that the $D_q$'s first grow at small $N$ and then decrease at large $N$ after passing through a maximum at a characteristic scale $N_c$. The position of the maximum moves to larger values of $N$ when $\lambda$ gets closer to $2$. The values of $N_c (\lambda)$ estimated from the non-monotonic behavior of the $D_q$'s are shown in Fig.~\ref{fig:dos}, indicating that the characteristic scale $\log(N_c)$ governing the finite-size behavior of the fractal exponents is well fitted by a power-law divergence of the form $\log(N_c) \propto (|\lambda| - 2)^{- \nu_{\rm GOE}}$ with $\nu_{\rm GOE} \approx 0.5$, and appears to be proportional to the correlation length $\xi(\lambda)$ extracted from the finite-size scaling analysis of Fig.~\ref{fig:upsilon}. 

\subsection{Convergence of the average density of states} \label{sec:dos}
In this section we investigate the convergence of the average DoS in the tails of the spectrum of the adjacency matrix of critical \ER graphs to the exact asymptotic behavior obtained in Ref.~\cite{Knowles} and given in Eq.~\eqref{eq:dos}. This analysis will allow us to obtain another complementary estimation of the characteristic scale that governs finite-size corrections. The numerical results are obtained using both exact diagonalizations (for sizes $N=2^n$ with $9 \le n \le 18$) and the numerical solution of the self-consistent cavity equations for the Green's function (for sizes $N=2^n$ with $12 \le n \le 28$). In both cases we have set $b=0.5$.

In the left panel of Fig.~\ref{fig:dos} we plot the average DoS in the 
interval $2<|\lambda| < \lambda_{\rm max}$ for several system sizes obtained from EDs (symbols) and the cavity method (solid lines) for $b=0.5$. We also plot the exact asymptotic estimation of Eq.~\eqref{eq:dos} obtained by counting the number of vertices of abnormally large degree corresponding to a given energy (dashed lines)~\cite{Knowles}, as well as the estimation  of the average DoS obtained from the approximate treatment of the cavity equations, Eq.~\eqref{eq:dosapprox} (dashed-dotted lines). 

The first important observation is that the results obtained using the cavity method 
are in excellent agreement with the ED ones, although the DoS is still very far from the asymptotic expression~\eqref{eq:dos} for the accessible system sizes.  We also note that the approximate DoS obtained from Eq.~\eqref{eq:dosapprox} provides in fact a reasonably good approximation. 
In order to characterize the finite-size corrections it is instructive to compute the relative distance between the measured DoS at finite $N$ from the asymptotic value $\rho^{\infty}$. 
In the right panel of Fig.~\ref{fig:dos} we plot $\Delta \rho = (\rho - \rho^{\infty})/\rho^{\infty}$ as a function of $n = \log_2 N$ for several values of the energy in the interval $2<|\lambda| < \lambda_{\rm max}$. 
The plot clearly shows that $\Delta \rho$ has a non-monotonic behavior as a function of $N$ characterized by a well-defined maximum that becomes higher and moves to larger values of $N$ as the energy is decreased. This implies that the finite-size corrections become stronger and stronger as the transition from the semilocalized phase and the fully delocalized one is approached, and are governed by a characteristic volume that grows when $\lambda$ gets close to the transition point at $|\lambda| = \lambda_{\rm GOE} = 2$. By determining the position of the maximum of $\Delta \rho$ for different values of $\lambda$ one thus obtains a direct estimation of the correlation volume $N_c (\lambda)$, which is shown in the right panel of Fig.~\ref{fig:dos}. It turns out that $N_c$
is  well fitted by a an exponential divergence at the transition point of the form $\log N_c (\lambda) \propto (\lambda - 2)^{-\nu_{\rm GOE}}$, with $\nu_{\rm GOE} \approx 0.64$.  The plot also indicates that such estimation of $N_c$ is roughly proportional to the one obtained from the non-monotonic behavior of the flowing fractal dimensions and from the finite-size scaling analysis of Fig.~\ref{fig:upsilon}.

\begin{figure*}
	 \hspace{-0.22cm}\includegraphics[width=0.51\textwidth]{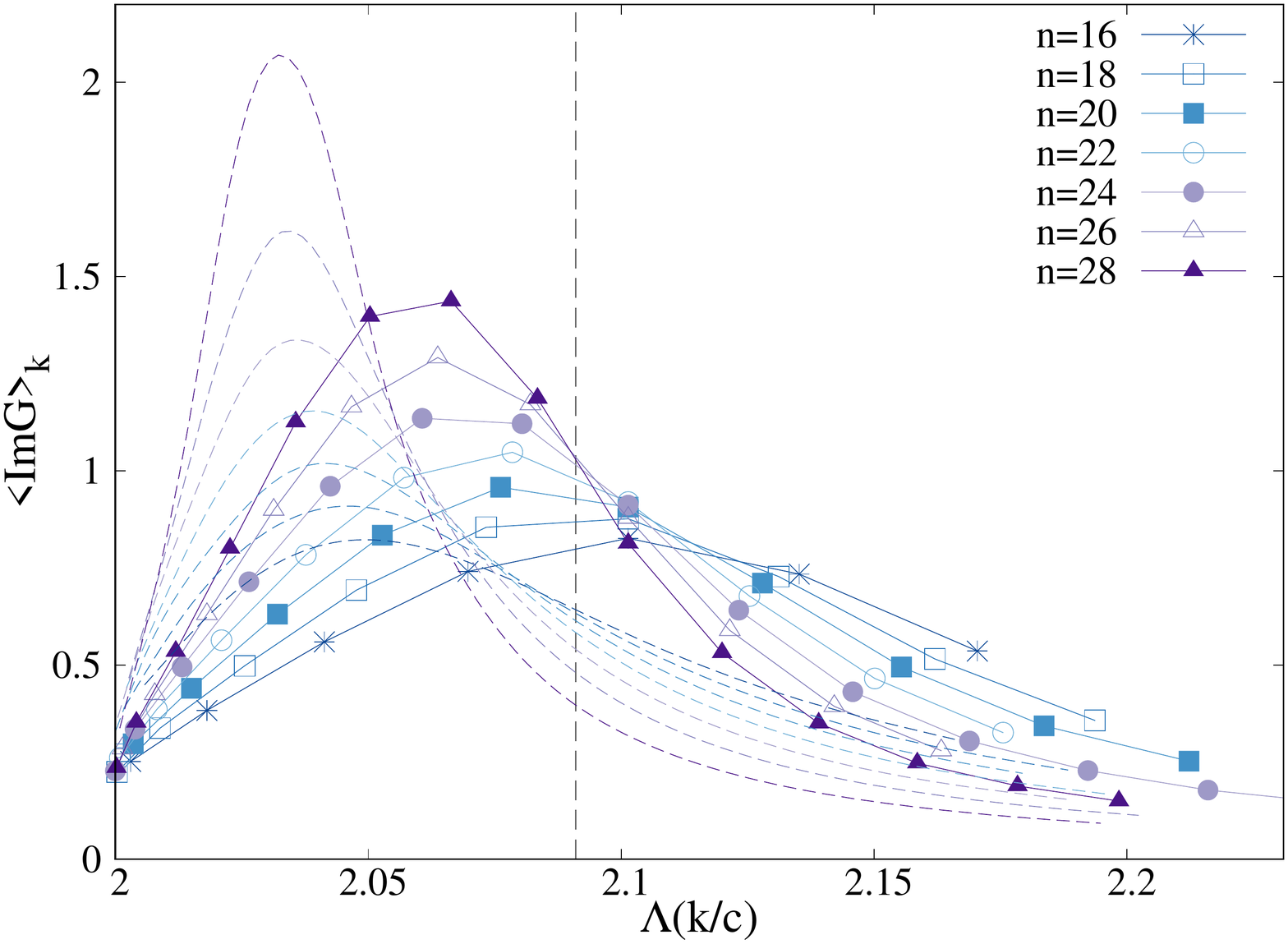}\hspace{-.35cm}
	 \includegraphics[width=0.51\textwidth]{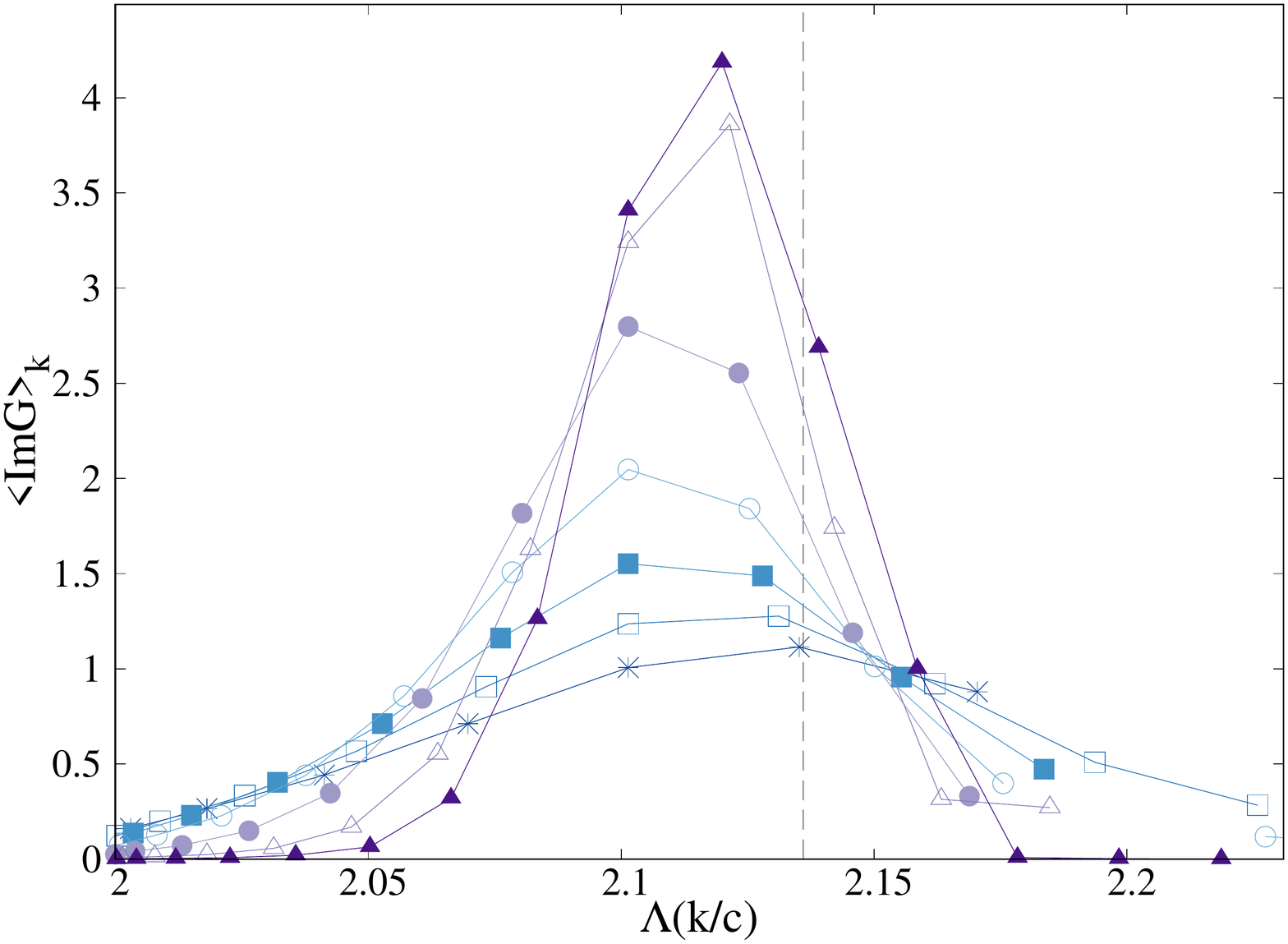} 
	 	\vspace{-0.2cm}
	\caption{(color online) Average DoS restricted to vertices of degree $k$, $\langle {\rm Im} G \rangle_k$, as a function of $\Lambda(k/c)$, in the tails of the spectrum of critical \ER graphs (with $b=0.5$). Different system sizes $N = 2^n$ (with $n$ from $16$ to $28$) correspond to different symbols and colors as indicated in the legend. The vertical dashed lines corresponds to the value of the corresponding energy, i.e. $\lambda=2.09$ (left panel) and $\lambda=2.135$ (right panel). The dashed curves in the left panel show to the value of $\langle {\rm Im}G \rangle_k$ obtained using the approximation of Eqs.~\eqref{eq:gapprox} and~\eqref{eq:dosapprox}.}
\label{fig:igk}
\end{figure*}

It is also instructive to study the evolution with the system size of the average DoS restricted to the nodes with degree $k$ (with $k> 2 c$), 
$\langle {\rm Im} G \rangle_k$. 
This quantity, which can be easily computed numerically from the solution of the  self-consistent cavity equations for the resolvent, is plotted in Fig.~\ref{fig:igk} for two values of the energy in the interval $|\lambda| \in (2, \lambda_{\rm max})$ as a function of $\Lambda(k/c)$ (i.e., the value of the energy which in the thermodynamic limit is associated with vertices of degree $k$) for several system sizes $n=\log_2 N$ with $16 \le n \le 28$. These plots show that the correlation volume $N_c$ also reflects in the finite-size behavior of $\langle {\rm Im} G \rangle_k$.  In fact, according to the rigorous results of Refs.~\cite{Knowles,Knowles1}, in the thermodynamic limit $\langle {\rm Im} G \rangle_k$ should approach a narrowly peaked function around $\lambda$ (vertical dashed lines), due to the bijection between resonant vertices of degree greater than $2c$ and eigenvalues larger than $2$. For the accessible system sizes, we observe that $\langle {\rm Im} G \rangle_k$ exhibits a maximum located around values of the $\Lambda(k/c)$ smaller than $\lambda$. The position of the maximum moves first slightly leftwards for $N < N_c (\lambda)$, and then slightly rightwards for $N > N_c (\lambda)$, while the function becomes more peaked as the system size is increased. In the left panel we also show the results  obtained for $\lambda=2.09$ using the approximation of Eqs.~\eqref{eq:gapprox} and~\eqref{eq:dosapprox} (dashed curves), which in fact describe qualitatively well the evolution of $\langle {\rm Im} G \rangle_k$ with the system size. The same approximation cannot be used for $\lambda = 2.135$ since the system enters in the localized regime ($N > N_{\rm loc} (\lambda)$) in which the approximation breaks down, as explained in Sec.~\ref{sec:cavity}.

All in all, the resulted presented above indicate the presence of a correlation volume, $N_c (\lambda)$, which diverges exponentially fast when $|\lambda| \to 2$ with an exponent close to $\nu_{\rm GOE} \approx 0.5$. Note that a similar divergence is also observed on the delocalized side of the Anderson model on the Bethe lattice~\cite{mirlin_fyodorov,fyodorov_mirlin,fyodorov_mirlin_sommers,fyod,mirlin1994,Zirn,tikhonov2019,gabriel,large_deviations,mirlinrrg}. In this case the volumic scaling is associated to the fact that the critical point is in the localized phase and the fractal dimensions exhibit a discontinuous jump at the critical point from $D_q = 0$ for $W \ge W_c$  to $D_q = 1$ for $W \to W_c^-$. For critical \ER graphs the situation is somehow reversed, in the sense that here the critical point at $\lambda=2$ is in the delocalized phase, with a finite jump of the fractal dimensions from $D_q = 1$ for $|\lambda| \le 2$ to $D_q < 1$ for $|\lambda| \to 2^+$, and the scaling in terms of an exponentially large correlation volume is found on the semilocalized side of the transition~\cite{LRP}.

\begin{figure*}
	 \hspace{-0.22cm}\includegraphics[width=0.341\textwidth]{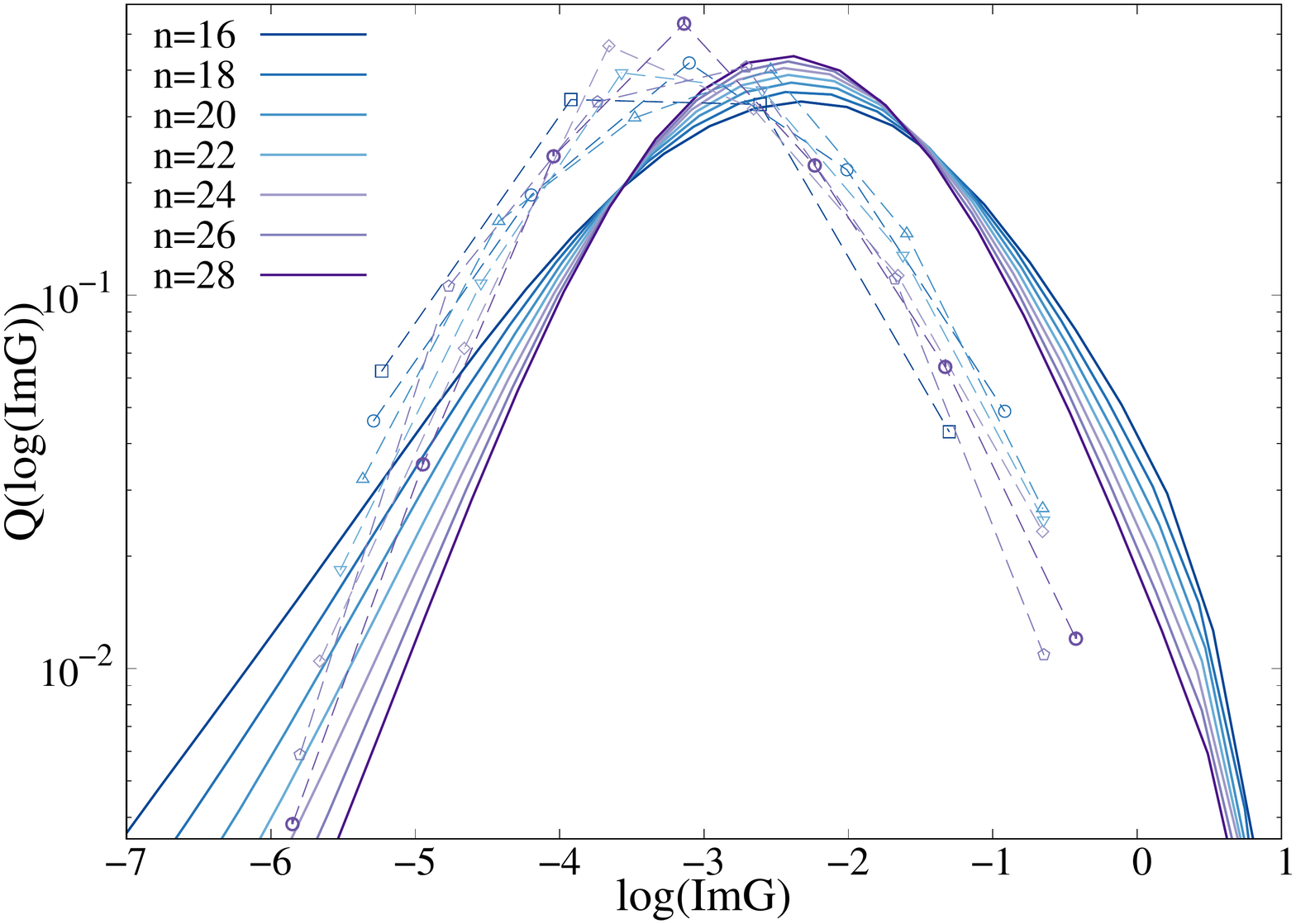}\hspace{-.2cm}
	 \includegraphics[width=0.341\textwidth]{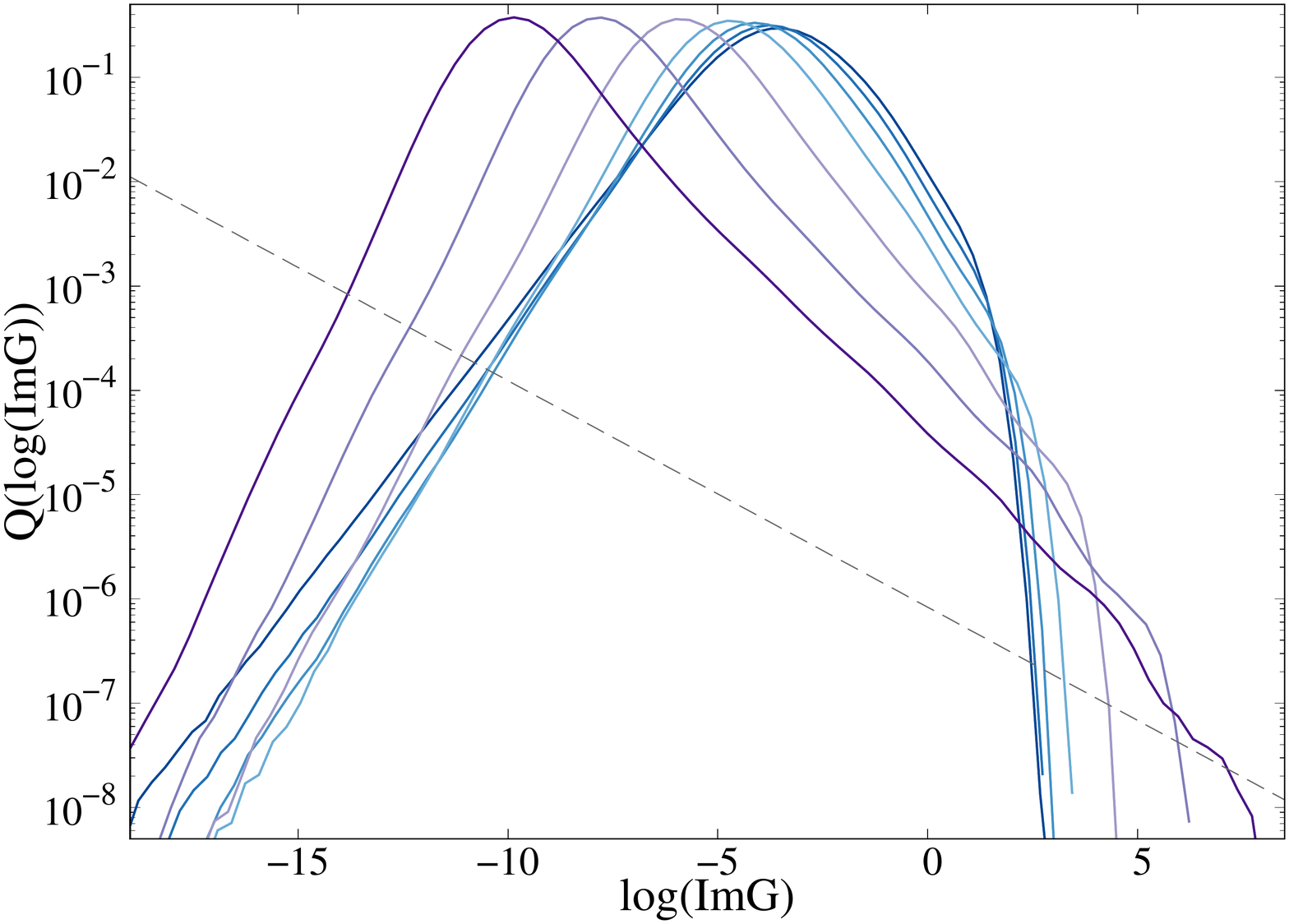} \hspace{-.4cm}
	 \includegraphics[width=0.341\textwidth]{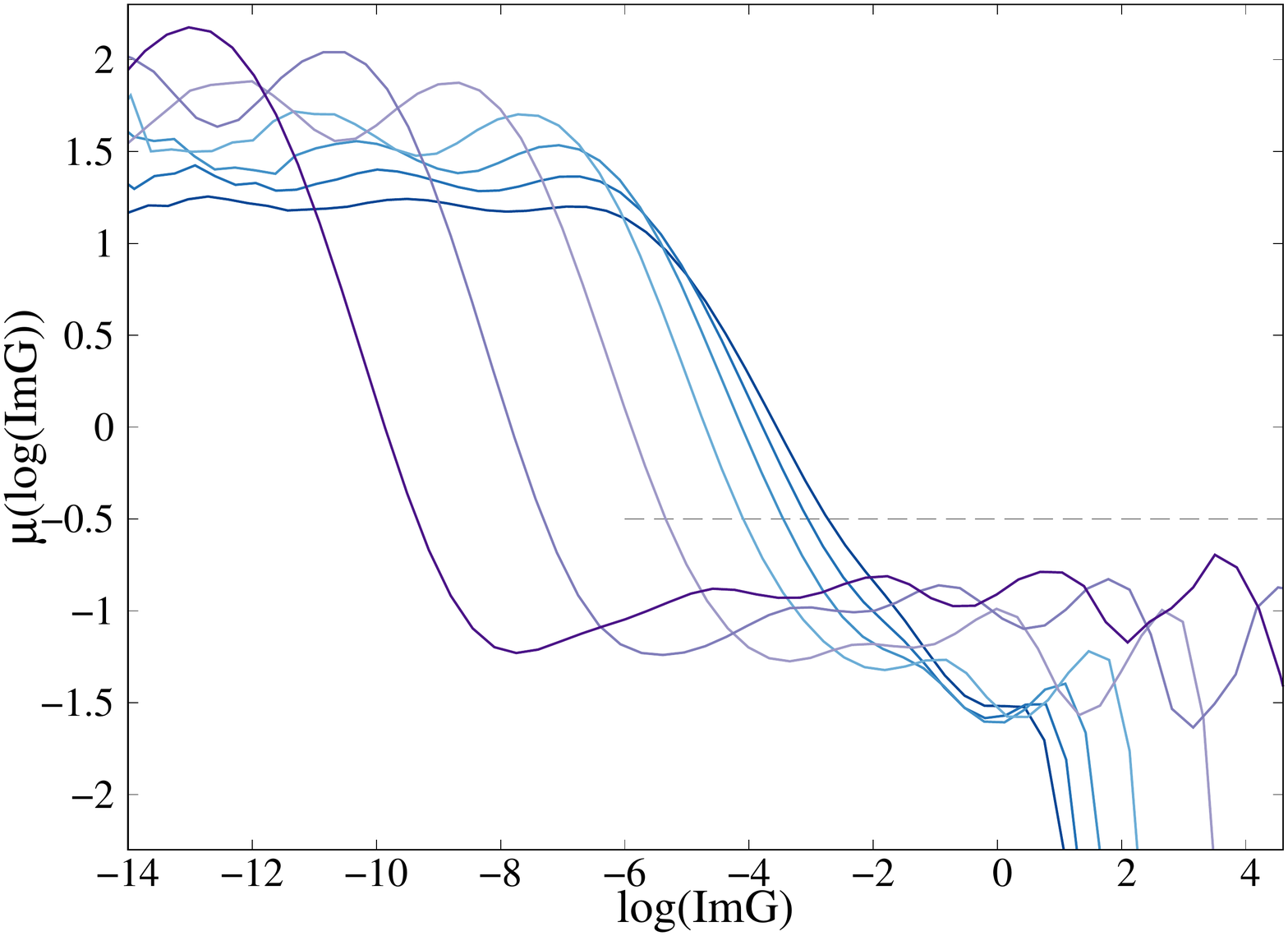}
	 	\vspace{-0.1cm}
	\caption{(color online) Probability distribution functions of $\log {\rm Im} G$ for critical \ER graphs (with $b=0.5$) for $\lambda=2.04$ (left panel) and $\lambda=2.135$ (right panel) and for several system sizes $N= 2^n$ with $n= 16, \ldots, 28$ as indicated in the legend. The dashed straight line in the middle panel represents the slope $\mu = -1/2$ of the standard localized phase. In the right panel we plot the local slope of the probability distribution $\mu(\log ({\rm Im} G))$, as defined in the text.}
\label{fig:Pldos}
\end{figure*}

\subsection{Statistics of the local density of states through the mobility edge}

The transition from the partially delocalized phase to the fully localized one can be also inspected by analyzing numerically the spectral statistics of the LDoS and of its correlations. Throughout this section we will consider critical \ER graphs with average degree $c = b \log N$ with $b=0.5$. 

In Fig.~\ref{fig:Pldos} we plot the probability distribution $Q(\log ({\rm Im} G))$ obtained solving the self-consistent cavity equations~\eqref{eq:green} and~\eqref{eq:greenF} for several system sizes $N=2^n$ and for two values of the energy respectively in the putative partially delocalized but non-ergodic phase ($\lambda=2.04$, left panel) and in the fully localized phase ($\lambda=2.135$, middle panel). The imaginary regulator $\eta$ is set here to a very small value ($\eta = 10^{-16}$), much smaller than the mean level spacing. For $\lambda=2.135$ the probability distribution of the LDoS seems to approach slowly but gradually the standard localized behavior as $N$ is increased: In particular one clearly observes the emergence of a power-law regime which becomes broader and broader as $N$ is increased and is characterized by an exponent which evolves  with $N$. The power-law establishes between the typical value of the LDoS (which drifts to smaller values when $N$ is increased) and a sharp cut-off (that drifts to larger values as $N$ is increased). In order to characterize the exponent of the power-law, in the right panel of Fig.~\ref{fig:Pldos} we plot the local slope of the distribution function, computed numerically as:
\[
\mu(\log ({\rm Im} G)) = \frac{\partial Q (\log ({\rm Im} G))}{\partial (\log ({\rm Im} G))} \, .
\]
In the standard localized regime the tails of the distribution of the LDoS are described by $Q({\rm Im} G) \propto \sqrt{\eta}/({\rm Im} G)^{3/2}$ (for ${\rm Im} G$ smaller than a cut-off proportional to $\eta^{-1}$), i.e. $\mu = -1/2$. The figure indeed shows that as $N$ is increased the region where $\mu$ is approximately constant becomes broader and the values of $\mu$ slowly increases towards the value $\mu= -1/2$ (dashed lines). 

This behavior must be contrasted with the one of the partially delocalized phase, shown in the left panel of Fig.~\ref{fig:Pldos}. For $\lambda=2.04$ one indeed observes (at least for the accessible system sizes) that the typical value and the cut-off of the distributions of the LDoS stay of order $1$ as $N$ is increased and, albeit an apparent power-law regime seems to set in for large enough $N$, the exponent $\mu$ is much smaller than $-1/2$ and decreases with $N$. We also show the approximate result for $Q (\log ({\rm Im} G))$ obtained from Eqs.~\eqref{eq:dosapprox}, \eqref{eq:gapprox}, and~\eqref{eq:qgapprox}, that in fact accounts reasonably well for the exact distributions in this regime.

\begin{figure*}
	 \hspace{-0.22cm}\includegraphics[width=0.49\textwidth]{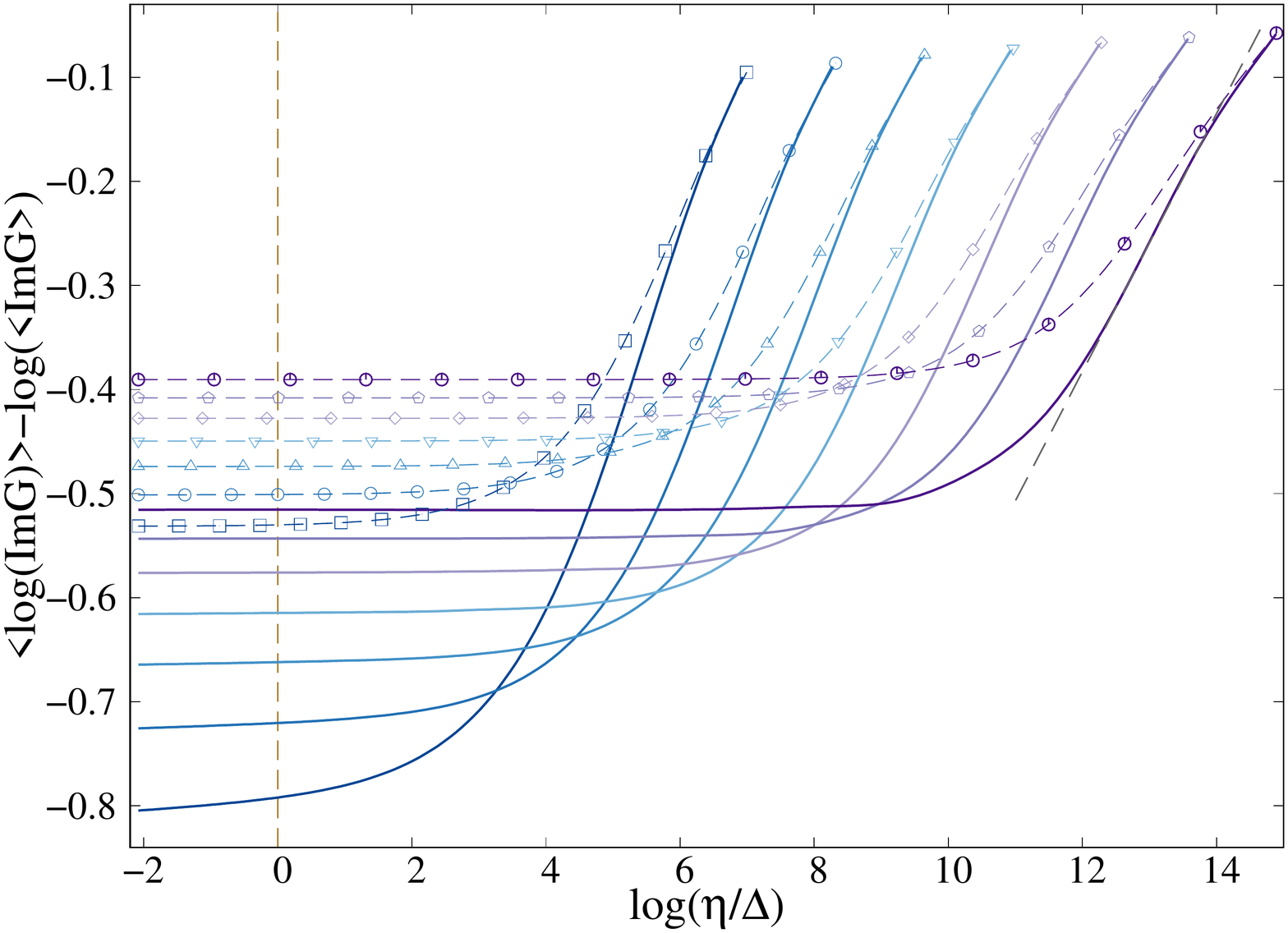}\hspace{-.2cm}
	 \includegraphics[width=0.49\textwidth]{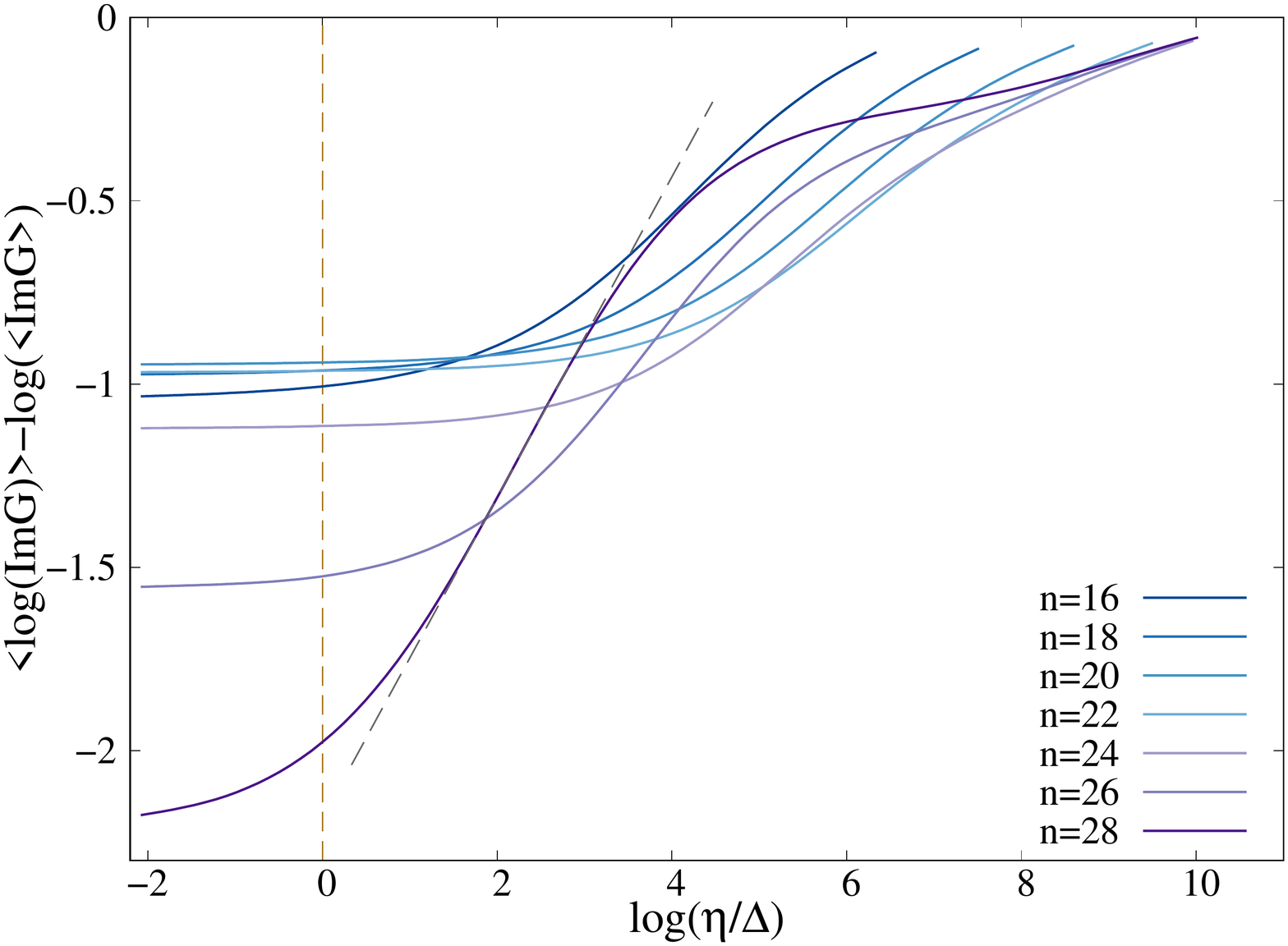} \vspace{.1cm}
	 
	 \hspace{-0.22cm}\includegraphics[width=0.49\textwidth]{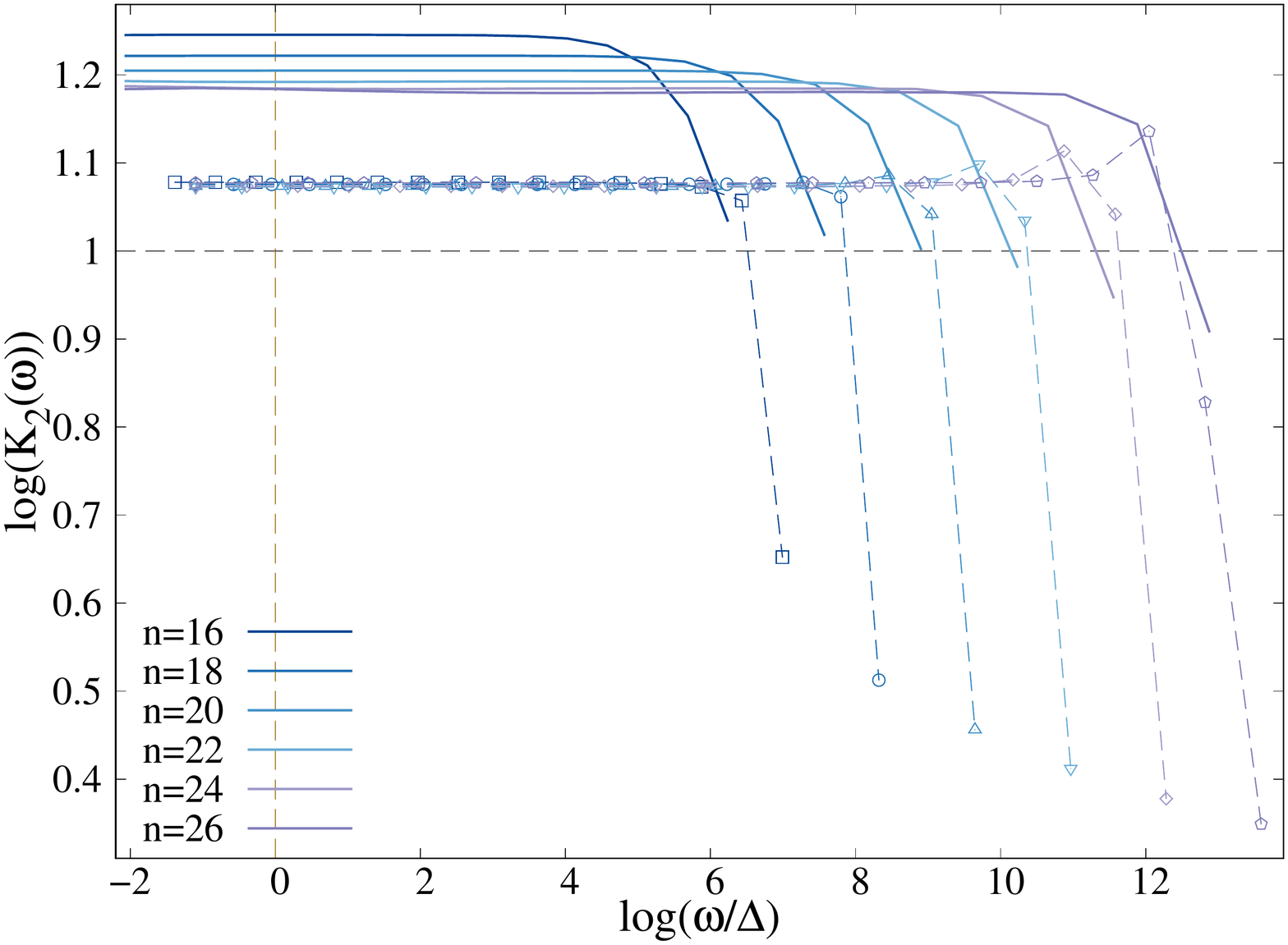}\hspace{-.0cm}
	 \includegraphics[width=0.49\textwidth]{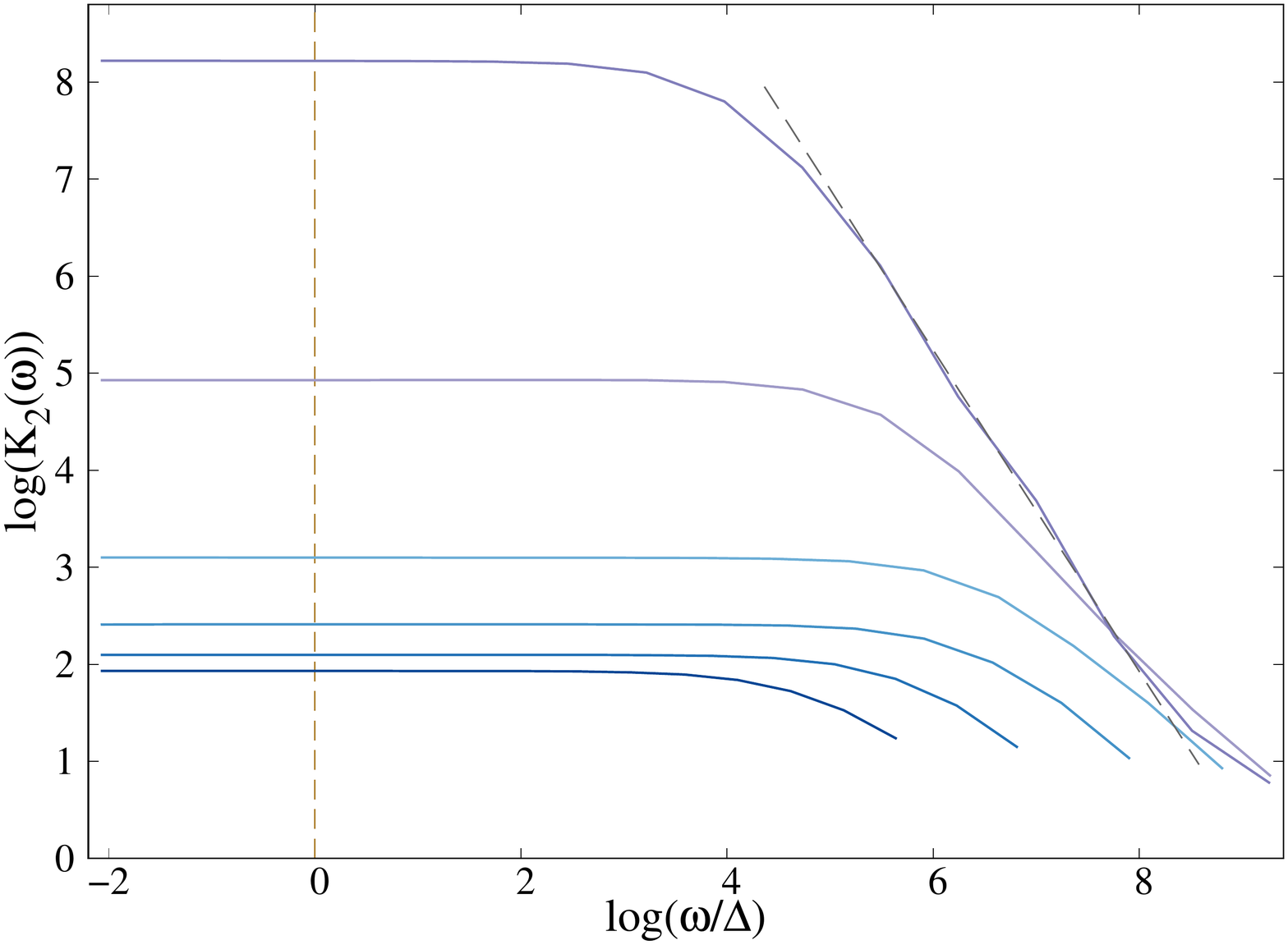} 
	 	\vspace{-0.1cm}
	\caption{(color online) Top panels: Logarithm of the typical DoS, $\langle \log {\rm Im} G \rangle - \log \langle {\rm Im} G \rangle$, as a function of the imaginary regulator $\eta$ divided by the mean level spacing $\Delta = \pi/(N \langle {\rm Im} G (\eta \to 0^+) \rangle)$, for to critical \ER graphs (with $b=0.5$) of size $N=2^n$ (with $n$ from $16$ to $28$ as indicated in the legend), for $\lambda = 2.07$ (top-left panel) and $\lambda = 2.135$ (top-right panel). The dashed straight lines correspond to a fit of the form $\rho_{\rm typ} \propto (\eta / \Delta)^\beta$ ($\beta \approx 0.124$ for $\lambda=2.07$ and $\beta \approx 0.438$ for $\lambda=2.135$ for the largest available system size $N=2^{28}$). The results obtained from the self-consistent solution of the cavity equations~\eqref{eq:green} and~\eqref{eq:greenF} are shown as continuous lines. The dashed lines represent the typical DoS obtained from the approximate treatment of the cavity equations, Eqs.~\eqref{eq:gapprox} and~\eqref{eq:dosapprox}. Bottom panels: Logarithm of  the overlap correlation function versus $\log (\omega / \Delta)$ for critical \ER graphs (with $b=0.5$) of size $N=2^n$, with $n=16, \ldots, 26$ (different colors correspond to different value of $n$). In the bottom-left panel $\lambda=2.07$ and in the bottom-right panel $\lambda=2.135$. The horizontal dashed line in the left panel represents the fully GOE behavior ($K_2(\omega)=1$). Continuous curves show the results obtained from the solution of the self-consistent cavity equations~\eqref{eq:green} and~\eqref{eq:greenF}, while the dashed lines are obtained from the approximate treatment of Eqs.~\eqref{eq:gapprox} and~\eqref{eq:dosapprox}. The dashed straight line in the right panel shows a power law fit of the form $K_2(\omega) \propto \omega^{-\theta}$ 
	(the exponent $\theta$ grows with $N$ and $\theta \approx 1.66$ for the largest available system size $N=2^{26}$).}
\label{fig:imgtyp}
\end{figure*}

It is also instructive to inspect the scaling behavior of the typical value of the LDoS as a function of the system size when the imaginary regulator is varied. In fact, as discussed in Refs.~\cite{kravtsov,facoetti,bogomolny,LRP} in the context of random matrix models, in the putative partially delocalized but non-ergodic phase eigenstates occupy a sub-extensive fraction of the total volume and spread over $N^{D}$ nearby energy levels hybridized by the off-diagonal perturbation. Assuming for simplicity that the mini-bands are locally compact (as in the Gaussian RP model~\cite{kravtsov,facoetti,bogomolny}) the with of the mini-bands, i.e. the Thouless energy $E_{\rm Th} \equiv \Gamma$, is given by the product of the number of sites over which the eigenvectors are delocalized times the typical distance between consecutive levels: $E_{\rm Th} \propto N^D \Delta = N^D/(N \rho)$. At this energy scale the spectral statistics displays a crossover from a behaviour characteristic of standard localized phases to a behaviour similar to the one of standard delocalized phase. AL occurs when the mini-bands' width formally becomes smaller than the mean level spacing, $E_{\rm Th} \sim \Delta$. At this point typically the localization centers are almost unaffected by the off-diagonal hybridization rates. (Conversely, full ergodicity is restored when the Thouless energy becomes of the order of the total spectral bandwidth, $E_{\rm Th} \sim O(1)$.) Hence, the scaling behavior of the local resolvent statistics encodes useful information on the structure of the local spectum and gives direct access to the support set of the mini-bands.

In Fig.~\ref{fig:imgtyp} we plot the logarithm of the typical value of the LDoS, defined as 
\[
\rho_{\rm typ} = e^{\langle \log {\rm Im} G \rangle}/\langle {\rm Im} G \rangle \, .
\]
We have computed $\rho_{\rm typ}$ numerically by solving the self-consistent cavity equations~\eqref{eq:green} and~\eqref{eq:greenF} 
for several values of the regulator $\eta$, for several system sizes $N=2^n$ (with $n$ from $16$ to $28$), and for two values of the energy respectively in the putative partially delocalized but non-ergodic phase ($\lambda = 2.07$, top-left panel) and in the fully localized phase ($\lambda= 2.135$, top-right  panel). The imaginary regulator is measured in units of the mean level spacing $\Delta = 1/(N \rho (N, \lambda)) = \pi / (N \langle {\rm Im} G (\eta \to 0^+)\rangle$). The curves corresponding to different size display a crossover at a well-defined energy scale from a plateau at small $\eta$ and a power-law of the form $\rho_{\rm typ} \propto (\eta / \Delta)^\beta$ at large $\eta$. As explained above the origin of such crossover scale is due to the fact that wave-functions close in energy are hybridized by the off-diagonal perturbation and form mini-bands. When $\eta$ is smaller than the width of the mini-bands $\rho_{\rm typ}$ has a delocalized-like behavior and is independent of the regulator. 
Conversely, when $\eta$ is larger than the energy spreading of the mini-bands one finds a behavior similar to that of the localized phase, where $\rho_{\rm typ}$ grows with $\eta$.

At large energy ($\lambda = 2.135$, top-right panel) the ratio $E_{\rm Th}/\Delta$ and the height of the plateau behave non-monotonically as they first increase for $N < N_c$, and than start to decrease for $N > N_c$. The characteristic size $N_c \approx 2^{23}$ for $\lambda = 2.135$ turns out to be precisely the one highlighted in Sec.~\ref{sec:dos}. At larger $N$ one clearly sees that the ratio $E_{\rm Th}/\Delta$ moves to smaller and smaller values and eventually for $N \gtrsim 2^{26}$ crosses the vertical dashed line, i.e. the Thouless energy becomes smaller than the mean level spacing. Concomitantly, the height of the plateau at small $\eta$ decreases rapidly with the system size. This behavior is fully consistent with that of a fully localized regime.  

At smaller energy, instead ($\lambda = 2.07$ in the putative partially delocalized but non-ergodic phase, top-left panel), the ratio $E_{\rm Th} / \Delta$ moves to larger and larger values as $N$ is increased. 
This behavior is compatible with the presence of mini-bands in the local spectrum, at least for the accessible system size. In the left panel we also show the approximate result for $\rho_{\rm typ}$ obtained using the approximate treatment of the cavity equations, Eqs.~\eqref{eq:gapprox} and~\eqref{eq:dosapprox}. Although this approximation clearly overestimates the typical DoS in the small $\eta$ regime, it captures very accurately the crossover energy scale.

Another insightful probe of the level statistics and of the statistics of wave-functions' amplitudes is provided by the spectral correlation function $K_2 (\omega)$ between eigenstates at different energy, which allows one to distinguish between ergodic, localized, and partially delocalized states~\cite{altshulerK2,mirlin,chalker,kravK2,kravtsov,khay,LRP}:
\begin{equation} \label{eq:K2}
	\begin{aligned}
K_2 (\omega) & = \left \langle \frac{N \sum_i{\rm Im} G_{ii} (\omega/2) \, {\rm Im} G_{ii} (-\omega/2)}{\sum_i{\rm Im} G_{ii} (\omega/2) \sum_i {\rm Im} G_{ii} (-\omega/2)} \right \rangle \, .
\end{aligned}
\end{equation}
For GOE matrices $K_2(\omega) = 1$ identically, independently of $\omega$ on the entire spectral bandwidth. In a standard metallic phase (e.g., in the extended phase of the Anderson tight-binding model in $d \ge 3$) $K_2(\omega)$ has a plateau at small energies, for $\omega < E_{\rm Th}$, followed by a fast-decay which is described by a power-law, with a system-dependent exponent~\cite{chalker}. The height of the plateau is larger than one, which implies an enhancement of correlations compared to the case of independently fluctuating Gaussian wave-functions. The Thouless energy which separates the plateau from the power-law decay stays finite in the thermodynamic limit and extends to larger energies as one goes deeply into the metallic phase, and corresponds to the  energy band over which GOE-like correlations establish~\cite{altshulerK2}. In a partially delocalized but non-ergodic phase the plateau is present only in a narrow energy interval, as $E_{\rm Th}$ shrinks to zero in the thermodynamic limit still staying much larger than the mean level spacing. Beyond $E_{\rm Th}$ eigenfunctions poorly overlap with each other and the statistics is no longer Wigner-Dyson and $K_2(\omega)$ decay to zero~\cite{kravtsov,LRP,khay}. 

Our numerical results are presented in the bottom panels of Fig.~\ref{fig:imgtyp}. The overlap correlation function is computed from the numerical solution of the self-consistent cavity equations 
for several values of the energy separation $\omega$, for several system sizes ($N=2^n$ with $16 \le n \le 26$), and for the same values of $\lambda$ as above (and setting $\eta$ to a very small value, $\eta = 10^{-16}$, much smaller than the mean level spacing). 

At small enough energy ($\lambda = 2.07$ in the putative partially delocalized but non-ergodic phase, bottom-left panel) $K_2(\omega)$ is constant for $\Delta < \omega < E_{\rm Th}$, reflecting the fact that the mini-bands are locally compact, as in the Gaussian RP model~\cite{kravtsov,LRP}. In agreement with the behavior of the typical DoS discussed above, we find that the ratio $E_{\rm Th}/\Delta$ moves to larger and larger values as $N$ is increased. At larger energy separation, $\omega \gg E_{\rm Th}$, eigenfunctions poorly overlap with each other, the statistics is no longer Wigner-Dyson and $K_2(\omega)$ decay fast to very small values. Again, we find that the approximate treatment of the cavity equations, Eqs.~\eqref{eq:gapprox} and~\eqref{eq:dosapprox}, provide a very accurate estimation of the Thouless energy.

In the fully localized phase ($\lambda = 2.135$, bottom-right panel) the ratio $E_{\rm Th}/\Delta$ displays a non-monotonic dependence on $N$, as discussed above. For $N \gg N_c \approx 2^{23}$ the Thouless energy eventually becomes smaller than the mean level spacing and a fully localized behavior is recovered. The plateau at small energy is followed by a fast decrease $K_2(\omega) \propto 1/\omega^{\theta}$.

\section{Statistics of the fluctuation of the largest eigenvalue} \label{sec:lmax}

In this section we analyze the statistics of the fluctuations of the largest (non trivial) eigenvalue of the laplacian of critical \ER graphs whose asymptotic distribution, as mentioned in the introduction and as discussed in Refs.~\cite{KnowlesL,KnowlesE} in great details, is given by a law that does not match with any previously known distribution and does not satisfy the conclusion of the Fisher–Tippett–Gnedenko theorem. (Note that we do not consider here the largest Perron-Frobenius eigenvalue of ${\cal H}$ associated to the flat eigenvector $1/\sqrt{N}(1, \ldots, 1)$, which is an outlier separated from the rest of the spectrum, see e.g. Ref.~\cite{ourselves}).

\begin{figure}
\includegraphics[width=0.48\textwidth]{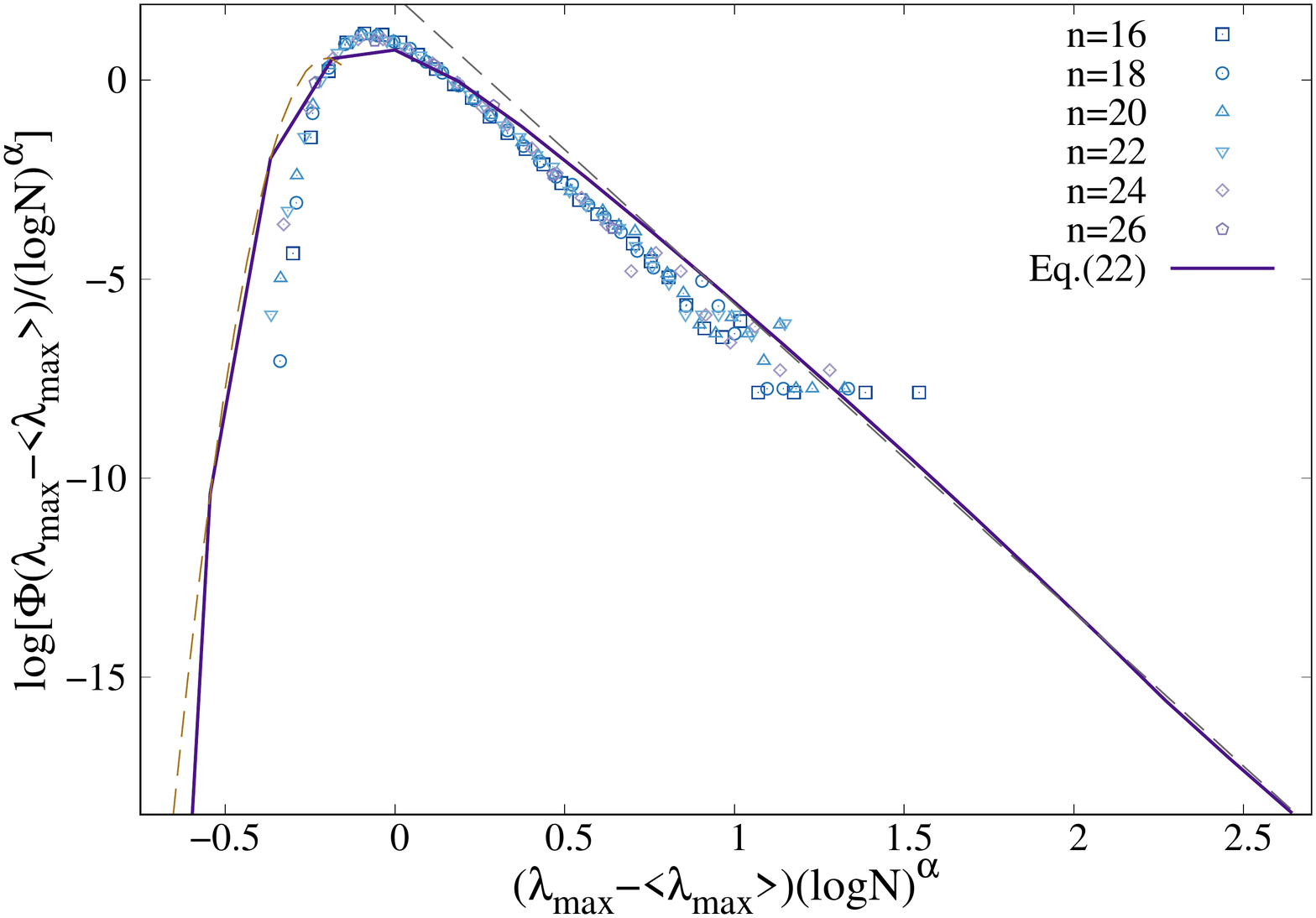}
	 	\vspace{-0.5cm}
	\caption{(color online) Probability distribution of the largest eigenvalue (besides the one associated to the flat eigenvector) of the adjacency matrix of critical \ER graphs with $b=0.5$. Different symbols and colors correspond to different system sizes $N=2^n$ as indicated in the legend. The data corresponding to different $N$ collapse on the same curve when $\lambda_{\rm max} - \langle \lambda_{\rm max} \rangle$ is multiplied by $(\log N)^\alpha$. The best collapse is achieved for $\alpha = 3/4$. The continuous curve correspond to the analytical expression obtained from Eqs.~\eqref{eq:Pkmax} and~\eqref{eq:Plmax} for large $N = 2^{40}$. The right tails of the distribution of the largest eigenvalue are well fitted by an exponential decay (dashed straight line on the positive side), while the left tails are much sharper, and are possibly Gaussian (dashed curve on the negative side).
\label{fig:Plmax}}
\end{figure}

As shown in~\cite{KnowlesL,KnowlesE}, $\lambda_{\rm max}$ corresponds to the largest degree of ${\cal H}$ and its fluctuations can be computed in terms of the fluctuations of the largest value of $N$ i.i.d. Poisson variables of average $c=b \log N$. The probability that $k_{\rm max} = k$ can be esily expressed in terms of the cumulative distribution of the degree probability $R(k) = \sum_{k^\prime = 0}^k P(k^\prime)$: 
\begin{equation} \label{eq:Pkmax}
\Psi(k_{\rm max}) = [R(k_{\rm max})]^N - [R(k_{\rm max}-1)]^N \, .
\end{equation}
By changing variable from $k_{\rm max}$ to $\lambda_{\rm max}$ via the bijection $\lambda_{\rm max} = \Lambda (k_{\rm max}/c)$ one immediately obtains the probability distribution function of the largest eigenvalue:
\begin{equation} \label{eq:Plmax}
\Phi(\lambda_{\rm max}) = c \, \Psi ( c \tilde{\kappa} (\lambda_{\rm max})) \, \tilde{\kappa}^\prime (\lambda_{\rm max}) \, .
\end{equation}
We have computed $\Phi(\lambda_{\rm max})$ for critical \ER graphs (with $b=0.5$) both analytically, using Eqs.~\eqref{eq:Pkmax} and~\eqref{eq:Plmax} for large $N \lesssim 2^{40}$, and numerically, using the Lanczos algorithm for the two largest eigenvalues of the adjacency matrix for $2^{16} \le N \le 2^{26}$.  The results are shown in Fig.~\ref{fig:Plmax}. We empirically find that the data corresponding to different $N$ nicely collapse on the same curve if $\lambda_{\rm max} - \langle \lambda_{\rm max} \rangle$ is multiplied by $(\log N)^\alpha$, with $\alpha = 3/4$. The right tails of the distribution are well represented by an exponential decay, while the left tails are much sharper, although there is no level repulsion with the eigenvectors on the left of $\lambda_{\rm max}$.

\section{Relationship with the out-of-equilibrium phase diagram of the quantum random energy model} \label{sec:qrem}

The QREM, is the quantum  version of  Derrida's  Random  Energy Model~\cite{rem} and provides the simplest toy model of mean-field spin glasses. For $n$ spin-$1/2$s it is defined by the following Hamiltonian:
\begin{equation}
\label{eq:HQREM}
{\cal H}_{\rm qrem} = E(\{ \hat{\sigma}_i^z \}) - \Gamma \sum_{i=1}^n \hat{\sigma}_i^x \, ,
\end{equation}
where $\Gamma$ is the transverse field, and $E(\{ \hat{\sigma}_i^z \})$ is a random operator diagonal in the $\{ \hat{\sigma}_i^z \}$ basis, which takes $2^n$ different values for the $2^n$ configurations of the $n$ spins in the $z$-basis, identically and independently distributed according to:
\[
P(E) = \frac{e^{-E^2/n}}{\sqrt{\pi n}} \, .
\]
With this choice of the scaling, the random many-body energies are with high probability contained in the interval $[-n \sqrt{\log 2}, +n \sqrt{\log 2}]$ in the thermodynamic limit. Hereafter we denote by $\varepsilon = E/n$ the intensive energy per spin corresponding to the extensive energy $E$. 

As discussed above, the QREM can be viewed as the simplest many-body model that displays AL in its Hilbert's space: If one chooses as a basis the tensor product of the simultaneous eigenstates of the operators $\sigma_i^z$, the Hilbert space of the many-body Hamiltonian is a $n$-dimensional hypercube of $N = 2^n$ sites and degree $n$. One can map a configuration of $n$ spins to a corner of the $n$-dimensional hypercube  by  considering $\sigma_i^z = \pm1$ as  the  top/bottom  face  of  the  cube's $i$-th  dimension. The random part of the Hamiltonian is by definition diagonal on this basis, and gives uncorrelated random energies on each site orbital of the hypercube:  At $\Gamma = 0$ the many-body eigenstates of Eq.~(\ref{eq:HQREM}) are simply product states of the form $\vert \sigma_1^z \rangle \otimes \vert \sigma_2^z \rangle \otimes \cdots \otimes \vert \sigma_n^z \rangle$, and the system is fully localized. The interacting part of the Hamiltonian acts as single spin flips on the configurations $\{ \sigma_i^z \}$, and plays the role the hopping rates connecting ``neighboring'' sites in the configuration space. The many-body quantum dynamics is then recast as a single-particle non-interacting tight-binding Anderson model for spinless electrons in a disordered potential living  on the $2^n$ corners of an hypercube in $n$ dimensions (and degree $n$), with the spin configurations being ``lattice sites'',  and the transverse field playing the role of the hopping amplitude between neighboring sites.

The out-of-equilibrium phase diagram of the QREM has been analyzed in great details in several recent papers~\cite{Laumann2014,Baldwin2016,qrem1,qrem2,qrem3,qrem4,qrem5}. At low enough transverse field, the DoS is controlled by the random on-site energies, $\rho_{\rm qrem} (\varepsilon) \simeq P(\varepsilon) = \sqrt{n/\pi} \, e^{-n \varepsilon^2}$, and strongly concentrate around zero energy density in the thermodynamic limit, as naturally expected for many-body systems. Using the same notation as before, one has that for $|\varepsilon|>0$ the DoS scales as $\rho_{\rm qrem} (\varepsilon) \propto N^{\tau(\varepsilon) - 1}$, with $\tau(\varepsilon) = 1 - \varepsilon^2/\log 2$. Hence, the vast majority of the states are found in the bulk of the spectrum that concentrates around $\varepsilon = 0$, while a small sub-extensive fraction of them are in the tails, in the interval $0 < |\varepsilon| < \sqrt{\log 2}$. 

As it is apparent from the analysis of Refs.~\cite{qrem1,qrem2,qrem3,qrem4,qrem5}, in the localized phase the local structure of an eigenvector of the QREM model is similar to that of the critical \ER graph described above: exponentially decaying around well-separated  localization centres associated with resonances of energy $\varepsilon$ of the eigenvector. In the QREM the localization centers arise from exponentially rare vertices with exceptionally large local values of the potential, while in the critical \ER graphs the localization centers correspond to exponentially rare vertices of abnormally large connectivity. The only difference between the two models is the specific geometrical structure of the underlying graph, since the hypercube contains much more short loops compared to the \ER graph. There are in fact $r!$ paths of length $r$ connecting two nodes of the hypercube which correspond to spin configuration that differ by  $r$ spin flips, 
but this can be essentially recast 
as an effective renormalization of the hopping amplitude.

The out-of-equilibrium phase diagram of the QREM is in fact qualitatively identical to the one proposed in Fig.~\ref{fig:PD} for critical \ER graphs~\cite{qrem1,qrem2,qrem3,qrem4,qrem5}: In the bulk of the spectrum, $|\varepsilon| \approx 0$, one finds a fully delocalized GOE-like phase; At very large energy, close to the spectral edges, $|\varepsilon| \in (\varepsilon_{\rm loc}, \sqrt{\log 2})$, one finds a fully Anderson localized phase in which eigenvectors are exponentially localized around a single resonance. Finally, at intermediate energies, $\varepsilon \in (0,\varepsilon_{\rm loc})$ one has an intermediate partially delocalized but non-ergodic phase in which distant localization centers on the hypercube partially hybridize due to the exponentially small tunneling rates between them, thereby producing multifractal eigenfunctions  which occupy a diverging volume, yet an exponentially vanishing fraction of the total Hilbert space, with $0<D_q<\tau(q)$. 

\section{Conclusions and perspectives} \label{sec:conclusions}

In this paper we have analyzed both analytically and numerically the spectral properties of the tails of the spectrum of the adjacency matrix of critical \ER graphs, i.e. when the average degree is of the order of the logarithm of the number of vertices. 

In a series of recent inspiring papers Alt, Ducatez, and Knowles have rigorously shown that these systems exhibit a ``semilocalized'' phase in the tails of the spectrum where the eigenvectors are exponentially localized on a sub-extensive fraction of nodes with anomalously large degree~\cite{Knowles,KnowlesL,Knowles1}. We have proposed two approximate analytical treatments to analyze this regime. The first is based on simple rules of thumb for localization and ergodicity, often referred to as the Mott's criteria for localization and ergodicity. The second approach  relies on an approximate treatment of the self-consistent cavity equation for the resolvent. Both approaches indicate that the semilocalized phase splits in fact in two different phases separated by a mobility edge. At large energy, close to the spectral edges, as already rigorously proven in~\cite{KnowlesL}, one finds a fully Anderson localized phase in which the eigenvectors are localized around a {\it unique} localization center and the statistics of the eiganvalues is described by the Poisson statistics. At intermediate energy, sandwiched between the fully delocalized GOE-like phase in the bulk, and the Anderson localized phase at the edges, we find a partially delocalized but non-ergodic phase, in which the eigenstates spread over {\it many} resonant localization centers close in energy due to the hybridization of the exponentially decaying part of the wave-functions. In this regime the exponentially small tunneling amplitudes between far away localization centers is counterbalanced by the number of localization centers towards which tunneling can occur, and the system exhibits mini-bands in the local spectrum. The level statistics is therefore of Wigner-Dyson type up to an energy scale which is much smaller than $1$ but stays much larger than the mean level spacing. 

We have presented a numerical study of the finite size scaling behavior of several observables related to the spectral statistics that supports the theoretical predictions and  allows us to characterize the critical properties of the two transitions: The transition from the fully delocalized phase to the semilocalized one is accompanied by a correlation volume that diverges exponentially fast when the transition point is approached from above, $|\lambda| \to 2^+$, with an exponent $\nu_{\rm GOE} \approx 0.5$. The transition from Poisson to Wigner-Dyson statistics occurring at the AL threshold is instead associated to an exponent $\nu_{\rm loc} \approx 1$. This analysis also highlights the differences with respect to  the standard AL on sparse random graphs induced by the disorder in the local potential. In fact in this case it is well established that the critical point belongs to the localized phase~\cite{mirlinrrg,efetov,efetov1,tikhonov2019,fyod,Zirn,Verb}, while in the present case the critical point is described by the Wigner-Dyson statistics, as also found in random matrix models of the RP type which feature an intermediate partially delocalized but non-ergodic phase~\cite{pino,LRP}.

Finally, we have characterized the statistics of the fluctuations of the largest eigenvalue, which are essentially controlled by the fluctuations of the largest degree in the network.

Since critical \ER graphs  provide an idealized representation of the topological features of the Hilbert space of generic interacting many-body systems, we believe the the results presented here might give new insights on the understanding of the mechanisms that produce localized and multifractal wave-functions even in more complex settings. In fact we put forward a direct correspondence between the phase diagram of critical \ER graphs and the out-of-equilibrium phase diagram of the QREM, which is the simplest model featuring a many-body localization transition. In this respect, it might be useful to generalize the approximate treatment of the cavity equations proposed in Sec.~\ref{sec:cavity} to similar situations in which wave-functions are localized around many resonant nodes, such as, for instance, in the QREM. 

Several important questions remain of course still open. The most important one is probably related to the possibility that the putative delocalized but non-ergodic phase is only a finite-size crossover and eventually disappears in the thermodynamic limit. In fact the estimation of the mobility edge based on the Mott criterion does not take into account neither the effect of the loops nor of higher order terms in the perturbative expansion, while the approximate treatment of the cavity equations is also based on a quite drastic simplification in which the local fluctuations of the degree are completely neglected. The finite-size scaling analysis of the observables related to the spectral statistics presented in Fig.~\ref{fig:level_statistics} is limited to too small sizes to rule out definitely this possibility. One might therefore wonder whether for very large sizes, i.e. $N \gg N_c (\lambda)$, eventually all the eigenvectors in the tails of the spectrum become fully localized. A similar crossover occurs for instance in the tight-binding Anderson model on random-regular graphs, where the existence of a genuine delocalized but non-ergodic phase in the infinite size limit has been the subject of an intense debate in the latest years and has been strongly questioned by recent works~\cite{noi,scardicchio1,ioffe1,ioffe3,bera2018,detomasi2020,refael2019,pinorrg,mirlinrrg,gabriel,tikhonov2019,large_deviations,Levy,mirlinreview,metz}. Another important aspect concerns the structure of the fractal states. In fact, throughout this paper we have assumed that the partially delocalized but non-ergodic phase is analogous to the one found in RP-type models with uncorrelated entries~\cite{kravtsov,kravtsov1,khay,LRP,facoetti,bogomolny}. On the other hand, there are several correlated random matrix models~\cite{kutlin,motamarri,tang} in which the structure of the fractal states is quite different and do not feature, for instance, the formation of mini-bands within which the Wigner-Dyson statistics is established. In the tails of critical \ER graphs the energies of the localization centers of a given degree (which depend mostly on the degrees of the neighbors~\cite{Knowles1,Knowles}) and the effective tunneling amplitudes between them (which depend mostly on their distances) are essentially uncorrelated. It is therefor natural to assume that RP models with iid elements provide the correct physical picture for the partially delocalized but non-ergodic wave-functions. Yet, it would be highly desirable to put our conclusions on a firmer and more rigorous ground and to provide more stringent numerical tests of the existence of the partially extended but non-ergodic phase and of its nature.

Another important open question is related to the critical behavior for $b > b_{\rm loc}$. Indeed our analysis indicates that the intermediate partially delocalized but non-ergodic phase only exists provided that $b$ is smaller than $b_{\rm loc} = b_\star/2$ (see Fig.~\ref{fig:PD}), while for $b \in (b_{\rm loc}, b_\star)$ one should observe a direct transition at $|\lambda| = 2$ from the fully delocalized phase in the bulk of the spectrum to a fully Anderson localized phase in the tails in which eigenvectors are exponentially localized around a unique localization node. In this case the critical properties of such transition might be different compared to the one observed at $b=0.5$ and discussed in Sec.~\ref{sec:numerics}, and it is natural to wonder whether at large $b$ one recovers the standard critical behavior of AL on sparse random graphs induced by a quenched random potential.

The limit $b \to 0$ is also puzzling for two reasons, and deserves special attention. On the one hand, the exponents $\tau$ and $D$, which are predicted to exhibit a finite jump from $1$ for $|\lambda|=2$ to $1 - b/b_\star$ and $1 - 2 b / b_\star$ respectively for $|\lambda| \to 2^+$, behave continuously at the transition from the fully delocalized GOE-like phase to the partially extended but non-ergodic phase for $b \to 0$, which could result in a modification of the critical properties compared to the $b>0$ case. On the other hand, we know from previous studies~\cite{rodgers88,biroli99,semerjian02} that for $c$ arbitrarily large but finite the spectrum of \ER graphs is characterized by Lifshitz tails due to extremely rare fluctuations of the local degrees associated to fully localized eigenvectors, which, however, does not match with the $b \to 0$ limit of the phase diagram of Fig.~\ref{fig:PD}. 

Possibly the most interesting perspective for future work is  to study how  the addition of some amount of disorder in the local potential affects the spectral properties of critical \ER graphs. On the one hand one might expect that  quenched on-site randomness might destabilize the partially delocalized phase by suppressing the effective tunneling rates between the far-away localization centers. On the other hand, since the Anderson tight-binding model on random graphs of fixed connectivity is already at the brink of developing a delocalized but non-ergodic phase~\cite{kravtsov1,khay,noi,scardicchio1,ioffe1,ioffe3,bera2018,detomasi2020,refael2019,pinorrg,mirlinrrg,gabriel,tikhonov2019,large_deviations,Levy,mirlinreview,metz}, the addition of strong fluctuations of the local degrees might in fact favour the formation of multifractal wave-functions. 

\begin{acknowledgments}
	I would like to warmly thank  I. M. Khaymovich, J. Alt, R. Ducatez, and A. Knowles for many enlightening and helpful discussions.
\end{acknowledgments}

\appendix

\section{Upper bound on the position of the mobility edge} \label{app}

In this appendix we revise the rules of thumb criteria for localization and ergodicity discussed in Sec.~\ref{sec:mott} attempting to provide an upper bound on the position of the mobility edge and for the support set of the mini-bands. 

In fact the matrix elements between the localization centers, Eq.~\eqref{eq:Gfsa}, decay exponentially fast with the distance with a very high rate. One might then argue that the amplitudes ${\cal G}_r (\lambda)$ are dominated by the pairs of closest resonant localization centers. 
In the following we repeat the reasoning of Sec.~\ref{sec:mott} but, instead of using the typical distance between all pairs of localization nodes, $r_{\rm typ} = \log N / \log c$, to estimate the typical value of the tunneling rates, we take instead the minimal distance between pairs of nearby localization centers. This should provide a upper bound for $|{\cal G}_r (\lambda)|$ and hence for the position of the mobility edge as well as for the exponent $D$.

In order to compute the minimal distance between pairs of localization centers at a given energy $\lambda$ we start by evaluating the probability $P_\lambda (r)$ to find a localization center (at energy $\lambda$) at distance $r$ from a given localization center (at the same energy) located at the origin. This is given by the probability that a ball of radius $r$ around the origin (which roughly contains $k c^{r-1}$ vertices) does not contain a localization center of energy $\lambda$ times the probability to find one localization center exactly at distance $r$ from the origin:
\[
\begin{aligned}
P_\lambda (r) & \approx \rho(\lambda) k c^{r-1} \left ( 1 - \rho(\lambda) \right)^{k c^{r-1}} \\
& \approx \rho(\lambda) \tilde{\kappa} (\lambda) c^r e^{- \tilde{\kappa} (\lambda) \rho(\lambda) c^r} \, ,
\end{aligned}
\]
where $k = c \tilde{\kappa} (\lambda)$ is the connectivity of the localization nodes which give rise to eigenstates of energy $\lambda$. From the expression above one immediately obtains the typical value of the distance between the two closest localization centers at energy $\lambda$ as:
\begin{equation} \label{eq:rtyp}
r_{\rm min} (\lambda) \approx - \frac{\log \left[ \tilde{\kappa} (\lambda) \rho(\lambda) \right]} {\log c} \, . 
\end{equation}
Hence, for critical \ER graphs $r_{\rm min}$ scales in the thermodynamic limit as $r_{\rm min} (\lambda) \propto (1 - \tau(\lambda)) \log N / \log (b \log N)$ minus a small $\lambda$-dependent correction proportional to $\log \tilde{\kappa} (\lambda)/\log (b \log N)$. The origin of this correction comes from the fact that the volume of a ball of radius $r$ around a localization center increases with the energy $\lambda$ (i.e. the degree $c \tilde{\kappa} (\lambda)$ of the node). Note that for $\rho \to 0$ one has that $r_{\rm min}$ coincides with the radius of the \ER graph $r_{\rm typ} = \log N / \log c$, while for $\rho \to 1$ the minimal distance becomes of order $1$, as expected.

\begin{figure*}
 \hspace{-0.22cm}\includegraphics[width=0.341\textwidth]{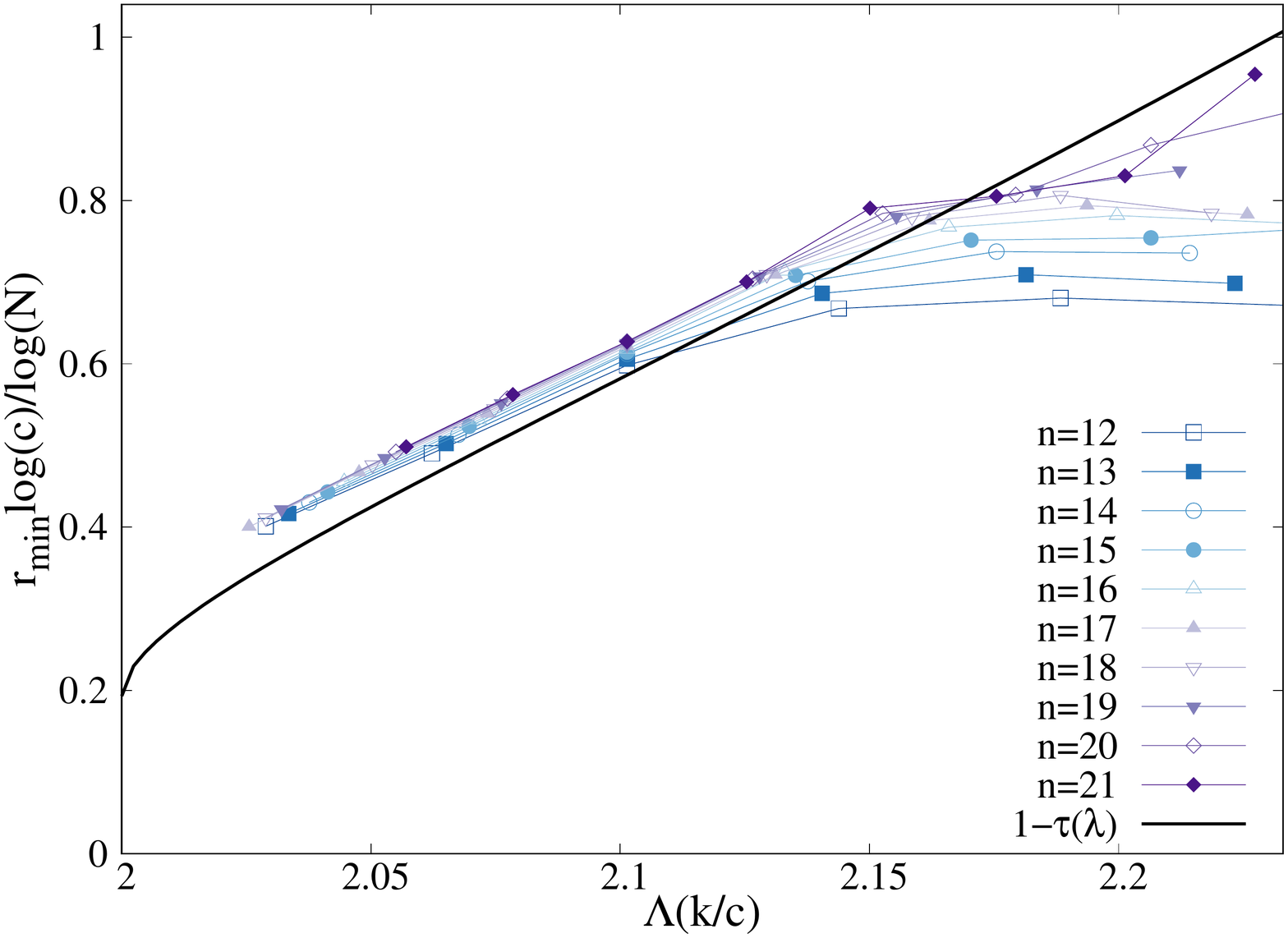}\hspace{-.3cm}
	 \includegraphics[width=0.341\textwidth]{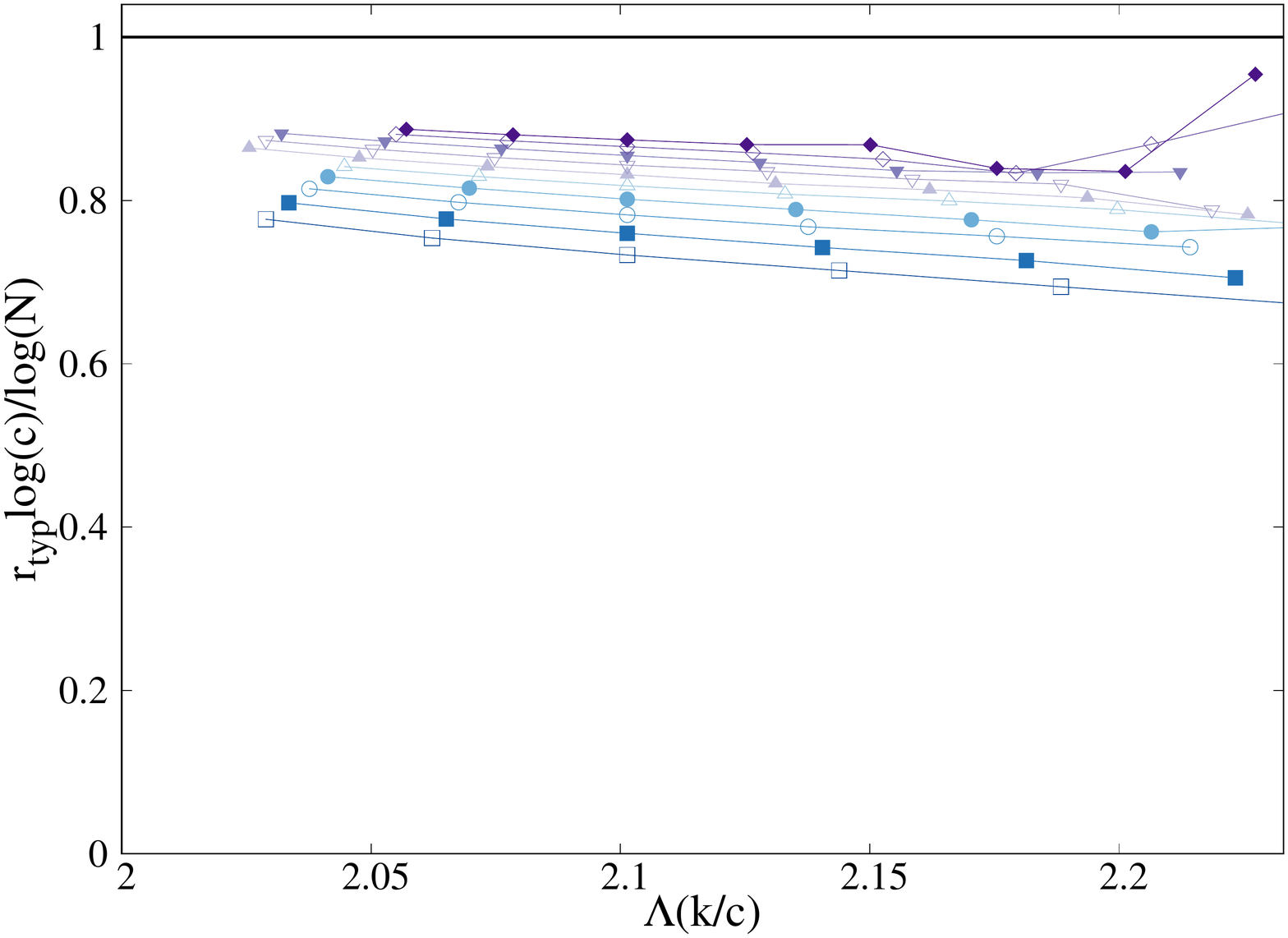} \hspace{-.3cm}
	 \includegraphics[width=0.341\textwidth]{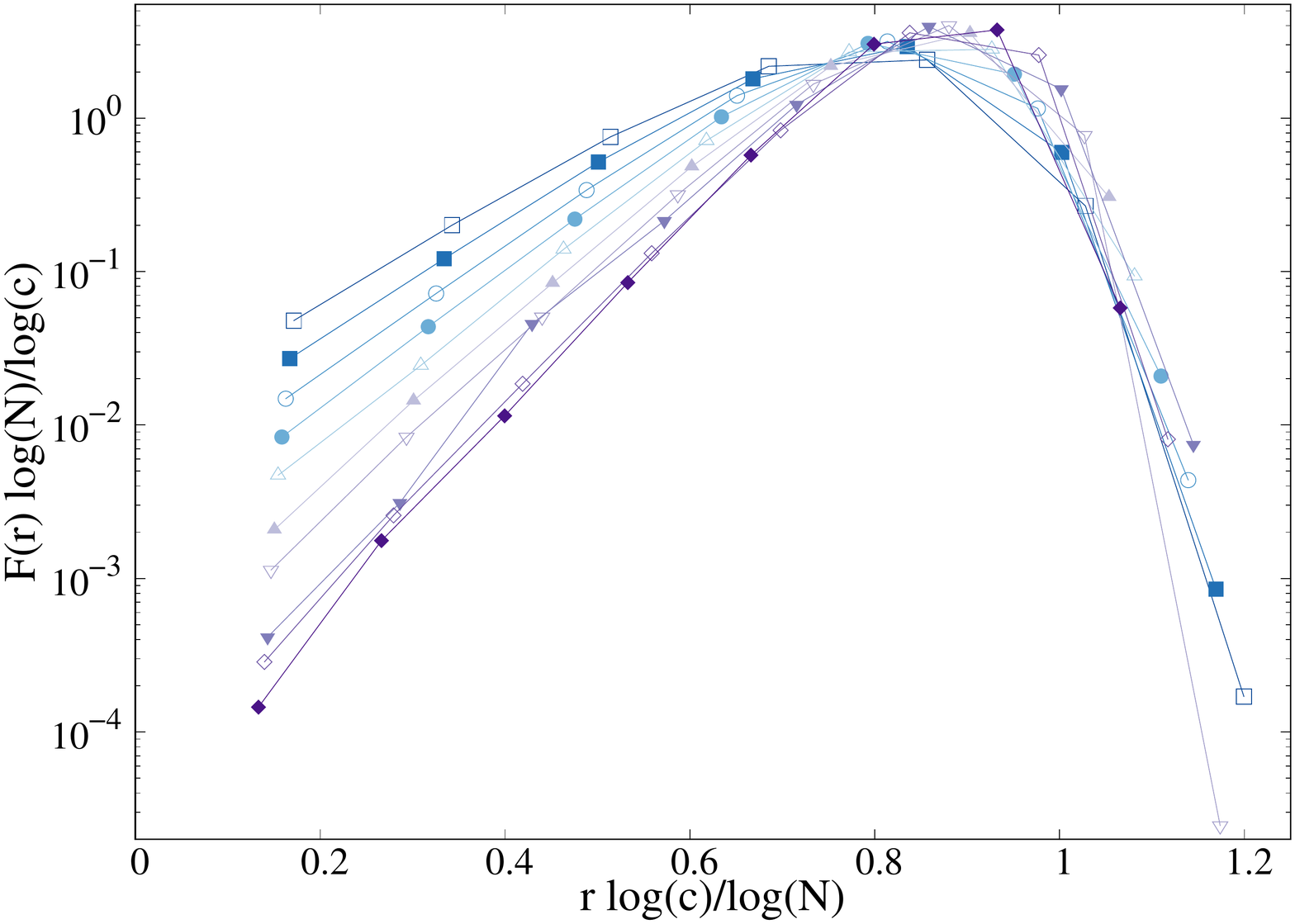}
	 	\vspace{-0.2cm}
	\caption{(color online) Scaling behavior of the distance between localization centers for critical \ER graphs with $b=0.5$. In the left panel we plot the typical value of the minimal distance between pairs of resonant localization centers of degree $k>2c$. The data corresponding to different system sizes (as indicated in the legend) are divided by $\log N/\log c$ and are plotted as a function of the energy of the corresponding eigenstates $\Lambda(k/c)$. The solid line correspond to $1 - \tau (\lambda)$, in agreement with the scaling given in Eq.~\eqref{eq:rtyp}. In the middle panel we plot the typical distance between all pairs of resonant localization centers of degree $k>2c$ divided by $\log N/\log c$ as a function of the energy of the corresponding eigenstates $\Lambda(k/c)$ for the same values of $N$ as before. Similar results are found for different values of $b$. In the right panel we plot the probability distributions of the distances between pairs of localization nodes of fixed energies. In particular we consider vertices of degree $n$ within graphs of size $N=2^{n}$, giving rise asymptotically to eigenvalues of energy $|\lambda| \approx 2.1014$ for $b=0.5$. The distance $r$ is rescaled by $\log N/\log c$.
\label{fig:RTYP}}
\end{figure*}

In order to check that the scaling obtained in Eq.~\eqref{eq:rtyp} is correct, we have computed numerically the typical value of the minimal distance between localization centers for critical \ER graphs with $b=0.5$ as explained below. We generate random instances of the adjacency matrix according to the probability distribution~\eqref{eq:H} and considered all the nodes of abnormally large connectivity $k > 2 c$. For any given node of connectivity $k$ (corresponding to a localization center of energy $\lambda = \Lambda( k /c)$), we measure the distance from its closest localization center of the same degree, and average this distance over all the nodes of degree $k$ and over different random realizations of the graph. The results are reported in the left panel of Fig.~\ref{fig:RTYP}, where this distance is plotted as a function of $\lambda = \Lambda(k/c)$. We see that the points corresponding to different sizes of the graph $n = \log_2 N$ collapse on the same curve corresponding to $1 - \tau(\lambda)$ when rescaled by the factor $\log N / \log c$, in agreement with Eq.~\eqref{eq:rtyp}. (Small deviations are observed for the smallest sizes at large $\lambda$, as explained above.) In the middle panel we also plot the average distance between all pairs of nodes of degree $k$ as a function of $\lambda$ and for several values of $N$. We see that the data points corresponding to different sizes approach $1$ when rescaled by the factor $\log N / \log c$, as expected. The origin of the finite-size corrections can be again understood recalling that nodes with abnormally large degree have $\tilde{\kappa}(\lambda)$ more neighbors at a given distance than the nodes with degree of order $c$. Finally, in the right panel of Fig.~\ref{fig:RTYP} we plot the whole probability distributions $F(r)$ of the distance between pairs of localization nodes of fixed energy for several sizes of the graph. In particular we focus on vertices of degree $n$ found within graphs of $N=2^n$ nodes. Via the bijection~\eqref{eq:Lambda} their energy corresponds asymptotically to $|\lambda| = \Lambda ( ( b \log 2)^{-1}) \approx 2.1014$ for $b=0.5$. One clearly observes that, upon rescaling the distance $r$ by $\log N / \log c$, the distributions are peaked around $1$, as expected, and become more narrow as $N$ is increased.

Inserting now the estimation of $r_{\rm min}$~\eqref{eq:rtyp} into Eq.~\eqref{eq:Gfsa}, the Mott criterion for full localization around a unique vertex yields $|G_{r_{\rm min} (\lambda)}(\lambda)| < (N \rho(\lambda))^{-1}$, i.e.
\[
\begin{aligned}
N^{\left[ \tau(\lambda) - (1 - \tau(\lambda)) \left( \frac{1}{2} + \frac{\log \lambda}{\log c}\right) \right] } < 0 \, .
\end{aligned}
\]
In the thermodynamic limit (and in the critical regime, $c = b \log N$) this condition is only fulfilled provided that $\tau (\lambda) < 1/3$. Using the asymptotic expression for the exponent $\tau$ given in Eq.~\eqref{eq:dos}, one then obtains a modified implicit equation for the mobility edge:
\begin{equation} \label{eq:loc_bound}
\tilde{\kappa} (\tilde{\lambda}_{\rm loc}) \left[ \log \tilde{\kappa} (\tilde{\lambda}_{\rm loc}) - 1 \right] = \frac{2}{3 b} - 1\, .
\end{equation}
Since $\tau (\lambda)$ is a decreasing function of $\lambda$ which tends to $1 - b / b_\star$ for $|\lambda| \to 2^+$, the existence of the partially delocalized but non-ergodic phase is only possible if $b < \tilde{b}_{\rm loc} = 2 b_\star/3 = 1/(\log 8 - 3/2)$. 

At this point one can proceed further and compute the escape rate of a particle sitting on a localization center using $r_{\rm min}$ instead of $r_{\rm typ}$ in the expression of the transition rate, and compare it to the spectral bandwidth at the same energy. The Fermi Golden Rule gives:
\[
\Gamma (\lambda) \approx 2 \pi N \rho (\lambda) |{\cal G}_{r_{\rm min} (\lambda)}(\lambda)|^2 \propto N^{2 \tau (\lambda) - 1} \, .
\]
Assuming again for simplicity that this energy scale coincides with the number of hybridized states within a mini-band times the mean level spacing, one obtains a direct estimation of the fractal exponent $D$ as $\Delta N^{D} \propto \Gamma$, with $\Delta = 1/(N \rho)$, yielding $D = 3 \tau - 1$ (note that $D=0$ at the localization threshold where $\tau(\tilde{\lambda}_{\rm loc}) = 1/3$).

Since the number of localization centers is at most equal to $N \rho$, one has that $D$ is at most equal to $\tau$, and $D (\lambda) = {\rm min} \{ \tau (\lambda), 3 \tau(\lambda) - 1 \}$. Hence, this argument predicts the existence of another transition at an energy $\lambda_{\rm ergo}$ within the tails of the spectrum from a phase, for $|\lambda| \in (2,\lambda_{\rm ergo}]$, where the wave-functions spread uniformly over {\it all} the $N^{\tau}$ localization centers (i.e. $D = \tau$), to a non-ergodic phase, for $|\lambda| \in (\lambda_{\rm ergo},\tilde{\lambda}_{\rm loc})$, in which the wave-functions only occupy a small fraction of the $N^\tau$ localization centers at that energy (i.e. $D$ is strictly smaller than $\tau$). The implicit equation for $\lambda_{\rm ergo}$ is given by the condition $\tau (\lambda_{\rm ergo}) = 1/2$, which is in fact the same condition that we obtained for the mobility edge when we used the typical distance between pairs of localization centers to evaluate the transition amplitudes, Eq.~\eqref{eq:loc}. Finally, this argument gives the following upper bound for the fractal exponent $D$ in the thermodynamic limit:
\[
D (\lambda) = \left \{
\begin{array}{ll}
\tau(\lambda) & \textrm{for~} 2 < |\lambda| < \lambda_{\rm ergo} \, , \\
3 \tau(\lambda) - 1 & \textrm{for~} \lambda_{\rm ergo} < |\lambda| < \tilde{\lambda}_{\rm loc} \, , \\
0 & \textrm{for~} |\lambda| > \tilde{\lambda}_{\rm loc} \, .
\end{array}
\right.
\]


\begin{thebibliography}{99}
\bibitem{sda} Note that this idea has also been exploited in the physics literature to characterise the Lifshitz tails of \ER graphs when the average degree $c$ is of order 1 in the context of the SDA~\cite{biroli99,semerjian02}.
\bibitem{notaRP} Perhaps the best example of that is provided by the generalization of the Rosenzweig-Porter random matrix model of Ref.~\cite{kravtsov} (see also Ref.~\cite{LRP}), where there is a whole region of the parameter space where wave-functions are delocalized but multifractal and strongly correlated, while the statistics of neighboring gaps is still described by the GOE ensemble. In this case one numerically finds that $\langle q \rangle$ converges to its GOE universal value $2/ \pi$ irrespective of the fact that wave-functions amplitudes are not uncorrelated Gaussian random variables of variance $1/N$.
\bibitem{remark} We have checked that computing the discrete logarithmic derivatives using $3$ or $7$ points does not modify significantly the results.
\end{thebibliography}
\end{document}